\documentclass[twocolumn,journal]{IEEEtran}

\IEEEoverridecommandlockouts

\usepackage{cite}
\usepackage{amsmath,amssymb,amsfonts}
\usepackage{algorithmic}
\usepackage{graphicx}
\usepackage{textcomp}
\usepackage{xr}
\def\BibTeX{{\rm B\kern-.05em{\sc i\kern-.025em b}\kern-.08em
    T\kern-.1667em\lower.7ex\hbox{E}\kern-.125emX}}

\usepackage{tikz}
\usetikzlibrary{spy}
\usetikzlibrary{calc,arrows.meta}
\usepackage{pgfplots}
\usetikzlibrary{plotmarks}
\usetikzlibrary{arrows.meta}
\usetikzlibrary{positioning,fit}
\usetikzlibrary{intersections}
\usetikzlibrary{patterns}
\usetikzlibrary{decorations.shapes}

\usepgfplotslibrary{patchplots}
\usepgfplotslibrary{colormaps}

\usepgfplotslibrary{fillbetween}

\usepackage{cancel}

\usepackage{pdfpages} %

\usepackage{multirow}
\usepackage{makecell}

\usepackage{algorithmic}
\usepackage[ruled,vlined]{algorithm2e}
\SetKwRepeat{Do}{do}{while}

\pgfplotsset{compat=1.16}
\newcounter{isum}
\pgfplotsset{summand/.initial=max}
\pgfmathdeclarefunction{sum}{4}{%
\begingroup%
\pgfkeys{/pgf/fpu,/pgf/fpu/output format=fixed}%
\edef\myfun{\pgfkeysvalueof{/pgfplots/summand}}%
\pgfmathsetmacro{\mysum}{0}%
\pgfmathsetmacro{\myx}{#1}%
\pgfmathsetmacro{\mystart}{#2}%
\pgfmathsetmacro{\mystep}{#3}%
\pgfmathtruncatemacro{\imax}{#4-1}%
\setcounter{isum}{0}%
\loop
\pgfmathsetmacro{\mysum}{\mysum+\myfun(\myx,\mystart+\value{isum}*\mystep)}%
\ifnum\value{isum}<\imax\relax
\stepcounter{isum}\repeat
\pgfmathparse{\mysum}%
\pgfmathsmuggle\pgfmathresult\endgroup%
}%

\pgfdeclarelayer{back1}
\pgfdeclarelayer{lay1}
\pgfdeclarelayer{back2}
\pgfdeclarelayer{lay2}
\pgfdeclarelayer{front1}
\pgfsetlayers{back2,lay2,back1,lay1,main,front1}

\newcommand{\exportFigures}{true}
\newcommand{\exportFiguresAsPNG}{false}

\ifthenelse{\equal{\exportFigures}{true}}
{
  \usepgfplotslibrary{external}
  \tikzexternalize[prefix=compiled_tikz_figures/,optimize command away=\includepdf]
  \ifthenelse{\equal{\exportFiguresAsPNG}{true}}
  {
    \tikzset
    {   png export/.style={
        external/system call={
        pdflatex \tikzexternalcheckshellescape -halt-on-error --extra-mem-top=10000000 -interaction=batchmode -jobname "\image" "\texsource" && pdftops -eps "\image.pdf" && convert -density 700 -transparent white "\image.pdf" "\image.png"
    }}}
  \tikzset{png export}
  }
  {}
}
{}

\usepackage{url}

\usepackage{bm}
\usepackage[caption=false,font=footnotesize]{subfig}
\usepackage[normalem]{ulem}

\usepackage{mathtools, cuted}
\usepackage{acronym}

\usepackage{booktabs}

\usepackage[latin1]{inputenc}
\usepackage{tikz}
\usetikzlibrary{shapes,arrows}
\usetikzlibrary{arrows.meta}
\usetikzlibrary{positioning}

\usepackage{ellipsis}
\usetikzlibrary{calc}
\usetikzlibrary{decorations.pathreplacing,decorations.markings,shapes.geometric}
\usetikzlibrary{decorations.pathmorphing}
\usetikzlibrary{fit}
\usetikzlibrary{pgfplots.groupplots}

\usepackage[most]{tcolorbox}
\tcbuselibrary{breakable}
\tcbset{every box/.style={enhanced,breakable}}
\tcbset{colframe=black,colback=red!10,enhanced,breakable,sharp corners}

\usepackage{enumitem} %

\usepackage{notation}

\usepackage{adjustbox}

\definecolor{alex}{RGB}{51,183,150}
\definecolor{erik}{RGB}{235,134,52}

\newcommand{\ticked}{$\text{\rlap{$\checkmark$}}\square$}
\newcommand{\unticked}{{$\square$}}
\newcommand{\tick}[1]{\ifthenelse{#1=1}{\ticked}{\unticked}}

\newcommand{\rmv}{\hspace*{-.3mm}}
\renewcommand{\Re}[1]{\ensuremath{\text{Re}\!\left[#1\right]}}%
\renewcommand{\Im}[1]{\ensuremath{\text{Im}\!\left[#1\right]}}%

\newcommand{\norm}[2]{\ensuremath{\lVert #1 \rVert^{#2}}}%

\newcommand{\minus}{\rmv - \rmv}

\newcommand{\s}{\hspace*{0.5pt}}

\newcommand{\pe}{p_{\text{E}}(u_n^{(j)}\rmv\rmv\rmv,q_n^{(j)})}
\newcommand{\perv}{p_{\text{E}}(\rv{u}_n^{(j)}\rmv\rmv\rmv, \rv{q}_n^{(j)})}

\newcommand{\pdrv}{p_{\text{D}}(\rv{u}_n^{(j)})}

\providecommand{\norm}[1]{\lVert#1\rVert}

\newcommand{\ist}{\hspace*{.3mm}}

\newcommand{\zd}{\ensuremath{{z_\mathrm{d}^{(j)}}_{\rmv\rmv\rmv\rmv\rmv\rmv\rmv  n,m}}}
\newcommand{\zu}{\ensuremath{{z_\mathrm{u}^{(j)}}_{\rmv\rmv\rmv\rmv\rmv\rmv\rmv n,m}}}
\newcommand{\zdr}{\ensuremath{{\rv{z}_\mathrm{d}^{(j)}}_{\rmv\rmv\rmv\rmv\rmv\rmv\rmv  n,m}}}
\newcommand{\zur}{\ensuremath{{\rv{z}_\mathrm{u}^{(j)}}_{\rmv\rmv\rmv\rmv\rmv\rmv\rmv n,m}}}
\newcommand{\zdOne}{\ensuremath{{z_\mathrm{d}^{(1)}}_{\rmv\rmv\rmv\rmv\rmv\rmv\rmv  n,m}}}

\newcommand{\zdZero}{\ensuremath{{z_\mathrm{d}^{(j)}}_{\rmv\rmv\rmv\rmv\rmv\rmv\rmv  0,m}}}
\newcommand{\zuZeroMax}{\ensuremath{{{z}_\mathrm{u}^{(j)}}_{\rmv\rmv\rmv\rmv\rmv\rmv\rmv 0, \text{max}}}}

\newcommand{\sdnrp}{\ensuremath{\omega}}
\newcommand{\sdnr}[1]{\ensuremath{\omega_n^{(j)\s #1}}}
\newcommand{\sdnrr}[1]{\ensuremath{\rv{\omega}_n^{(j)\s #1}}}
\newcommand{\sdnrt}[1]{\ensuremath{\tilde{\omega}_n^{(j)\s #1}}}

\newcommand{\noise}{{w}}

\mathchardef\Re="023C
\mathchardef\Im="023D

\newlength{\figureheight}
\newlength{\figurewidth}
\graphicspath{{./figures/}}

\definecolor{mycolor01}{rgb}{0.00000,0.00000,1.00000}
\definecolor{mycolor02}{rgb}{0.133,0.545,0.133}
\definecolor{mycolor03}{rgb}{0.50000,0.00000,0.50000}
\definecolor{mycolor04}{rgb}{1.00000,0.83984,0.00000}
\definecolor{mycolor05}{rgb}{0.92969,0.50781,0.92969}
\definecolor{mycolor06}{rgb}{1.00000,0.64453,0.00000}
\definecolor{mycolor07}{rgb}{0.50000,0.50000,0.50000}
\definecolor{mycolor08}{rgb}{1.00000,0.00000,0.00000}
\definecolor{mycolor09}{rgb}{0.2510 ,0.8784, 0.8157}
\definecolor{mycolor10}{rgb}{0.54297,0.00000,0.00000}
\definecolor{mycolor11}{rgb}{0.6445 , 0.1641,0.1641}

\tikzset{
	every picture/.style={solid} %
}

\makeatletter
  
\tikzset{
  nomorepostactions/.code={\let\tikz@postactions=\pgfutil@empty},
  decmark/.style 2 args={decoration={markings,
    mark= between positions 0 and 1 step (1/6)*\pgfdecoratedpathlength with{%
        \tikzset{#2,every mark}\tikz@options
        \pgftransformresetnontranslations
        \pgfuseplotmark{#1}%
      },  
    },
    postaction={decorate},
    /pgfplots/legend image post style={
        mark=#1,mark options={#2},every path/.append style={nomorepostactions}
    },
  },
  markbeginend/.style 2 args={decoration={markings,
		mark= between positions 0 and 1 step (1)*\pgfdecoratedpathlength with{%
			\tikzset{#2,every mark}\tikz@options
			\pgfuseplotmark{#1}%
		},  
	},
	postaction={decorate},
	/pgfplots/legend image post style={
		mark=#1,mark options={#2},every path/.append style={nomorepostactions}
	},
  },
  markend/.style 2 args={decoration={markings,
		mark= at position \pgfdecoratedpathlength with{%
			\tikzset{#2,every mark}\tikz@options
			\pgfuseplotmark{#1}%
		},  
	},
	postaction={decorate},
	/pgfplots/legend image post style={
		mark=#1,mark options={#2},every path/.append style={nomorepostactions}
	},
  },
  posmark/.style 2 args={decoration={markings,
		mark= at position #2 with{%
			\tikzset{solid,every mark}\tikz@options
			\pgftransformresetnontranslations
			\pgfuseplotmark{#1}%
		},  
	},
	postaction={decorate},
	/pgfplots/legend image post style={
		mark=#1,mark options={solid},every path/.append style={nomorepostactions}
	},
  },
}

\makeatother

\pgfplotsset{
resultStyle1/.style={mark=none, line width=0.5pt, mycolor01, decmark={oplus}{solid}},
resultStyle2/.style={mark=none, line width=0.5pt, mycolor02, decmark={+}{solid}},%
resultStyle3/.style={mark=none ,line width=0.5pt, mycolor03, decmark={triangle}{solid}},
resultStyle4/.style={mark=none, line width=0.5pt, mycolor06, decmark={star}{solid}},
resultStyle5/.style={mark=none, line width=0.5pt, mycolor08, decmark={o}{solid}},
resultStyle6/.style={mark=none, line width=0.5pt, mycolor05, decmark={square}{solid}}, 
resultStyle7/.style={mark=none, line width=0.5pt, mycolor09, decmark={diamond}{solid}}, 
resultStyle8/.style={mark=none, line width=0.5pt, mycolor11, decmark={diamond}{solid}}, 
resultStyleBase/.style={mark=none, line width=0.5pt,}, 
compareStyle1/.style={mark=none, line width=0.5pt, mycolor01},
compareStyle2/.style={mark=none, line width=0.5pt, mycolor02},%
compareStyle3/.style={mark=none ,line width=0.5pt, mycolor03},
compareStyle4/.style={mark=none, line width=0.5pt, mycolor06},
compareStyle5/.style={mark=none, line width=0.5pt, mycolor08},
compareStyle6/.style={mark=none, line width=0.5pt, mycolor05}, 
}
  
  \pgfplotsset{
        compat=newest,
        simple style group/.style={
                label style={font=\scriptsize},
                legend style={font=\scriptsize},
                tick label style={font=\scriptsize},
                nodes near coords style={font=\scriptsize},
                title style={font=\scriptsize},
                scale only axis,
                grid style={dotted},
                mark options={solid}, %
        },
        simple style/.style={
                label style={font=\scriptsize},
                legend style={font=\scriptsize},
                tick label style={font=\scriptsize},
                nodes near coords style={font=\scriptsize},
                title style={font=\scriptsize},
                width=\figurewidth,
                height=\figureheight,
                at={(0\figurewidth,0\figureheight)},
                scale only axis,
                grid style={dotted},
                mark options={solid}, %
        },
        base style/.style={
                label style={font=\scriptsize},
                legend style={font=\scriptsize},
                tick label style={font=\scriptsize},
                nodes near coords style={font=\scriptsize},
                title style={font=\scriptsize},
                width=\figurewidth,
                height=\figureheight,
                at={(0\figurewidth,0\figureheight)},
                scale only axis,
                cycle list={
                {mark=none, line width=0.5pt, mycolor01, solid},
                {mark=none, line width=0.5pt, mycolor02, dash dot},
                {mark=none ,line width=0.5pt, mycolor03, densely dashed},
                {mark=none, line width=0.5pt, mycolor04, dash dot dot},
                {mark=x   , line width=0.5pt, mycolor05},
                {mark=.   , line width=0.7pt, mycolor06}, 
                {mark=square,only marks, mark size = 0.8pt, mycolor07,
                mark options = {line width = 0.4pt}},
                {mark=x,     only marks, mark size = 1.3pt, mycolor08,
                mark options = {line width = 0.4pt}},
                {mark=o,     only marks, mark size = 0.8pt, mycolor09,
                mark options = {line width = 0.4pt}},
                {mark=o, mycolor10},
                },
                grid style={dotted},
                xmajorgrids,
                ymajorgrids,
                mark options={solid}, %
        },
        std graph style new/.style={
                xlabel style={yshift=1mm},
                ylabel style={yshift=-1.5mm},
                yticklabel style={xshift=1mm},
        },
        color lines style/.style={
                cycle list={
                    {mark=none, mycolor01, decmark={oplus}{solid} },
                    {mark=none, mycolor02, decmark={+}{solid} }, 
                    {mark=none, mycolor03, decmark={triangle}{solid} }, 
                    {mark=none, mycolor04, decmark={star}{solid} }, 
                    {mark=none, mycolor05, decmark={o}{solid} },
                    {mark=none, mycolor06, decmark={square}{solid} },
                },
        },
        meas graph style/.style={
                xlabel style={yshift=1mm},
                ylabel style={yshift=-1mm},
                xmajorgrids,
                ymajorgrids,
                mark repeat = 1,
                mark phase = 0,
                cycle list={
                    {color=black, only marks, mark=*, mark size=0.5pt, mark options={solid, black}},
                    {color=red, only marks, mark=*, mark size=0.1pt, line width=0.25pt},
                },
                ylabel={},
        }, 
        ci graph style/.style={
                xlabel style={yshift=1mm},
                ylabel style={yshift=-1.5mm},
                yticklabel style={xshift=1mm},
                mark repeat = 1,
                mark phase = 0,
                ymin=1e-3,
                ymax=100,
                ytick = {100, 50, 10, 1, 0.1, 0.01, 1e-3, 1e-4},
                yticklabels = {$0$, $50$, $90$, $99$, $99.9$, $99.99$, $99.999$, $99.9999$},
                y dir=reverse,
        },     
        bp coeff style/.style={
               scale only axis=true,
               width=0.225*.9\linewidth,
               height=0.225*.9\linewidth,
               scale only axis,
               xmin=-4.000,
               xmax=4.000,
               xlabel={$\ell${\color{white}$\aod$}},
               ticklabel style={font=\footnotesize},
               ymin=0.000, ymax=0.9,
               ylabel={$c_\ell$},
               xlabel style={font=\footnotesize},
               ylabel style={font=\footnotesize},
               major tick length=2pt%
        },
        bp graph style/.style={        
               scale only axis=true,
               width=0.35*1.1\linewidth,
               height=0.225*.9\linewidth,
               scale only axis,
               xmin=-3.14, xmax=3.14,
               xlabel={$\aod${\color{white}$\ell$}},
               ticklabel style={font=\footnotesize},
               xtick={-3.14,-1.57,0.0,1.57,3.14},
               xticklabels={$-\pi$,$-\tfrac{\pi}{2}$,$0$,$\tfrac{\pi}{2}$,$\pi$},
               ymin=0.000, ymax=3,
               ylabel={Beampattern},
               xlabel style={font=\footnotesize}, ylabel style={font=\footnotesize},
               major tick length=2pt
        },
        peb graph style/.style={        
               width=0.66\linewidth,%
               scale only axis,
               point meta min=-2.583,
               point meta max=-0.300,
               axis on top,
               xmin=0.000,
               xmax=12.000,
               xlabel={x in meter},
               y dir=reverse,
               ymin=0.000,
               ymax=8.000,
               ylabel={y in meter},
               ytick={7.0,6.0,...,0.0},
               xtick={0.0,1.0,...,12.0},
               yticklabels={$1$,$2$,$3$,$4$,$5$,$6$,$7$,$8$},
               xlabel style={font=\scriptsize,yshift=0.125cm},
               ylabel style={font=\scriptsize,yshift=-0.125cm},
               ticklabel style={font=\scriptsize},
               unit vector ratio*=1 1 1,
               yticklabel pos=left,
               major tick length=2pt,
               colormap={mymap}{[1pt] rgb(0pt)=(1,1,1); rgb(1pt)=(0.858903,0.984776,0.839302); rgb(2pt)=(0.777958,0.94143,0.649487); rgb(3pt)=(0.755504,0.864264,0.463393); rgb(4pt)=(0.777509,0.754439,0.310168); rgb(5pt)=(0.820314,0.619497,0.21003); rgb(6pt)=(0.854796,0.471879,0.170327); rgb(7pt)=(0.851327,0.326629,0.183322); rgb(8pt)=(0.784671,0.198575,0.225774); rgb(9pt)=(0.637629,0.0993149,0.259577); rgb(10pt)=(0.400067,0.0343393,0.229819); rgb(11pt)=(0,0,0)},
               colorbar style={ylabel={Position Error Bound in centimeter (logscale)}, ytick={-0.4,-0.82,...,-2.92}, yticklabels={$39.8$, $15.1$, $5.8$, $2.2$, $0.8$, $0.3$},ylabel style={yshift=0.5mm,font=\scriptsize,scale=0.8},width=2.0mm,xshift=-4.25mm,ticklabel style={font=\scriptsize},major tick length=0pt}, %
               colormap access=piecewise constant
        },
        peb ellipses/.style={color=white, line width=0.4pt, forget plot}
    }

\tikzset{naming/.style={align=center,font=\small}}
\tikzset{antenna/.style={insert path={-- coordinate (ant#1) ++(0,0.25) -- +(135:0.25) + (0,0) -- +(45:0.25)}}}
\tikzset{station/.style={naming,draw,shape=dart,shape border rotate=90, minimum width=10mm, minimum height=10mm,outer sep=0pt,inner sep=3pt}}
\tikzset{mobile/.style={naming,draw,shape=rectangle,minimum width=12mm,minimum height=6mm, outer sep=0pt,inner sep=3pt}}
\tikzset{radiation/.style={{decorate,decoration={expanding waves,angle=90,segment length=4pt}}}}

\tikzset{
  pobl/.style={
    inner sep=0pt, outer sep=0pt, fill=#1,
  },
  pobl gron/.style n args={2}{
    pobl=#1, rounded corners=#2,
  },
  pics/person/.style n args={3}{
    code={
      \node (-corff) [pobl=#1, minimum width=.25*#2, minimum height=.375*#2, rotate=#3, pic actions] {};
      \node (-pen) [minimum width=.3*#2, circle, pobl=#1, outer sep=.01*#2, anchor=south, rotate=#3, pic actions] at (-corff.north) {};
      \node (-coes dde) [pobl gron={#1}{1pt}, anchor=north west, minimum width=.12125*#2, minimum height=.25*#2, rotate=#3, pic actions] at (-corff.south west) {};
      \node [pobl=#1, anchor=north, minimum width=.12125*#2, minimum height=.15*#2, rotate=#3, pic actions] at (-coes dde.north) {};
      \node (-coes chwith) [pobl gron={#1}{1pt}, anchor=north east, minimum width=.12125*#2, minimum height=.25*#2, rotate=#3, pic actions] at (-corff.south east) {};
      \node [pobl=#1, anchor=north, minimum width=.12125*#2, minimum height=.15*#2, rotate=#3, pic actions] at (-coes chwith.north) {};
      \node (-braich dde) [pobl gron={#1}{.75pt}, minimum width=.075*#2, minimum height=.325*#2, outer sep=.0064*#2, anchor=north west, rotate=#3, pic actions] at (-corff.north east)  {};
      \node [pobl=#1, minimum width=.05*#2, minimum height=.2*#2, outer sep=.0064*#2, anchor=north west, rotate=#3, pic actions] at (-corff.north east) {};
      \node (-braich chwith) [pobl gron={#1}{.75pt}, minimum width=.075*#2, minimum height=.325*#2, outer sep=.0064*#2, anchor=north east, rotate=#3, pic actions] at (-corff.north west) {};
      \node [pobl=#1, minimum width=.0375*#2, minimum height=.2*#2, outer sep=.0064*#2, anchor=north east, rotate=#3, pic actions] at (-corff.north west) {};
      \node (-fit person) [fit={(-pen.north) (-braich dde.east) (-coes chwith.south) (-braich chwith.west)}] {};
    },
  },
  pics/SBS/.style={code={
      \begin{scope}[local bounding box=#1]
      \fill [pic actions/.try] (-1,0) -- (-1/2,3) -- (1/2, 3) -- (1,0) -- cycle;
      \fill [pic actions/.try] (-1/16,2) rectangle (1/16,4);
      \fill [pic actions/.try] (0,4) circle [radius=1/4];
      \foreach \i in {-1,1}
        \fill [shift=(90:4), xscale=\i]
          \foreach \r in {1,3/2,2}{
            (-45:\r) arc (-45:45:\r) -- (45:\r-1/10)
            arc(45:-45:\r-1/10) -- cycle
          };
       \end{scope}
  }},
}

\newcommand*\mref[2]{\cite[#1~\ref{M-#2}]{Supplement}}

\IEEEoverridecommandlockouts

\begin{document}
\allowdisplaybreaks
\frenchspacing

\title{\huge A Graph-based Algorithm for Robust Sequential Localization\\Exploiting Multipath for Obstructed-LOS-Bias Mitigation}
\author{\normalsize \IEEEauthorblockN{Alexander Venus$^{1,2}$,~\IEEEmembership{\normalsize Student Member,~IEEE}, Erik Leitinger$^{1,2}$,~\IEEEmembership{\normalsize Member,~IEEE}, Stefan Tertinek$^{3}$,\\[-0mm] and Klaus Witrisal$^{1,2}$,~\IEEEmembership{\normalsize Member,~IEEE}\\[0mm]}
\thanks{The financial support by the Christian Doppler Research Association, the Austrian Federal Ministry for Digital and Economic Affairs and the National Foundation for Research, Technology and Development is gratefully acknowledged.}
\small{{$^1$Graz University of Technology, Austria}, {$^3$NXP Semiconductors, Austria},}\\[0mm]
\small{{$^2$Christian Doppler Laboratory for Location-aware Electronic Systems}}\\[0mm]
}

\maketitle

\renewcommand{\baselinestretch}{1}\small\normalsize

\vspace{-13mm}

\begin{abstract}

This paper presents a factor graph formulation and particle-based \ac{spa} for robust sequential localization in multipath-prone environments. The proposed algorithm jointly performs data association, sequential estimation of a mobile agent position, and adapts all relevant model parameters.
We derive a novel non-uniform \ac{fa} model that captures the delay and amplitude statistics of the multipath radio channel. This model enables the algorithm to indirectly exploit position-related information contained in the \acp{mpc} for the estimation of the agent position without using any prior information such as floorplan information or training data.  
Using simulated and real measurements in different channel conditions, we demonstrate that the algorithm can provide high-accuracy position estimates even in fully obstructed line-of-sight (OLOS) situations and show that the performance of our algorithm constantly attains the \ac{pcrlb}, %
facilitating the additional information contained in the presented \ac{fa} model. The algorithm is shown to provide robust estimates in both, dense multipath channels as well as channels showing specular, {resolved} \acp{mpc}, significantly outperforming state-of-the-art radio-based localization methods.

\end{abstract}


\IEEEpeerreviewmaketitle

\acresetall

\newlength{\textwidthav}
\setlength{\textwidthav}{1\textwidth}

\section{Introduction}\label{sec:introduction}

\begin{figure}[t]
	\vspace{-1mm}
	\centering
	\scalebox{1.1}{\includegraphics{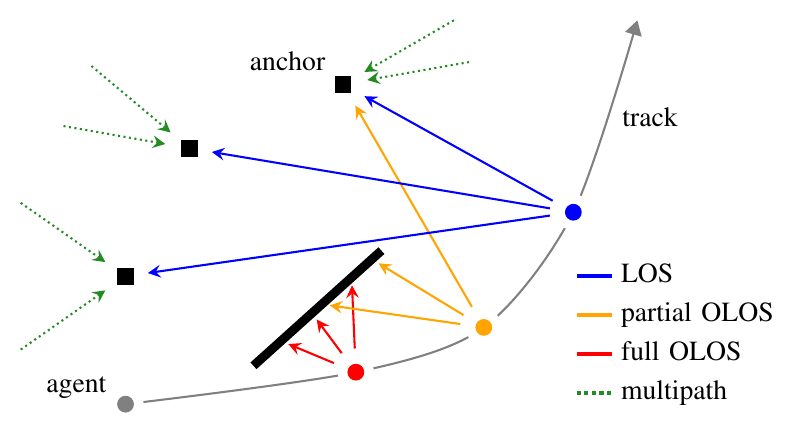}}
	\setlength{\abovecaptionskip}{0pt}
	\setlength{\belowcaptionskip}{0pt}
		\caption{A mobile agent is moving alongside the anchors on an example trajectory. Due to an obstacle, the \ac{los} to all anchors is not always available. There occur \textit{partial} and \textit{full} \ac{olos} situations. Multipath propagation may occur, but there is no prior information about the surrounding environment.}\label{fig:eye_catcher}
		\vspace{-6mm}
	\end{figure}

Localization of mobile agents using radio signals in environments such as indoor or urban territories is still a challenging task\cite{WitrisalSPM2016Copy,LeitingerJSAC2015,Mendrzik2019,WangShenTWC2020}. These environments are characterized by strong multipath propagation and frequent obstructed line-of-sight (OLOS)\acused{olos} situations, which can prevent the correct extraction of the \acf{los} component (see Fig.~\ref{fig:eye_catcher}). %
{Radio channels resulting from multipath propagation are commonly represented as a superposition of a finite number of specular \acp{mpc} \cite{RichterPhD2005, DardariProcIEEE2009,ShenTIT2010,AdityaProc2018}. However, cluttered environments with closely-spaced reflecting objects or with diffuse scatters (such as walls covered by shelves or irregular object shapes), along with the finite bandwidth of the measurement equipment, cause dense multipath propagation, which cannot be resolved into specular \acp{mpc} anymore \cite{RichterPhD2005,Karedal2007,KulmerPIMRC2018,JiangOJAP2022}.}

There exist many safety- and security-critical applications, such as autonomous driving \cite{Karlsson2017}, medical services \cite{KoEMBMag2010}, or keyless entry systems \cite{Kalyanaraman2020}, where robustness of the position estimate\footnote{We define robustness as the percentage of cases in which a system can achieve its given potential accuracy. I.e., a robust sequential localization algorithm can keep the agent's track in a very high percentage of cases, even in challenging environments.} is of critical importance.
\subsection{State-of-the-Art Methods} \label{sec:sota}

New localization and tracking approaches within the context of 5G localization \cite{ContiCommunMag2021} that take advantage of large measurement apertures as \ac{uwb} systems \cite{DardariProcIEEE2009, TaponeccoTWC2011} or mmWave systems \cite{RusekSPM2013} seek to mitigate the effect of multipath propagation \cite{GiffordTSP2022} (commonly referred to as ``NLOS propagation'') and \ac{olos} situations \cite{WymeerschIEEE2012, AdityaProc2018}, or even take advantage of \acp{mpc} by exploiting inherent position information, turning multipath from impairment to an asset \cite{LeitingerICC2014,LeitingerJSAC2015,GentnerTWC2016,LeitingerTWC2019, ShahmansooriTWC2018}.
Prominent examples of such approaches are multipath-based methods that estimate \acp{mpc} associate them to virtual anchors representing the locations of the mirror images of an anchor on reflecting surfaces \cite{PedersenJTAP2018}. The locations of virtual anchors are assumed to be known a priori \cite{LeitingerGNSS2016} or estimated jointly with the position of agents using \ac{mpslam} \cite{LeitingerTWC2019,GentnerTWC2016,KimTWC2020}. 
{Jointly estimating the positions of virtual anchors and agents allows \ac{mpslam} to provide high-accuracy position estimates, even in \ac{olos} situations, or to localize the agent with only a single anchor\cite{KulmerTWC2018}. However, it requires specular, resolved \acp{mpc}, which are consistent with the virtual anchor model \cite{KrekovicTSP2020}.} 
Other methods exploit cooperation among individual agents\cite{WymeerschProc2009,KulmerTWC2018,SharmaTSP2019,WangShenTWC2020}, or perform robust signal processing against multipath propagation and clutter measurements in general. 
The latter comprise heuristics \cite{DardariProcIEEE2009,ChianiProc2018}, machine learning-based approaches \cite{WymeerschIEEE2012,WangWCOMLetter2021,MaranoJSAC2010,StahlkeSensors2021} as well as Bayesian methods\cite{MeyerFUSION2018,Yu2020,VenusRadar2021}, and hybrids thereof \cite{Bartoletti2015, Mazuelas2018, ContiProcIEEE2019}. 
Heuristic methods, such as searching for the first amplitude to exceed a threshold value, are fast and easily implementable but suffer from low accuracy as well as a high probability of outage in low \ac{snr} regions \cite{DardariProcIEEE2009}. 
In recent years, machine learning methods have grown increasingly popular. Early approaches \cite{MaranoJSAC2010,WymeerschIEEE2012} extract specific features from the radio channel applying model-agnostic supervised regression methods on these features. While these approaches potentially provide high accuracy estimates at low computational demand (after training), they suffer from their dependence on a large representative measurement database and can fail in scenarios that are not sufficiently represented by the training data. This is why recent algorithms facilitate deep learning and auto-encoding based methods to directly operate on the received radio signal and reduce the dependence on training data \cite{LiShenMILCOM2021,StahlkeSensors2021,HuangTMC2022}. 
Multipath-based localization \cite{GentnerTWC2016,LeitingerICC2017,LeitingerTWC2019,LeitingerICC2019,KimTWC2020,Yu2020,VenusRadar2021}, multiobject-tracking \cite{BarShalomTCS2009,MeyerProc2018, MeyWilJ21}, and parametric channel tracking \cite{LiTWC2022} are applications that pose common challenges, such as uncertainties beyond Gaussian noise, like missed detections and clutter, an uncertain origin of measurements, and unknown and time-varying number of objects to be localized and tracked. These challenges can be well addressed by Bayesian inference leveraging graphical models to perform joint detection and estimation. 
Since the measurement models of these applications are nonlinear, most methods typically rely on sampling techniques such as recursive Monte Carlo sampling or particle filtering, or use linearized Gaussian models \cite{ArulampalamTSP2002, DurrantWhyte2006}.
Similarly, the \ac{pda} algorithm \cite{BarShalomTCS2009,BarShalom1995} represents a low-complexity Bayesian method for robust localization and tracking with extension to multiple-sensors \ac{pda} \cite{JeoTugTAES2005} and \ac{pdaai} \cite{LerroACC1990,LeitingerICC2019}. 
All these methods can be categorized as ``two-step approaches'', in the sense that they do not operate on the received sampled radio signal, but use extracted measurements provided by a preprocessing step, providing a high level of flexibility and a significant reduction of computational complexity. In contrast, ``direct positioning approaches" such as \cite{LeitingerICC2014,ZhaStaJosWanGenDamWymHoeTAES2020,KropfreiterFUSION2021} directly exploit the received sampled signal, which can lead to a better detectability of low-\ac{snr} features, yet, they are computationally very demanding. 

\subsection{Contributions} \label{sec:contributions}

In this paper, we propose a particle-based \ac{spa} that sequentially estimates the position of a mobile agent by utilizing the position-related information contained in the \ac{los} component as well as in \acp{mpc}\footnote{Throughout this paper, \acp{mpc} denote all components of the received signal that are caused by the transmit signal, except the \ac{los} component, i.e. ``\ac{nlos} components".}. %
The proposed algorithm jointly performs probabilistic data association and estimation of the mobile agent state \cite{MeyerProc2018,LeitingerTWC2019} together with all relevant model parameters, employing the \ac{spa} on a factor graph \cite{KschischangTIT2001}. Similar to other two-step approaches, it uses signal component delays and amplitudes estimated by a snapshot-based parametric \acf{ceda} as measurements. 
The proposed algorithm adapts in an online manner the time-varying component \ac{snr} \cite{LeitingerICC2019} as well as the detection probability of the \ac{los} \cite{LeitingerICC2017,SoldiTSP2019}. 
To this end, we propose a novel detection probability model that allows for both an exhaustive representation of the detection space and a smooth estimate of the \ac{snr}.
The algorithm exploits a novel non-uniform ``\ac{fa} model"\footnote{Typically the \ac{fa} or clutter model for delay measurements is chosen to be a uniform distribution inside the observation region of the sensor \cite[Sec. I-C]{MeyerProc2018} \cite[Sec. 2.5.2]{BarShalom1995}. Since we do not distinguish between \acp{fa} and \acp{mpc}, the resulting distributions of delay and amplitude measurements are non-uniform with respect to delay.%
 that %
explicitly models measurements originating from \acp{mpc}. 
More specifically, the introduced model represents the non-uniform distribution of delay measurements and corresponding {delay-dependent distribution of amplitude measurements} caused by \acp{mpc} and \acp{fa} in a joint manner. 
We refer to this part of the model using the terms ``\ac{nlos} model" or ``\ac{nlos} measurement model" throughout the paper. 
The presented NLOS model is derived from a stochastic radio signal model, which represents \acp{mpc} by their \ac{dps}, also referred to as power delay profile in the literature \cite{Karedal2007,RichterPhD2005,SteinboeckJTAP2013,WitrisalJWCOML2016}}.
Additionally, the model couples \ac{mpc} measurements to the \ac{los} measurement by a jointly inferred bias state. %
This enables the algorithm to utilize the position-related information contained in the \acp{mpc} %
without inferring specific map information, which can increase the accuracy and robustness of the agent's position estimate in challenging environments, characterized by strong multipath propagation and temporary \ac{olos} situations.
The proposed algorithm is able to operate without any prior information (no floorplan information or training data are needed). It is demonstrated to provide robust estimates for specular, resolved multipath as well as dense, non-resolvable multipath, while offering sub-second runtime\footnote{The runtimes were determined on PC, see Sec.~\ref{sec:execution_time} for details.} even in environments characterized by strong multipath propagation and, thus, a high number of measurements. 
The contributions of this paper are as follows.
\setlist[itemize]{leftmargin=6mm}
\begin{itemize}
 \item We derive a novel non-uniform \ac{nlos} model that is adapted to the distribution of the \ac{mpc} delays and amplitudes corresponding to a stochastic radio signal model \cite{Karedal2007,SteinboeckJTAP2013,WitrisalJWCOML2016} and verify its potential in a numerical study.
 \item We present a new factor graph and corresponding \ac{spa} in order to efficiently infer the marginal posterior distributions of all state variables of the introduced joint probabilistic model. %
 \item We show that the proposed algorithm is capable of overcoming even fully-\ac{olos} situations and providing \ac{crlb}-level position accuracy using both synthetic and real radio signal measurements.
 \item We analyze the influence of the individual features of our algorithm and compare it to a particle-based variant of the multi-sensor \ac{pdaai} algorithm, to the \ac{mpslam} algorithm presented in \cite{LeitingerTWC2019,LeitingerICC2019}, and to the \ac{pcrlb} \cite{Tichavsky1998}.
\end{itemize}
This work advances over the preliminary account of our %
conference publication \cite{VenusRadar2021} (and that of the related work \cite{Yu2020}) by (i) applying an accurate, adaptive model for the joint distribution of delay and amplitude measurements instead of using heuristical models, (ii) sequentially inferring all parameters of the \ac{nlos} model together with the agent instead of using predetermined constants, (iii) improving the convergence behavior using a modified, ``decoupled" \ac{spa} (see Sec.~\ref{sec:spa}), (iv) demonstrating the performance of the proposed algorithm using simulated \textit{radio signals} as well as real radio measurements obtained by (v) applying a \ac{ceda}, (vi) comparing to the \ac{mpslam} algorithm \cite{LeitingerTWC2019,LeitingerICC2019} and (vii) providing the \ac{pcrlb} as a performance benchmark.
\section{Notations and Definitions}
 Column vectors and matrices are denoted by boldface lowercase and uppercase letters. \Acp{rv} are displayed in san serif, upright font, e.g., $\rv{x}$ and $\RV{x}$ and their realizations in serif, italic font, e.g. , $x$ and $\V{x}$; $\tilde{x}$ denotes the true value of ${x}$. The same notation applies for stochastic processes $\rv{x}(t)$ and their realizations $x(t)$. $f({x})$ and $p({x})$ denote, respectively, the \ac{pdf} or \ac{pmf} of a continuous or discrete \ac{rv} $\rv{x}$. $(\cdot)^{\mathrm{T}}$, $(\cdot)^\ast$, and $(\cdot)^{\text{H}}$ denote matrix transpose, complex conjugation and Hermitian transpose, respectively. 
$ \norm{\cdot}{} $ is the Euclidean norm. $ \vert\cdot\vert $ represents the cardinality of a set. $ \mathrm{diag}\{\V{x}\} $ denotes a diagonal matrix with entries in $ \V{x} $. %
$\M{I}_{[\cdot]}$ is an identity matrix of dimension given in the subscript. $[\V{X}]_{n,n}$ denotes the $n$th diagonal entry of $ \V{X} $. %
Furthermore, %
${1}_{\mathbb{A}}(\V{x})$ denotes the indicator function that is ${1}_{\mathbb{A}}(\V{x}) = 1$ if $\V{x} \in \mathbb{A}$ and 0 otherwise, for $\mathbb{A}$ being an arbitrary set and $\mathbb{R}^{\text{+}}$ is the set of positive real numbers. %
We predefine the following \acp{pdf} with respect to $\rv{x}$: The truncated Gaussian \ac{pdf} is
\begin{equation} \label{eq:eq:truncated_gaussian_pdf}
	f_\text{TN}(x; \mu , \sigma , \lambda) = \frac{1}{Q(\frac{\lambda - \mu}{\sigma}) \sqrt{2\pi} \sigma} e^{\frac{-(x-\mu)^2}{2\,\sigma^2}}  {1}_{\mathbb{R}^{\text{+}}}(x \minus \lambda)
\end{equation}
with mean $\mu$, standard deviation $\sigma$, truncation threshold $\lambda$ and $Q(\cdot)$ denoting the Q-function \cite{Kay1998}. Accordingly, the Gaussian \ac{pdf} is $f_\text{N}(x; \mu , \sigma) =f_\text{TN}(x; \mu , \sigma , \text{-}\s\infty)$. The truncated Rician \ac{pdf} is \cite[Ch. 1.6.7]{BarShalom2002EstimationTracking} 
\begin{equation} \label{eq:truncated_rice_pdf}
	f_\text{TRice}(x;\rmv s ,\rmv u , \lambda) = \frac{1}{Q_1(\frac{u}{s}, \frac{\lambda}{s})}\frac{x}{s^2} e^{\frac{-(x^2+u^2)}{2\,s^2}} I_0(\frac{x\, u}{s^2}) {1}_{\mathbb{R}^{\text{+}}}(x \minus \lambda)
\end{equation}
with non-centrality parameter $u$, scale parameter $s$ and truncation threshold $\lambda$. $I_0(\cdot)$ is the 0th-order modified first-kind Bessel function and $Q_1(\cdot,\cdot)$ denotes the Marcum Q-function \cite{Kay1998}. The truncated Rayleigh \ac{pdf} is \cite[Ch. 1.6.7]{BarShalom2002EstimationTracking}
\vspace{1mm}
\begin{equation} \label{eq:truncated_rayleigh_pdf}
	f_\text{TRayl}(x; s , \lambda) = \frac{x}{s^2}\, e^{\frac{-(x^2-\lambda^2)}{2\, s^2}}  {1}_{\mathbb{R}^{\text{+}}}(x - \lambda)
\end{equation}
\vspace{1mm}
with scale parameter $s$ and truncation threshold $\lambda$. This formula corresponds to the so-called Swirling I model\cite{BarShalom2002EstimationTracking}. %
Finally, we define the uniform \ac{pdf} $f_\mathrm{U}(x;a,b) = 1/(b-a) {1}_{[a,b]}(x)$ and the uniform \ac{pmf} $f_\mathrm{UD}(x;\mathcal{X}) = 1/|\mathcal{X}| {1}_{\mathcal{X}}(x)$.

\section{Radio Signal Model}\label{sec:signal_model}
\begin{table*}[bt] \vspace*{-3mm}
	\renewcommand{\baselinestretch}{1}\small\normalsize
	\setlength{\tabcolsep}{3pt} %
	\renewcommand{\arraystretch}{1.1} %
	\footnotesize
	\centering
	\caption{Summary and description of all unobserved \acp{rv} of the system model.}\label{tbl:rvs}
	\begin{tabular}{r||c|c||c|c|c|c||c||c}
		\textbf{Description} & \textbf{agent state} & \textbf{rise time} & \makecell{\textbf{normalized}\\\textbf{amplitude}} & \textbf{DNR} & \textbf{NLOS bias} & \textbf{fall time} & \makecell{\textbf{LOS}\\\textbf{probability}} & \makecell{\textbf{association}\\\textbf{variable}} \\ 
		\textbf{Symbol} & $\RV{x}_n$ & $\rv{\gamma}_{\text{r}\s n}$ & $\rv{u}_n^{(j)}$ & $\sdnrr{}$ & $\rv{b}_n^{(j)}$ & $\rv{\gamma}_{\text{f}\s n}^{(j)}$  & $\rv{q}_n^{(j)}$ & $\rv{a}_n^{(j)}$\\ 
		\textbf{Type} & continuous & continuous & continuous & continuous & continuous & continuous & \textit{discrete} & \textit{discrete} \\ %
		\textbf{State Transition} & Markov & Markov & Markov & Markov & Markov & Markov & Markov & \textit{independent} \\ 
		\textbf{Anchor Relation} & \textit{common} & \textit{common} & separate & separate & separate & separate & separate & separate \\ \midrule
		\textbf{Description} & \multicolumn{2}{c||}{\textbf{augmented agent state}} & \multicolumn{4}{c||}{\textbf{anchor state}} & \\ 
		\textbf{Symbol} & \multicolumn{2}{c||}{$\bar{\RV{x}}_n  = [{\RV{x}}_n^\text{T}\; \rv{\gamma}_{\text{r}\s n}]^\text{T}$} & \multicolumn{4}{c||}{ $\RV{y}_n^{(j)}  = [\rv{u}_n^{(j)} \sdnrr{} \rv{b}_n^{(j)} \rv{\gamma}_{\text{f}\s n}^{(j)}]^\text{T}$} &  \\ 
	\end{tabular}
	\vspace{-2.5mm}
\end{table*}
At each discrete time $n$, the mobile agent at position $\tilde{\bm{p}}_n$ transmits a signal $s(t)$\vspace{0.3mm} and each anchor $j \rmv\rmv \in \rmv\rmv \{1,\s...\s,J\}$ at anchor position $\bm{p}_{\text{A}}^{(j)} = [p_{\text{Ax}}^{(j)} \; p_{\text{Ay}}^{(j)}]^\text{T}$ acts as a receiver. 
The complex baseband signal received at the $j$th anchor is modeled as
\begin{align}\label{eq:signal_cont} \vspace{-10mm}
\rmv\rmv\rmv\rmv\rmv\rmv \rv{r}_n^{(j)}(t)\rmv\rmv = \rmv\rmv \tilde{\alpha}_{n,0}^{(j)} s\big(t\minus \tilde{\tau}_{n,0}^{(j)}\big) \rmv  + \rmv\rmv\rmv \sum_{k = 1}^{\tilde{K}_n^{(j)}} \tilde{\alpha}_{n,k}^{(j)} s\big(t\minus \tilde{\tau}_{n,k}^{(j)}\big)
 \rmv +  \rv{\noise{}}_{n}^{(j)}(t)\ist.
\end{align}
The first and second term describe the \ac{los} component and the sum of $\tilde{K}_n^{(j)}$ specular \acp{mpc} with their corresponding complex amplitudes $\tilde{\alpha}_{n,k}^{(j)} \in \mathbb{C}$ and delays $\tilde{\tau}_{n,k}^{(j)} \in \mathbb{R}^{\text{+}}$, respectively. The delays are related to respective distances via $\tilde{\tau}_{n,k}^{(j)} = \tilde{d}_{n,k}^{(j)}/ \s c$ with $c$ being the speed of light. The third term represents an \ac{awgn} process $\rv{\noise{}}_{n}^{(j)}(t)$ with double-sided power spectral density $\tilde{N}_{0}^{(j)}/\s 2$.  The \ac{los} distance is geometrically related to the agent position via $\tilde{d}_{n,0}^{(j)} \triangleq d^{(j)}_{\text{LOS}\s}(\tilde{\bm{p}}_n)$ with $d^{(j)}_{\text{LOS}\s}(\tilde{\bm{p}}_n) = \norm{\tilde{\bm{p}}_n - \bm{p}_\text{A}^{(j)}}{}$.
We assume time synchronization between all anchors and the mobile agent{\footnote{{Note that state-of-the art \ac{uwb} ranging devices (e.g., NXP SR040/SR150 or Qorvo DW1000/DW3000) can provide synchronized channel impulse responses (CIRs) using a two-way ranging protocol \cite{KulmerICC2017}.}}}. However, our algorithm can be extended to an
unsynchronized system along the lines of \cite{GentnerTWC2016,LeitingerJSAC2015,EtzlingerTSP2017}.

The signal $\rv{r}_n^{(j)}(t)$ in \eqref{eq:signal_cont} is uniformly sampled with sampling frequency $ f_{\text{s}}$ at corresponding sampling interval $T_{\text{s}} \rmv\rmv  = \rmv \rmv 1/f_{\text{s}}$ and $N_{\text{s}}$ %
samples are collected, yielding a duration $T = N_{\text{s}}\, T_{\text{s}}$. %
By stacking the samples, we obtain the discrete time signal vector 
\vspace*{-2mm}
\begin{equation}\vspace*{-2mm}
\RV{r}_{n}^{(j)} = \tilde{\alpha}_{n,0}^{(j)} \V{s}(\tilde{\tau}_{n,0}^{(j)}) +\sum_{k = 1}^{\tilde{K}_n^{(j)}} \tilde{\alpha}_{n,k}^{(j)} \V{s}(\tilde{\tau}_{n,k}^{(j)}) + \RV{\noise{}}_{n}^{(j)}
\label{eq:signal_model_sampled}
\end{equation}
where $\bm{s}(\tau) \triangleq [s(-N_\text{s}/2 \cdot T_\text{s}-\tau) \,\,\, \cdots \,\,\, s((N_\text{s}-1)/2\cdot T_\mathrm{s} - \tau)]^\text{T}\in \mathbb{C}^{N_{\text{s}}\times1}$ is the stacked signal vector containing the samples of the transmit signal $s(t)$. The measurement noise vector $\RV{\noise{}}_{n}^{(j)} \in \mathbb{C}^{N_{\text{s}} \times 1} $ is a zero-mean, circularly-symmetric complex Gaussian random vector with covariance matrix %
$\tilde{\sigma}^{(j)\s 2} \M{I}_{N_{\text{s}}}$ and noise variance $\tilde{\sigma}^{(j)\s 2} = \tilde{N}_{0}^{(j)}/T_{\text{s}}$. The \acp{mpc} arise from reflection or scattering by unknown objects, since we assume that no map information is available. %

For a very large number of \acp{mpc} $ \tilde{K}_n^{(j)} $ and limited bandwidth of $s(t)$, the \acp{mpc} cannot be resolved anymore. Hence, the \acp{mpc} are described by a zero-mean, circularly-symmetric complex Gaussian stochastic process $\rv{\nu}_{\text{D}\s n}^{(j)}(\tau) $\cite{BelloTCS1963, FleuryTIT2000, LeitingerAsilomar2020}. The corresponding discrete time signal vector reads
\begin{equation} \label{eq:signal_model_sampled_stochastic}\vspace*{0mm}
	\RV{r}_{n}^{(j)} = \tilde{\alpha}_{n,0}^{(j)} \V{s}(\tilde{\tau}_{n,0}^{(j)}) + \int \rmv \rmv  \V{s}(\tau) \rv{\nu}_{\text{D}\s n}^{(j)}(\tau) \, \textrm{d} \tau + \RV{\noise{}}_{ n}^{(j)}
\end{equation}
\vspace{1mm}
with the second term denoting the dense multipath component \cite{RichterPhD2005,Karedal2007,SteinboeckJTAP2013,WitrisalJWCOML2016}. %
Assuming uncorrelated scattering for $\rv{\nu}_{\text{D}\s n}^{(j)}(\tau)$ \cite{BelloTCS1963, Karedal2007}, the noise covariance matrix of $\RV{r}_{n}^{(j)}$ %
is given by
\vspace{1mm}
\begin{equation} \label{eq:cov_dmc}
	\bm{C}_{\text{N}\s n}^{(j)} = \int  {S}_{\text{D}\s n}^{(j)}(\tau)\,\bm{s}(\tau)\, \bm{s}(\tau)^\text{H}\, \mathrm{d} \tau  +  \tilde{\sigma}^{(j)\s 2} \,\V{I}_{N_{\text{s}}} %
\end{equation}
\vspace{1mm}
where ${S}_{\text{D}\s n}^{(j)}(\tau)$ is the \ac{dps}. Using \eqref{eq:cov_dmc}, the \ac{snr} of the \ac{los} component is defined as\footnote{Note that the presented SNR model takes into account the interference between the LOS component and the dense multipath component \cite{WitrisalJWCOML2016}. In the absence of the dense multipath component this reduces to the familiar \ac{snr} $|\tilde{\alpha}^{(j)}_{n,0}|^2 \norm{\V{s}(\tilde{\tau}_{n,0}^{(j)})}{2} / \tilde{\sigma}^{(j)\s 2}$.} $ \mathrm{SNR}^{(j)}_{n} = |\tilde{\alpha}^{(j)}_{n,0}|^2 \V{s}(\tilde{\tau}_{n,0})^\text{H} \V{C}_\text{N\s n}^{(j) -1} \V{s}(\tilde{\tau}_{n,0})$ and the according normalized amplitude is $\tilde{u}^{(j)}_{n} \triangleq \mathrm{SNR}^{(j)\s\frac{1}{2}}_{n}$.

\vspace{-2mm}
\subsection{Delay Power Spectrum (DPS) Model}~\label{sec:dps_model}
\vspace{-2mm}
We choose to model the \ac{dps} ${S}_{\text{D}\s n}^{(j)}(\tau) $ as \cite{Karedal2007} 
\vspace{1mm}
\begin{align} \label{eq:dps_fun} 
     \tilde{S}_{\text{D}\s n}^{(j)}(\tau) &\triangleq {S}_{\text{D}}(\tau  ;  \tilde{\bm{p}}_n ,  \tilde{\Omega}_n^{(j)}\rmv\rmv\rmv, \tilde{\bm{\zeta}}_{\text{S}\s n}^{(j)} )
	\nonumber\\  & =  
	 \tilde{\Omega}_n^{(j)}\, \frac{\tilde{\gamma}_{\text{f}\s n}^{(j)} \rmv\rmv  +  \tilde{\gamma}_{\text{r}\s n}}{{\tilde{\gamma}_{\text{f}\s n}^{(j)\s 2}}} \Big(1 \rmv\,{-}\, e^{\rmv\rmv-\frac{\Delta_\text{b}(\tau;\cdot)}{\tilde{\gamma}_{\text{r} \s n}}}\Big)\, e^{\rmv-\frac{\Delta_\text{b}(\tau;\cdot)}{\tilde{\gamma}_{\text{f}\s n}^{(j)}}}\, \rmv\rmv {1}_{\mathbb{R}^{\text{+}}}(\Delta_\text{b}(\tau;\cdot))
	 \\[-7mm]\nonumber
\end{align}
which is a double exponential function with $\tilde{\Omega}_n^{(j)}$ being the \ac{dps} power. The rise time $\tilde{\gamma}_{\text{r}\s n}$ and fall time $\tilde{\gamma}_{\text{f}\s n}^{(j)}$ are shape parameters. The distance difference $\Delta_\text{b}(\tau;\cdot)$ is given by
\begin{equation} \label{eq:delta_fun} 
	\Delta_\text{b}(\tau;\tilde{\bm{p}}_n, \tilde{b}_n^{(j)} ) = c\, \tau - d_{\text{LOS}\s}^{(j)}(\tilde{\bm{p}}_n) - \tilde{b}_n^{(j)} %
	\nonumber
\end{equation}
where $\tilde{b}_n^{(j)}$ is the \ac{nlos} bias, which denotes the difference between the \ac{los} distance $d_{\text{LOS}\s}^{(j)}(\tilde{\bm{p}}_n)$ and the ``onset distance". $\tilde{\bm{\zeta}}_{\text{S}\s n}^{(j)} = [ \tilde{b}_n^{(j)}\, \tilde{\gamma}_{\text{f}\s n}^{(j)}\, \tilde{\gamma}_{\text{r}\s n}]^\text{T}$ collects the \ac{nlos} shape parameters for each time $n$ and anchor $j$. 
Experimental evidence motivates this model: The \ac{dps} typically exhibits an exponentially decaying tail \cite{RichterPhD2005, Karedal2007} and a smooth onset \cite{Karedal2007,JakobsenTAP2014_DynBirthDeath_2014}. In particular, when the \ac{los} power is excluded, as is done in \eqref{eq:signal_model_sampled_stochastic}. Note that $\tilde{\gamma}_{\text{r}\s n}$ is mainly determined by the signal bandwidth and onset-density of \acp{mpc}. For homogeneous deployment environments the on-set density is well modeled as being invariant. Therefore, $\tilde{\gamma}_{\text{r}\s n}$ is assumed to be the same for all anchors.

For inference, we also define the normalized \ac{dps} $\bar{S}_{\text{D}}(d,\tilde{\bm{p}}_n,\rmv\tilde{\bm{\zeta}}_{\text{S}\s n}^{(j)}) \rmv = \rmv {S}_{\text{D}}(d / c \,  ;  \tilde{\bm{p}}_n , \rmv \tilde{\Omega}_n^{(j)}\rmv\rmv\rmv\rmv\rmv,  \tilde{\bm{\zeta}}_{\text{S}\s n}^{(j)} ) \rmv\,/\,\rmv \tilde{\Omega}_n^{(j)}$ and the \ac{dnr} $\sdnrt{} = \norm{\bm{s}({\tau})}{} \,\tilde{\Omega}_n^{(j)\s \frac{1}{2}} /\s\s \tilde{\sigma}^{(j)}$, where the \ac{dnr} denotes the square-root power ratio between the dense multipath component and \ac{awgn}.

The proposed algorithm utilizes the position information contained in $ {S}_{\text{D}\s n}^{(j)}(\tau)$ to improve the position estimate without explicitly exploiting map information.

\subsection{Parametric Channel Estimation} \label{sec:channel_estimation}

By applying a suitable snapshot-based \acf{ceda}\cite{RichterPhD2005,ShutWanJos:CSTA2013,BadiuTSP2017,HanFleuRao:TSP2018} to the {observed} discrete signal vector $\V{r}_{n}^{(j)}$, one obtains, at each time $n$ and anchor $j$, a number of $M_n^{(j)}$ measurements denoted by ${\bm{z}^{(j)}_{n,m}}$ with $m \in \mathcal{M}_n^{(j)} = \{1,\,...\,,M_n^{(j)}\}$. Each  ${\bm{z}^{(j)}_{n,m}} = [\zd~ \zu]^\text{T}$ contains a distance measurement $\zd \in [0,~d_\text{max}]$, with maximum distance $d_\text{max} = c\, T$, %
 and a normalized amplitude measurement $\zu$. 
 The \ac{ceda} decomposes the discrete signal vector $\V{r}_n^{(j)}$ into individual, decorrelated components according to \eqref{eq:signal_model_sampled}, reducing the number of dimensions (as ${M}_n^{(j)}$ is usually much smaller than $N_{\text{s}}$). It thus can be said to compress the information contained in $\V{r}_n^{(j)}$ into $ \bm{z}^{(j)}_{n} = [{\bm{z}^{(j)\text{T}}_{n,1}}  ... \, {\bm{z}^{(j)\text{T}}_{n,M_n^{(j)}}}]^\text{T}$. 
 See the supplementary material~\mref{Sec.}{sec:app_ceda} \vspace{0.3mm} for further details. 
 The stacked vector $\bm{z}_n = [\bm{z}^{(1)\, \text{T}}_{n} \rmv\rmv\rmv ... \, \bm{z}^{(J)\,\text{T}}_{n}]^\text{T}$ is used as noisy measurement by the proposed algorithm.

\section{System Model} \label{sec:system_model}

We consider a mobile agent to be moving along an unknown trajectory as depicted in Fig.~\ref{fig:eye_catcher}. The current state of the agent is described by the state vector $\RV{x}_n = [\RV{p}_n ^\text{T}\; \RV{v}_n^\text{T} ]^\text{T}$, which is composed of the agent's position $\RV{p}_n = [\rv{p}_{\text{x}\s n}\; \rv{p}_{\text{y}\s n}]^\text{T}$ and velocity $\RV{v}_n = [\rv{v}_{\text{x}\s n}\; \rv{v}_{\text{y}\s n}]^\text{T}$.
We also introduce the following additional state variables, which represent all \ac{rv} inferred along with $\RV{x}_n$: First, we define the augmented agent state $\bar{\RV{x}}_n = [{\RV{x}}_n^\text{T}\; \rv{\gamma}_{\text{r}\s n}]^\text{T} $, which collects all \acp{rv} that are \textit{common} for all anchors. Second, we define the anchor state $\RV{y}_n^{(j)}  = [\rv{u}_n^{(j)} \sdnrr{} \rv{b}_n^{(j)} \rv{\gamma}_{\text{f}\s n}^{(j)}]^\text{T}$ collecting all \textit{continuous} \acp{rv}, which are modeled \textit{separately} for each anchor. Third, there are two \textit{discrete} \acp{rv} $\rv{q}_n^{(j)}$ and $\rv{a}_n^{(j)}$, which denote the LOS probability and association variable, respectively, and are modeled \textit{separately} for all anchors. For the sake of clarity, all \acp{rv} constituting the system model are summarized and described in Table~\ref{tbl:rvs}. %

 At each time $n$ and for each anchor $j$ the \ac{ceda} provides the currently observed measurement vector $\bm{z}_n^{(j)}$, with fixed ${M}^{(j)}_n$, according to Sec.~\ref{sec:channel_estimation}. Before the measurements %
are observed, they are random and represented by the vector  ${\RV{z}^{(j)}_{n,m}} = [\zdr~\zur]^\text{T}$. In line with Sec.~\ref{sec:channel_estimation} we define the nested random vectors $\RV{z}^{(j)}_{n} = [{\RV{z}^{(j)\text{T}}_{n,1}} \, ... \, {\RV{z}^{(j)\text{T}}_{n,\rv{M}_n^{(j)}}}]^\text{T}$ and $\RV{z}_n = [\RV{z}^{(1)\, \text{T}}_{n} ... \, \RV{z}^{(J)\,\text{T}}_{n}]^\text{T}$. Also the number of measurements $\rv{M}^{(j)}_n$ is a \ac{rv}. The vector containing all measurement numbers is defined as  $\RV{M}_n = [\rv{M}_n^{(1)}\, ... \, \rv{M}_n^{(J)}]^\text{T}$.

Each measurement $\RV{z}_{n,m}^{(j)}$ either originates from the \ac{los} or is due to an \ac{mpc}. It is also possible that a measurement $\RV{z}_{n,m}^{(j)}$ did not originate from any physical component, but from \acp{fa} of the \ac{ceda}. The presented model only distinguishes between ``LOS measurements'' originating from the \ac{los} and ``\ac{nlos} measurements'', i.e., measurements due to \acp{mpc} or \acp{fa}.

\subsection{\ac{los} Measurement Model} \label{sec:los_model}
The \ac{los} \ac{lhf} of an individual distance measurement $\zdr$ is given by
\begin{equation}
 f_\text{L}(\zd | \bm{p}_n, u_n^{(j)}) \triangleq f_\text{N}(\zd;\, d_\text{LOS}^{(j)}(\bm{p}_n),\, \sigma_{\text{d}}(u^{(j)}_{n}))
\end{equation}
with mean $d^{(j)}_{\text{LOS}}(\RV{p}_n)$ and variance $\sigma^2_\text{d}(\rv{u}^{(j)}_{n})$. The variance is determined based on the Fisher information given by
$ \sigma_{\text{d}}^{2} (\rv{u}^{(j)}_{n}) =   c^2 / ( 8\,  \pi^2 \, \beta_\text{bw}^2 \, \rv{u}^{(j)\s 2}_{n} ) $, where $\beta_\text{bw}$ is the root mean squared bandwidth \cite{WitrisalSPM2016Copy,LeitingerJSAC2015} and $\rv{u}_n^{(j)}$ is the normalized amplitude at anchor $j$. The \ac{los} \ac{lhf} of the normalized amplitude measurement $\zur$ is modeled as\footnote{The presented model describes the distribution of the amplitude estimates of a complex baseband signal in \ac{awgn} obtained using maximum likelihood estimation and generalized likelihood ratio test detection \cite{Kay1993,Kay1998,LerroACC1990}.} \cite{LeitingerICC2019,LiTWC2022}
\begin{equation} \label{eq:los_amplitude}
	 f_\text{L}(\zu| u^{(j)}_{n}) \triangleq f_\text{TRice}(\zu; \sigma_\mathrm{u} (u^{(j)}_{n}) ,u^{(j)}_{n}, \gamma)
\end{equation}
with $f_\text{TRice}(\cdot)$ being a truncated Rician PDF \eqref{eq:truncated_rice_pdf}. $\gamma$ is the detection threshold of the \ac{ceda},  which is a constant to be chosen. As for the distance \ac{lhf}, the scale parameter is determined based on the Fisher information given as 
$\sigma_{\mathrm{u}}^{2} (\rv{u}^{(j)}_{n}) = 1/2 +\rv{u}^{(j)\s 2}_n\, 1/{(4\s N_{\text{s}})}$. Note that this expression reduces to $1/2$ if the \ac{awgn} noise variance $\sigma^{(j)\s 2}$ is assumed to be known or $N_{\text{s}}$ to grow indefinitely (see \cite{LiTWC2022} for a detailed derivation). Note that for \eqref{eq:los_amplitude} the Marcum-Q function in \eqref{eq:truncated_rice_pdf} represents the detection probability $\pdrv$ (see Sec.~\ref{sec:pnl}).

\subsection{\ac{nlos} Measurement Model} \label{sec:nlos_model}
\begin{figure}[t]
	\centering
	\captionsetup[subfloat]{farskip=3pt,captionskip=0pt} 
	
	\setlength{\figurewidth}{0.32\textwidth}
	\setlength{\figureheight}{0.18\textwidth}
	\hspace*{-3mm}\subfloat[\label{fig:single_nlos_ampl_like}]{\scalebox{1.1}{\includegraphics{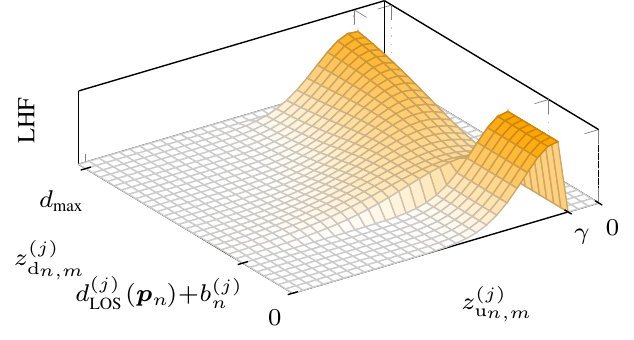}}}
	\setlength{\figurewidth}{0.41\textwidth}
	\setlength{\figureheight}{0.12\textwidth}
	\def\datapath{./figures/nlos_delay_lhf_thres}
	\hspace*{-2mm}\subfloat[\label{fig:nlos_like_gamma}]{\scalebox{1.1}{\includegraphics{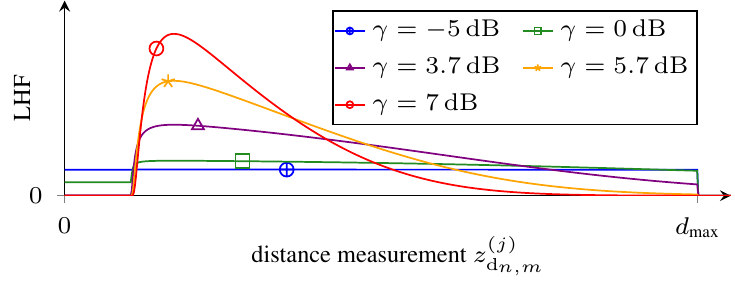}}}
	
	\caption{Graphical representation of \protect\subref{fig:single_nlos_ampl_like} the amplitude \ac{nlos} {LHF} $f_\text{NL}(\zu| \zd,\rmv \bm{p}_n,\bm{\zeta}_n^{(j)})$ and \protect\subref{fig:nlos_like_gamma} the distance \ac{nlos} {LHF} $f_\text{NL}(\zd| \bm{p}_n,\bm{\zeta}_n^{(j)} )$ for different values of the detection threshold $\gamma$ at $\sdnr{2} = 20\, \text{dB}$, $d_{\text{LOS}\s}^{(j)}( \bm{p}_n \rmv)  +   b_n^{(j)} = 7\, \text{m}$, $\gamma_{\text{f} \s n}^{(j)} = 6$ m, $\gamma_{\text{r}\s n} = 0.5$ m and $d_\text{max} = 30$ m.
	}
	\label{fig:single_measurment_like}
	\vspace{-3mm}
\end{figure}

The \ac{nlos} \ac{lhf} of an individual normalized amplitude measurement $\zur$  is given as
\begin{align} \label{eq:nlos_amplitude_lhf}
	 &f_\text{NL}(\zu| \zd, \bm{p}_n, {\bm{\zeta}}_n^{(j)}) \nonumber\\
	 &\hspace{1.7cm} \triangleq  f_\text{TRayl}(\zu; s_\text{u} (\zd,\bm{p}_n, {\bm{\zeta}}_n^{(j)}), \gamma) \\[-6mm]\nonumber  
\end{align}
where $f_\text{TRayl}(\cdot)$ is a truncated Rayleigh PDF \eqref{eq:truncated_rayleigh_pdf} and
\begin{equation} \label{eq:nlos_scale_function}
	s^2_\text{u} (\zd,\V{p}_n,\V{\zeta}_n^{(j)}) =  \frac{1}{2}\rmv(\sdnr{2}\, \bar{S}_{\text{D}}(\zd,\V{p}_n, {\V{\zeta}}_{\text{S}\s n}^{(j)}) + 1)
\end{equation}
is the \ac{nlos} scale function. 
We used $\RV{\zeta}_{\text{S}\s n}^{(j)} = [\rv{b}_n^{(j)}\; \rv{\gamma}_{\text{f}\s n}^{(j)}\; \rv{\gamma}_{\text{r}\s n}]^\text{T}$ and ${\RV{\zeta}}_n^{(j)} = [\sdnrr{}\; {\RV{\zeta}}_{\text{S}\s n}^{(j)\s \text{T}}]^\text{T}$ for notational brevity. 
See the supplementary material \mref{Sec.}{sec:app_nlos_model} for details about the derivation of \eqref{eq:nlos_amplitude_lhf}. %
The shape of \eqref{eq:nlos_amplitude_lhf} with respect to $\zu $ and $ \zd$ is shown in Fig.~\ref{fig:single_nlos_ampl_like}.
 The \ac{nlos} \ac{lhf} of the distance measurement $\zdr$ is given by
\begin{align} \label{eq:nlos_delay_lhf} \nonumber
	 &f_\text{NL}(\zd| \bm{p}_n,\bm{\zeta}_n^{(j)} ) \\\nonumber
	&\vspace{3mm}= Q_0(\bm{p}_n,\bm{\zeta}_n^{(j)})^{-1}\int_\gamma^\infty \rmv\rmv\rmv f_\text{TRayl}(u; s_\text{u} (\zd,\bm{p}_n,\bm{\zeta}_n^{(j)}) , \gamma) \,\mathrm{d} u \\
	&= Q_0(\bm{p}_n,\bm{\zeta}_n^{(j)})^{-1}\, \text{exp}\Big({-} {\frac{\gamma^2}{2\, s_\text{u}^2 (\zd,\bm{p}_n,\bm{\zeta}_n^{(j)}) }}\Big)\\[-7mm]\nonumber
\end{align}
where $Q_0(\bm{p}_n,\bm{\zeta}_n^{(j)}) = \int_{0}^{d_\text{max}} \text{exp}({{{-}\,\gamma^2 / (2\, s_\text{u}^2 (d,\bm{p}_n,\bm{\zeta}_n^{(j)}) }} ) ) \, \mathrm{d} \s d $ is the normalization constant ensuring integration to $1$. The exponential term in \eqref{eq:nlos_delay_lhf} corresponds to the %
probability that at time $n$ for anchor $j$ a \ac{nlos} measurement at distance $\zd$ is generated. The shape of \eqref{eq:nlos_delay_lhf} with respect to $\zd$ for different values of $\gamma$ is shown in Fig.~\ref{fig:nlos_like_gamma}. Note that \eqref{eq:nlos_delay_lhf} approaches a uniform \ac{pdf} when $\gamma$ or ${\sdnrr{}}$ approach zero.

The presented \ac{nlos} measurement model is valid independently of the \ac{dps} model chosen in \eqref{eq:dps_fun}. However, \eqref{eq:dps_fun} is a reasonable choice as it is physically motivated \cite{Karedal2007} and is of moderate computational complexity.

\subsection{Data Association Model} \label{sec:association_model}
\begin{figure}[t]
	\centering
	
	\setlength{\figurewidth}{0.425\textwidth}
	\setlength{\figureheight}{0.09\textwidth}

	\setlength{\belowcaptionskip}{0pt}
	\captionsetup[subfloat]{farskip=3pt,captionskip=0pt} 
	\hspace*{-4mm}\subfloat[\label{fig:single_delay_like}]{{\includegraphics{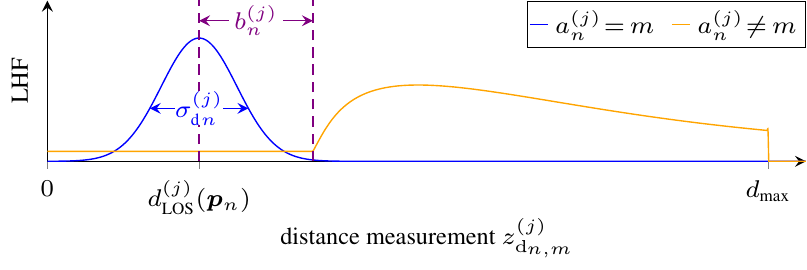}}}
	
	\hspace*{-5mm}
	\subfloat[\label{fig:single_amplitude_like}]{{\includegraphics{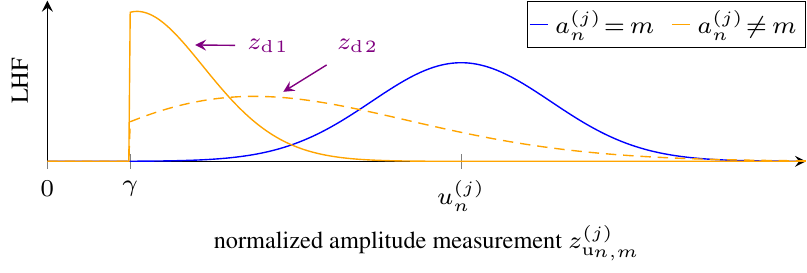}}}
	\caption{Graphical representation of the \protect\subref{fig:single_delay_like} the overall distance \ac{lhf} $f(\zd | {\bm{\zeta}}_{\text{E}\s n}^{(j)}\rmv )$ and \protect\subref{fig:nlos_like_gamma} overall amplitude \ac{lhf} $f(\zu | \zd, \rmv {\bm{\zeta}}_{\text{E}\s n}^{(j)}\rmv )$, all at fixed ${\bm{\zeta}}_{\text{E}\s n}^{(j)}$ in line with Fig.~\protect\ref{fig:single_measurment_like}. In \protect\subref{fig:single_amplitude_like} we also fix $\zd$ to $z_{\mathrm{d} 1} \rmv = \rmv 0 \, \mathrm{m}$ or $z_{\mathrm{d} 2} \rmv = \rmv 8 \, \mathrm{m}$.%
	}
	\vspace*{-5mm}
	\label{fig:like_da}
\end{figure}

At each time $n$ and for each anchor $j$, the measurements, i.e., the components of $\RV{z}_n^{(j)}$ are subject to data association uncertainty. Thus, it is not known which measurement $\RV{z}_{n,m}^{(j)}$ originated from the \ac{los}, or which one is due to an ``\ac{nlos} measurement'', i.e., measurements originating from \acp{mpc} or \acp{fa}. Based on the concept of \ac{pda} \cite{BarShalom1995}, we define the association variable $\rv{a}^{(j)}_{n}$ as %
\begin{equation}\label{eq:association}
	{a}^{(j)}_{n} \rmv \rmv = \rmv 
	\begin{cases} 
		m \rmv\rmv \in \rmv\rmv \mathcal{M}_n^{(j)}  \rmv\rmv \rmv\rmv, &
		\text{$\bm{z}_{n,m}^{(j)}$ is the LOS measurement in $\bm{z}_n^{(j)}$} \\
		0 \, , & \text{no LOS measurement in $\bm{z}_n^{(j)}$}
	\end{cases}\, .
\end{equation}
Assuming the number of \ac{nlos} measurements to follow a uniform distribution (so called ``non-parametric model''), the joint \ac{pmf} of $\rv{a}^{(j)}_{n}$ and $\rv{M}_n^{(j)}$ can be shown to be\cite{BarShalom1995}
\begin{align}  \label{eq:prior_association}
	p(a^{(j)}_{n}\rmv\rmv\rmv, M_n^{(j)} | u_n^{(j)}\rmv\rmv\rmv,q_n^{(j)}) =
	\begin{cases} 
		\frac{\pe}{M_n^{(j)}\, M_\text{max}} \, , &  a^{(j)}_{n} \in \mathcal{M}_n^{(j)} \\[3mm] \frac{1 - \pe \rmv\rmv\rmv}{M_\text{max}}, \rmv\rmv & a^{(j)}_{n} = 0
	\end{cases}\\[-7mm]\nonumber
\end{align}
where $\pe$ is the probability that there is a \ac{los} measurement for the current set of measurements defined in Sec. \ref{sec:pnl} and $M_\text{max}$ is an irrelevant constant. %
Incorporating $\rv{a}^{(j)}_{n}$ into the model, we define the overall distance \ac{lhf} as 
\begin{equation} \label{eq:single_delay_like}
	f(\zd | {\bm{\zeta}}_{\text{E}\s n}^{(j)})\rmv = \rmv
	\begin{cases} 
		 \rmv f_\text{L}(\zd | \bm{p}_n,\rmv u_n^{(j)})  ,  \rmv\rmv &  a^{(j)}_{n} = m \\
	     \rmv f_\text{NL}(\zd| \bm{p}_n, \rmv\bm{\zeta}_n^{(j)} ) , \rmv\rmv  & a^{(j)}_{n} \neq m
	\end{cases}\,%
\end{equation}
where we used ${\RV{\zeta}}_{\text{E}\s n}^{(j)} = [\RV{p}_n^\text{T}\; \rv{u}_n^{(j)}\rmv\rmv\rmv\; \rv{a}_n^{(j)}\rmv\; {\RV{\zeta}}_n^{(j)\s \text{T}}]^\text{T}$ for brevity. The shape of \eqref{eq:single_delay_like} is depicted in Fig.~\ref{fig:single_delay_like}.
Further, the overall amplitude \ac{lhf} is given by
\begin{align}
	f(\zu |& \zd, \rmv {\bm{\zeta}}_{\text{E}\s n}^{(j)}\rmv )\rmv \nonumber \\
	&\vspace{0mm} =\begin{cases} 
	   \rmv f_\text{L}(\zu| u^{(j)}_{n})  , \rmv\rmv\rmv &  a^{(j)}_{n} = m \\
	\rmv	f_\text{NL}(\zu| \zd, \bm{p}_n\rmv,\bm{\zeta}_n^{(j)} ) , \rmv\rmv\rmv  & a^{(j)}_{n} \neq m 
	\end{cases} \label{eq:single_ampl_like} 
\end{align}
which is shown in Fig.~\ref{fig:single_amplitude_like}. 
Using the common assumption of the measurements to be independent for different values of $m$ \cite{MeyerProc2018}, the joint \ac{lhf} for all measurements per anchor $j$ and time $n$ is
\begin{equation} \label{eq:overall_lhf}
	f(\bm{z}_n^{(j)} | {\bm{\zeta}}_{\text{E}\s n}^{(j)}) \rmv = \rmv\rmv\rmv\rmv \rmv  \prod_{m=1}^{\,\,M_{n}^{(j)}} \rmv\rmv\rmv 	f(\zu | \zd, \rmv {\bm{\zeta}}_{\text{E}\s n}^{(j)}\rmv )	\,	f(\zd | {\bm{\zeta}}_{\text{E}\s n}^{(j)})\, .
\end{equation}

\subsection{LOS Existence Probability Model}\label{sec:pnl}

We model the \ac{los} existence probability given in \eqref{eq:prior_association} as $\perv = \pdrv\, \rv{q}_n^{(j)}$. The probability of detection $\pdrv$ is the probability that at time step $n$ and anchor $j$ the agent generates a radio signal component whose amplitude is high enough so that it leads to an \ac{los} measurement. It is modeled by the counter probability of a Rician cumulative distribution function (CDF) given as
\vspace{1mm}
\begin{align}
	\pdrv =  Q_1\Bigg(\frac{\rv{u}^{(j)}_{n}}{\sigma_\mathrm{u} (\rv{u}^{(j)}_{n})}, \frac{\gamma}{\sigma_\mathrm{u} (\rv{u}^{(j)}_{n})}\Bigg)\\[-5mm]\nonumber
\end{align} 
by assuming that the proposed algorithm is applied after a generalized likelihood ratio test detector. $\rv{q}_n^{(j)}$ is the probability of the event that the \ac{los} is \textit{not} obstructed, which is referred to as \ac{los} probability in the following, and acts as a prior probability to the detection event. According to \cite{PapaICASSP2015,LeitingerICC2017,SoldiTSP2019}, we model $\rv{q}_n^{(j)}$ as discrete \ac{rv} that takes its values from a finite set $\mathcal{Q} = \{\lambda_1,\, ... \,, \lambda_Q\}$, where $\lambda_i \in (0,1]$. The \ac{los} probabilities for different sensors $j$ are assumed to be independent. 
The proposed \ac{los} existence probability model $\perv$ correctly incorporates the detection process into the system model via $\pdrv$ excluding a detection of measurements with $\zu$ below $\gamma$ and it can cope with {amplitude model mismatch} by correcting the amplitude-related probability of detection with $\rv{q}_n^{(j)}$. {With respect to implementation (see Sec.~\ref{sec:particle_filter}) this means that our model allows for smooth sequential inference of slow amplitude variations (e.g., due to path loss) via $\pdrv$, while $\rv{q}_n^{(j)}$ ensures a complete representation of the probability space, covering rapid amplitude variations (e.g., due to \ac{olos}).}

\subsection{State Transition model}

We model the evolution of $\bar{\RV{x}}_n$ and $\RV{y}_n^{(j)}$ and $\rv{q}_n^{(j)}$ over time $n$ as independent first-order Markov processes, which are defined by the joint state transition \ac{pdf} 
\begin{align}\nonumber 
	&f(\bar{\bm{x}}_n, {\bm{y}}_n,  q_{n}^{(j)}|\bar{\bm{x}}_{n\minus 1}, {\bm{y}}_{n\minus 1},  q_{n\minus 1}^{(j)})\\
	&\vspace{3mm}= f(\bar{\bm{x}}_n |\bar{\bm{x}}_{n\minus 1}) \prod_{j=1}^{J} f({\bm{y}}_n^{(j)}| {\bm{y}}_{n\minus 1}^{(j)})\, p(q_n^{(j)}| q_{n\minus 1}^{(j)})\, . \label{eq:statetrans}\\[-5mm]\nonumber
\end{align}
with $f(\bar{\bm{x}}_n |\bar{\bm{x}}_{n\minus 1}) $ and $ f({\bm{y}}_n^{(j)}| {\bm{y}}_{n\minus 1}^{(j)})$ being the respective state transition \acp{pdf}. 
For the {discrete} \ac{rv} $\rv{q}_n^{(j)}$ the first-order Markov process model results in a conventional Markov chain, with $[\bm{Q}^{(j)}]_{i,k} = p(q_n^{(j)} = \lambda_i| q_{n \minus 1}^{(j)} = \lambda_k)$ being the elements of the transition matrix. 

\section{Problem Formulation and Factor Graph}  \label{sec:factor_graph}
In this section we formulate the sequential estimation problem of interest and present the joint posterior and the factor graph underlying the proposed algorithm.
\subsection{Problem Statement}  

The problem considered is the sequential estimation of the agent state ${\RV{x}}_n$. This is done in a Bayesian sense by calculating the \ac{mmse} \cite{Kay1993} of the augmented agent state
\begin{equation}\label{eq:mmse}
{	\hat{\bar{\bm{x}}}^\text{MMSE}_{n} \,\triangleq \int \rmv \bar{\bm{x}}_{n} \, f(\bar{\bm{x}}_{n} | \bm{z} )\, \mathrm{d}\bar{\bm{x}}_{n} }\,.
\end{equation}
with $\hat{\bar{\bm{x}}}^{\text{MMSE}}_n = [\hat{\bm{x}}^{\text{MMSE T}}_n \, \hat{{\gamma}}^{\text{MMSE}}_{\text{r}\s n}]^\text{T}$ and $\hat{{\bm{x}}}^{\text{MMSE}}_n = [\hat{\bm{p}}^{\text{MMSE T}}_n $ $ \hat{\bm{v}}^{\text{MMSE {T}}}_n]$ and $\V{z} = [\V{z}^\text{T}_{{1}}\, ... \, \V{z}^\text{T}_{n}]^\text{T}$. Furthermore, we also calculate 
\begin{align}
	\hat{\bm{y}}^{(j)\s\text{MMSE}}_{n} &\,\triangleq \int \rmv \bm{y}_n^{(j)} \, f(\bm{y}_n^{(j)} | \bm{z} )\, \mathrm{d} \bm{y}_n^{(j)}, \label{eq:mmseAmpl}\\
	\hat{q}^{(j)\s\text{MMSE}}_{n} &\,\triangleq \sum_{\lambda_i \in \mathcal{Q}} \rmv \lambda_i \, p(q_n^{(j)} = \lambda_i | \bm{z} ) \label{eq:mmseplos}
\end{align}
with $\hat{\bm{y}}_n^{(j)\s \text{MMSE}}  = [\hat{u}_n^{(j)\s \text{MMSE}}\, \hat{\omega}_n^{(j)\s \text{MMSE}}\, \hat{b}_n^{(j)\s \text{MMSE}}\, \hat{\gamma}_{\text{f}\s n}^{(j)\s \text{MMSE}}]^\text{T}$.
In order to obtain \eqref{eq:mmse}, \eqref{eq:mmseAmpl}, and \eqref{eq:mmseplos}, {the respective marginal posterior \acp{pdf}} need to be calculated. Since direct marginalization of the joint posterior \ac{pdf} is computationally infeasible\cite{MeyerProc2018}, we perform message passing by means of the \ac{spa} rules on the factor graph that represents a factorized version of the joint posterior of our statistical model discussed in Sec.~\ref{sec:system_model}.
\subsection{Joint Posterior and Factor Graph}  
{For each $n$, let 
$\RV{y}_n = [\RV{y}^{(1)\s \text{T}}_{n} \vspace{0.5mm}\, ... \, \RV{y}^{(J) \s \text{T}}_{n}]^\text{T}$,
$\RV{a}_n = [\rv{a}^{(1)}_{n}\, ... \, \rv{a}^{(J)}_{n}]^\text{T}$, and 
$\RV{q}_n = [\rv{q}^{(1)}_{n}\, ... \, \rv{q}^{(J)}_{n}]^\text{T}$.} Furthermore, let
$\RV{z} = [\RV{z}^\text{T}_{{1}}\, ... \, \RV{z}^\text{T}_{n}]^\text{T}$, 
$\bar{\RV{x}} = [\bar{\RV{x}}^\text{T}_{0}\, ... \, \bar{\RV{x}}^\text{T}_{n}]^\text{T}$, 
$\RV{a} = [\RV{a}^\text{T}_{1}\, ... \, \RV{a}^\text{T}_{n}]^\text{T}$, 
$\RV{y} = [\RV{y}^\text{T}_{0}\, ... \, \RV{y}^\text{T}_{n}]^\text{T}$, 
$\RV{q} = [\RV{q}^\text{T}_{0}\, ... \, \RV{q}^\text{T}_{n}]^\text{T}$, 
and $\RV{M} = [\RV{M}^\text{T}_{1}\, ... \, \RV{M}^\text{T}_{n}]^\text{T}$.
We now assume that the measurements $\bm{z}$ are observed and thus fixed.
Applying Bayes' rule {as well as some commonly used independence assumptions}\cite{MeyerProc2018,LeitingerGNSS2016} the joint posterior for all states up to time $n$ and all $J$ anchors can be derived up to a constant factor as %
\begin{align} \label{eq:factorization1}
  &f(\bar{\bm{x}}, \rmv\bm{a}, \rmv\bm{y}, \rmv\bm{q}, \rmv\bm{M} | \bm{z} ) %
\nonumber\\
 &\propto  f(\bm{z} |  \bar{\bm{x}},\rmv\bm{a}, \rmv\bm{y}, \rmv\bm{q} ) \, f(\bar{\bm{x}},\rmv\bm{a}, \rmv\bm{y}, \rmv\bm{q})
  \nonumber\\ &
 = f(\bm{z} |  {\bar{\bm{x}}},\rmv\bm{a}, \rmv\bm{y}, \rmv\bm{q} ) \, f(\rmv\bm{a} | \bm{y}, \rmv \bm{q})\, f(\rmv \bar{\bm{x}}) \, p(\rmv\bm{q}) \, f(\rmv \bm{y}) \nonumber \\
&\propto f( \bar{\bm{x}}_0) \rmv\rmv \prod^{J}_{j= 1} p( q_0^{(j)}) \,f( \bm{y}_0^{(j)})  \rmv
  \rmv\rmv\prod^{n}_{n'= 1}  \rmv\rmv\rmv \Upsilon( \bar{\bm{x}}_{n'} | \bar{\bm{x}}_{n'\minus1})\, \Phi( \bm{y}_{n'}^{(j)} | \bm{y}_{n'\minus1}^{(j)}) \nonumber \\& ~~~\times \Psi( q_{n'}^{(j)} | q_{n'\minus1}^{(j)}) \,  \bar{g}( \bm{z}_{n'}^{(j)} ; \bm{p}_{n'}, \bm{y}^{(j)}_{n'} ,  a^{(j)}_{n'},  q^{(j)}_{n'})
\end{align}
where we introduced the state-transition functions  $\Upsilon(\bar{\bm{x}}_n|\bar{\bm{x}}_{n-1}) \triangleq f(\bar{\bm{x}}_n|\bar{\bm{x}}_{n-1})$, $\Phi(\bm{y}_n^{(j)}|\bm{y}_{n-1}^{(j)}) \triangleq f(\bm{y}_n^{(j)}|\bm{y}_{n-1}^{(j)})$, and $\Psi(q_n^{(j)}| q_{n \minus 1}^{(j)}) \triangleq p(q_n^{(j)}| q_{n \minus 1}^{(j)})$. We also introduced the pseudo \acl{lhf} 
$\bar{g}(\bm{z}_{n}^{(j)}; \bm{p}_{n},\rmv\rmv \bm{y}^{(j)}_{n} \rmv\rmv\rmv\rmv\rmv, \rmv a^{(j)}_{n}\rmv\rmv\rmv\rmv, \rmv q_n^{(j)})
\rmv\rmv \triangleq h(a_{n}^{(j)} ; \bm{y}_n^{(j)},q_n^{(j)})\,
g( \bm{z}_{n}^{(j)} ; \bm{p}_n, \rmv \bm{y}^{(j)}_{n}\rmv\rmv\rmv\rmv\rmv , \rmv a^{(j)}_{n})$. %
Finally, we define $g( \bm{z}_{n}^{(j)} ; \breve{\bm{\zeta}}_n^{(j)}) \triangleq  f(\bm{z}_n^{(j)} |\breve{\bm{\zeta}}_n^{(j)})$ and 
\vspace{1mm}
\begin{align}  \label{eq:h}
	h(a^{(j)}_{n}\rmv; \bm{y}_n^{(j)}\rmv\rmv\rmv,q_n^{(j)}) &\propto p(a^{(j)}_{n}\rmv\rmv\rmv, M_n^{(j)} | u_n^{(j)}\rmv\rmv\rmv,q_n^{(j)}) \nonumber\\
	&=
	\begin{cases} 
		\frac{\pe}{M_n^{(j)}} \, , &  a^{(j)}_{n} \in \mathcal{M}_n^{(j)} \\[3mm] {1 - \pe}, \rmv\rmv & a^{(j)}_{n} = 0
	\end{cases}
\end{align}
\vspace{1mm}
neglecting the constant terms in \eqref{eq:prior_association}.
Note that $\bm{M}$ vanishes in \eqref{eq:factorization1} as it is fixed and thus constant, being implicitly defined by the measurements $\bm{z}$.
Furthermore note that unlike \cite{BarShalom1995,LerroACC1990,MeyerProc2018,JeoTugTAES2005,LeitingerTWC2019,LeitingerICC2019,MeyWilJ21,LiTWC2022} in our model the \ac{nlos} \acp{lhf} \eqref{eq:nlos_amplitude_lhf} and \eqref{eq:nlos_delay_lhf} are both functions of \acp{rv} and, thus, cannot be neglected.
The joint posterior \ac{pdf} in \eqref{eq:factorization1} is represented by the factor graph shown in Fig.~\ref{fig:factor_graph}.

\section{Sum-Product Algorithm} \label{sec:algorithm}

 \subsection{Marginal Posterior and Sum-Product Algorithm (SPA)}  \label{sec:spa}
\begin{figure}[!t]

	\centering
		\hspace{-0cm}%
		{\includegraphics{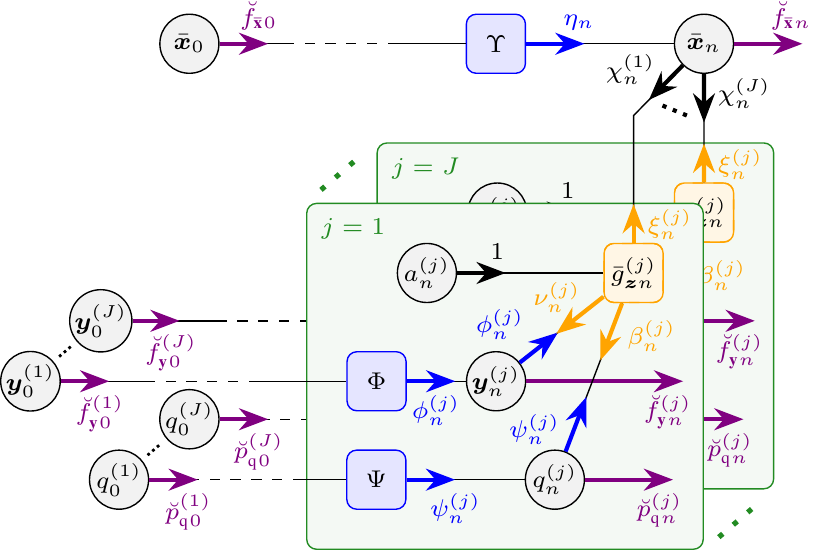}}
	\setlength{\abovecaptionskip}{0pt}
	\setlength{\belowcaptionskip}{0pt}
	\caption{Factor graph representing the factorization of the joint posterior \ac{pdf} in \eqref{eq:factorization1} as well as the respective messages according to the SPA (see Sec.~\ref{sec:spa}). The following short notations are used: %
		$ \eta_n                                         \rmv\triangleq\rmv  \eta( \bar{\bm{x}}_n)                                    $,  
		$ \phi_n^{(j)}                                   \rmv\triangleq\rmv  \phi( \bm{y}_n^{(j)})                                    $, 
		$ \psi_n^{(j)}                                   \rmv\triangleq\rmv  \psi(q_n^{(j)})                                          $, 
		$ \xi^{(j)}_n                                    \rmv\triangleq\rmv  \xi^{(j)}(\bar{\bm{x}}_n)                                $, 
		$ \chi^{(j)}_n                                   \rmv\triangleq\rmv  \chi^{(j)}(\bar{\bm{x}}_n)                               $, 
		$ \nu_n^{(j)}                                    \rmv\triangleq\rmv  \nu(\bm{y}_n^{(j)})                                      $, 
		$ \beta_n^{(j)}                                  \rmv\triangleq\rmv  \beta(q_n^{(j)})                                         $, 
		$ \chi^{(j)}_n                                   \rmv\triangleq\rmv  \chi^{(j)}(\bar{\bm{x}}_n)                               $, 
		$ \breve{f}_{\bar{\text{\textbf{x}}}\s n}        \rmv\triangleq\rmv  \breve{f}_{\bar{\text{\textbf{x}}}}(\bar{\bm{x}}_{n})    $, 
		$ \breve{f}_{\text{\textbf{y}}\s n}^{(j)}        \rmv\triangleq\rmv  \breve{f}_{\text{\textbf{y}}}(\bm{y}_{n}^{(j)})          $, 
		$ \breve{p}_{\text{q}\s n}^{(j)}                 \rmv\triangleq\rmv  \breve{p}_{\text{q}}(q_{n}^{(j)})                        $.    
	}\label{fig:factor_graph}
	
	\vspace{-3mm}
\end{figure}

The marginal posterior can be calculated efficiently by passing messages on the factor graph according to the \ac{spa}\cite{KschischangTIT2001}.
For the proposed algorithm, we specify not to send messages backward in time. This makes the factor graph in Fig.~\ref{fig:factor_graph} an acyclic graph. For acyclic graphs the \ac{spa}
yields \textit{exact results} for the marginal posteriors \cite{KschischangTIT2001}.
At time $n$, the following calculations are performed for all $J$ anchors. The prediction messages are given as 
\begin{align}\vspace{-1mm} \label{eq:message1}
\eta( \bar{\bm{x}}_n) &= \int {\Upsilon}( \bar{\bm{x}}_n | \bar{\bm{x}}_{n\minus 1})\, \breve{f}_{\text{\textbf{x}}}(\bar{\bm{x}}_{n\minus 1})\,\mathrm{d}\bar{\bm{x}}_{n\minus 1} \\[0pt]  \label{eq:message2}
 \phi( \bm{y}_n^{(j)}) &= \int \Phi( \bm{y}_n^{(j)} | \bm{y}_{n\minus1}^{(j)})\, \breve{f}_{\text{\textbf{y}}}(\bm{y}_{n\minus 1}^{(j)})\,\mathrm{d}\bm{y}_{n\minus 1}^{(j)} \\[-2pt]  \label{eq:message3}
 \psi(q_n^{(j)}) &= \sum_{q_{n\minus 1}^{(j)}=1}^{N_q} \Psi( q_n^{(j)} | q_{n\minus 1}^{(j)})\, \breve{p}_{\text{q}}(q_{n\minus 1}^{(j)})
\end{align}
where $\breve{f}_{\bar{\text{\textbf{x}}}}(\bar{\bm{x}}_{n\minus 1})$, $\breve{f}_{\text{\textbf{y}}}(\bm{y}_{n\minus 1}^{(j)})$ and $ \breve{p}_{\text{q}}(q_{n\minus 1}^{(j)})$ are messages of the previous time $n\minus 1$. 
The measurement update messages are given by
\begin{align}
\xi^{(j)}(\bar{\bm{x}}_n) &= \rmv\rmv \int \rmv\rmv \phi(\bm{y}_n^{(j)}) \rmv\rmv\rmv \sum_{q_{n}^{(j)}=1}^{N_q} \rmv\rmv\rmv \psi (q_n^{(j)}) \rmv\rmv\rmv\rmv\rmv\rmv\rmv\rmv    \nonumber\\ & \hspace{5mm}\times  \sum_{a_{n}^{(j)}=0}^{M_n^{(j)}} %
\bar{g}(\bm{z}_{n}^{(j)}; \bm{p}_{n},\rmv\rmv \bm{y}^{(j)}_{n} \rmv\rmv\rmv\rmv\rmv, \rmv a^{(j)}_{n}\rmv\rmv\rmv\rmv, \rmv q_n^{(j)}) 
\,\mathrm{d} \bm{y}_{n}^{(j)} \label{eq:messageUpdateX}\\[-2pt]
\label{eq:messagechi}
\chi^{(j)}(\bar{\bm{x}}_n) &= \eta( \bar{\bm{x}}_n) \prod_{j^{\prime}=1}^{J}  \xi^{(j^{\prime})}(\bar{\bm{x}}_n) /  \xi^{(j)}(\bar{\bm{x}}_n)\\[-2pt]
 \nu(\bm{y}_n^{(j)}) &= \rmv\rmv\rmv\rmv\rmv \rmv\rmv  \sum_{q_{n}^{(j)}=1}^{N_q} \rmv\rmv\rmv\rmv \psi(q_n^{(j)}) \rmv\rmv \rmv \int \rmv \rmv\rmv \chi^{(j)}(\bar{\bm{x}}_n) \rmv\rmv  \rmv\rmv \rmv\rmv %
 \nonumber\\ & \hspace{5mm}\times  \sum_{a_{n}^{(j)}=0}^{M_n^{(j)}} %
 \bar{g}(\bm{z}_{n}^{(j)}; \bm{p}_{n},\rmv\rmv \bm{y}^{(j)}_{n} \rmv\rmv\rmv\rmv\rmv, \rmv a^{(j)}_{n}\rmv\rmv\rmv\rmv, \rmv q_n^{(j)}) 
  \,\mathrm{d}\bar{\bm{x}}_{n} \label{eq:messageUpdateY}\\[-2pt]
\label{eq:mesageend}
 \beta(q_n^{(j)}) &= \rmv\rmv\rmv\rmv \int\rmv\rmv\rmv\rmv\rmv\rmv\rmv\rmv \int \rmv\rmv\rmv \phi( \bm{y}_n^{(j)}) \chi^{(j)}(\bar{\bm{x}}_n)\rmv\rmv\rmv\rmv\rmv\rmv %
 \nonumber\\ & \hspace{5mm}\times  \sum_{a_{n}^{(j)}=0}^{M_n^{(j)}} %
 \bar{g}(\bm{z}_{n}^{(j)}; \bm{p}_{n},\rmv\rmv \bm{y}^{(j)}_{n} \rmv\rmv\rmv\rmv\rmv, \rmv a^{(j)}_{n}\rmv\rmv\rmv\rmv, \rmv q_n^{(j)}) 
  \,\mathrm{d}\bar{\bm{x}}_{n} \,\mathrm{d} \bm{y}_{n}^{(j)}.
\end{align}
\ifthenelse{0=1}{
\begin{table}[b] \vspace*{-3mm}
	\renewcommand{\baselinestretch}{1}\small\normalsize
	\setlength{\tabcolsep}{6pt} %
	\renewcommand{\arraystretch}{1} %
	\footnotesize
	\centering
	\caption{messages}\label{tbl:messages}
	\begin{tabular}{c | c }
		prediction & update\\
		\midrule
\hspace{-10mm}\parbox[l]{0.5\textwidth}{
\begin{align} \label{eq:message1}
	\eta( \bar{\bm{x}}_n) &= \int {\Upsilon}( \bar{\bm{x}}_n | \bar{\bm{x}}_{n\minus 1})\, \breve{f}_{\text{\textbf{x}}}(\bar{\bm{x}}_{n\minus 1})\,\mathrm{d}\bar{\bm{x}}_{n\minus 1} \\[0pt]  \label{eq:message2}
	\phi( \bm{y}_n^{(j)}) &= \int \Phi( \bm{y}_n^{(j)} | \bm{y}_{n\minus1}^{(j)})\, \breve{f}_{\text{\textbf{y}}}(\bm{y}_{n\minus 1}^{(j)})\,\mathrm{d}\bm{y}_{n\minus 1}^{(j)} \\[-2pt]  \label{eq:message3}
	\psi(q_n^{(j)}) &= \sum_{q_{n\minus 1}^{(j)}=1}^{N_q} \Psi( q_n^{(j)} | q_{n\minus 1}^{(j)})\, \breve{p}_{\text{q}}(q_{n\minus 1}^{(j)})
	\\[-2pt] \label{eq:messagechi} 
	\chi^{(j)}(\bar{\bm{x}}_n) &= \eta( \bar{\bm{x}}_n) \prod_{j^{\prime}=1}^{J}  \xi^{(j^{\prime})}(\bar{\bm{x}}_n) /  \xi^{(j)}(\bar{\bm{x}}_n)
\end{align}
}	
 &
\parbox[r]{0.5\textwidth}{
\begin{align}
	\xi^{(j)}(\bar{\bm{x}}_n) &= \rmv\rmv\rmv\rmv \int \rmv\rmv\rmv\rmv \phi(\bm{y}_n^{(j)}) \rmv\rmv\rmv\rmv\rmv\rmv\rmv  \sum_{q_{n}^{(j)}=1}^{N_q} \rmv\rmv\rmv\rmv\rmv\rmv \psi (q_n^{(j)}) \rmv\rmv\rmv\rmv\rmv\rmv\rmv\rmv   \sum_{a_{n}^{(j)}=0}^{M_n^{(j)}} \rmv\rmv\rmv\rmv\bar{g}_{\bm{z} n}^{(j)}(\cdot)  \,\mathrm{d} \bm{y}_{n}^{(j)} \label{eq:messageUpdateX}\\[-2pt]
	\nu(\bm{y}_n^{(j)}) &= \rmv\rmv\rmv\rmv\rmv \rmv\rmv  \sum_{q_{n}^{(j)}=1}^{N_q} \rmv\rmv\rmv\rmv \psi(q_n^{(j)}) \rmv\rmv \rmv \int \rmv \rmv\rmv \chi^{(j)}(\bar{\bm{x}}_n) \rmv\rmv  \rmv\rmv \rmv\rmv \sum_{a_{n}^{(j)}=0}^{M_n^{(j)}} \bar{g}_{\bm{z} n}^{(j)}(\cdot)  \,\mathrm{d}\bar{\bm{x}}_{n} \label{eq:messageUpdateY}\\[-2pt]
	\label{eq:mesageend}
	\beta(q_n^{(j)}) &= \rmv\rmv\rmv\rmv \int\rmv\rmv\rmv\rmv\rmv\rmv\rmv\rmv \int \rmv\rmv\rmv \phi( \bm{y}_n^{(j)}) \chi^{(j)}(\bar{\bm{x}}_n)\rmv\rmv\rmv\rmv\rmv\rmv \sum_{a_{n}^{(j)}=0}^{M_n^{(j)}}\rmv\rmv\rmv %
	\bar{g}_{\bm{z} n}^{(j)}(\cdot)  \,\mathrm{d}\bar{\bm{x}}_{n} \,\mathrm{d} \bm{y}_{n}^{(j)}.
\end{align}
}
 \\
	\end{tabular}
	\vspace{-2.5mm}
\end{table}
}{}
Finally, we calculate the marginal posteriors as
$f(\bar{\bm{x}}_n| \bm{z}) \propto  \breve{f}_{\bar{\text{\textbf{x}}}}(\bar{\bm{x}}_{n}) \rmv\rmv =  \rmv\rmv \eta( \bar{\bm{x}}_n) \prod_{j=1}^{J} \xi^{(j)}(\bar{\bm{x}}_n)$, 
$f(\bm{y}_n^{(j)}| \bm{z}) \propto \breve{f}_{\text{\textbf{y}}}(\bm{y}_{n}^{(j)}) \rmv\rmv =  \phi(\bm{y}_n^{(j)}) \, \nu(\bm{y}_n^{(j)})$
and %
$p(q_n^{(j)}| \bm{z}) \propto \breve{p}_{\text{q}}(q_{n}^{(j)}) =  \psi(q_n^{(j)})\, \beta(q_n^{(j)}) \vspace*{1mm}$.

We additionally compare the performance of the above optimum \ac{spa} to that of a suboptimal message passing algorithm, which we refer to as ``decoupled SPA". Inspired by \cite{LeitingerICC2017}, we replace \eqref{eq:messagechi} by $\chi^{(j)}(\bar{\bm{x}}_n) = \eta( \bar{\bm{x}}_n)$ neglecting the mutual dependency of the uncertainties of individual anchor states $\RV{y}_n^{(j)}$. We demonstrate this modified algorithm to lead to improved numerical stability for a low number of particles. 
Hence, the particle-based implementation discussed in section ~\ref{sec:particle_filter} addresses the decoupled variant of the presented \ac{spa}. 
\subsection{Implementation Aspects}
\subsubsection{Particle-based Implementation} \label{sec:particle_filter}

Since the integrals involved in the calculations of the messages and beliefs \eqref{eq:message1}-\eqref{eq:mesageend} cannot be obtained analytically, we use a computationally efficient sequential particle-based message passing implementation that provides approximate computation. Our implementation uses a ``stacked state" \cite{MeyerJSIPN2016}, comprising the augmented agent state as well as the anchor states of all anchors $\mathcal{J} = \{1,..., J\}$. %

\newcommand{\wy}{{{}{w}_{{\mathsf{\mathbf{y}}}}}}
\newcommand{\wx}{{{}{w}_{{\bar{\mathsf{\mathbf{x}}}}}}}
\newcommand{\wyobar}{{{}\bar{w}_{{\mathsf{\mathbf{y}}}}}}
\newcommand{\wxobar}{{{}\bar{w}_{{\bar{\mathsf{\mathbf{x}}}}}}}
\newcommand{\wybarhat}{{\hat{\bm{w}}_{\bar{\mathsf{\mathbf{y}}}}}}
\newcommand{\wyhat}{{\hat{w}_{{\mathsf{\mathbf{y}}}}}}
\newcommand{\wxhat}{{\hat{w}_{{\bar{\mathsf{\mathbf{x}}}}}}}
\newcommand{\wyprime}{{{}{w}^{\s\prime}_{{\mathsf{\mathbf{y}}}}}}
\newcommand{\wxprime}{{{}{w}^{\s\prime}_{{\bar{\mathsf{\mathbf{x}}}}}}}
\newcommand{\wyprimeprime}{{{}{w}^{\s\prime\prime}_{{\mathsf{\mathbf{y}}}}}}
\newcommand{\wxprimeprime}{{{}{w}^{\s\prime\prime}_{{\bar{\mathsf{\mathbf{x}}}}}}}
\newcommand{\myn}{n}

\begin{enumerate}%
	\item[i)] \emph{Prediction}: The beliefs $\breve{f}_{\bar{\text{\textbf{x}}}}(\bar{\bm{x}}_{n-1})$ and 
	$\breve{f}_{\text{\textbf{y}}}(\bm{y}_{n-1}^{(j)})$ for all $ j \rmv\in\rmv \mathcal{J}$ calculated at the previous time step $ n-1 $, respectively, are represented by $ I $ particles and corresponding weights, i.e., 
	$\{ \bar{\RV{x}}_{n-1}^{[i]}, \wx{}_{n-1}^{[i]}\}_{i=1}^{I} $ and  $\{ {\RV{y}}_{n-1}^{(j)\s [i]}, \wy{}_{n-1}^{(j)\s [i]}\}_{i=1}^{I} $ for all $ j \rmv\in\rmv \mathcal{J}$. 
	Weighted particles 	$\{ \bar{\RV{x}}_{n}^{\prime\s  [i]}, \wxprime{}_{n}^{[i]}\}_{i=1}^{I} $ and  $\{ {\RV{y}}_{n}^{\prime\s (j)\s [i]}, \wyprime{}_{n}^{(j)\s [i]}\}_{i=1}^{I}$ for all $ j \rmv\in\rmv \mathcal{J}$, representing the messages $\eta( \bar{\bm{x}}_n)$ and $\phi( \bm{y}_n^{(j)})$ in \eqref{eq:message1} and \eqref{eq:message2} are determined as follows: For each particle $\bar{\RV{x}}_{n-1}^{[i]}$ and ${\RV{y}}_{n-1}^{(j)\s [i]}$ with  $ i \in \{1,\dots, I\} $,  one particle $\bar{\RV{x}}_{n}^{\prime\s [i]}$ and ${\RV{y}}_{n}^{\prime\s (j)\s [i]}$ with corresponding weights $  \wxprime{}_{n}^{[i]} =  \wx{}_{n-1}^{[i]} $ and $  \wyprime{}_{n}^{(j)\s [i]} =  \wy{}_{n-1}^{(j)\s [i]} $ is drawn from %
	$  f(\bar{\bm{x}}_n |\bar{\bm{x}}_{n\minus 1}^{[i]})$ and $f({\bm{y}}_n^{(j)}| {\bm{y}}_{n\minus 1}^{(j)\s [i]}) $ for all $ j \rmv\in\rmv \mathcal{J}$.%
	\item[ii)] \emph{Measurement Update}: The non-normalized weights representing the messages $\xi^{(j)}(\bar{\bm{x}}_n) $ and $\nu(\bm{y}_n^{(j)}) $ in \eqref{eq:messageUpdateX} and \eqref{eq:messageUpdateY} are calculated by \vspace*{-2mm}
	\begin{align}
	\wxprimeprime{}_{n}^{(j)\s [i]} &= \wyprime{}_{n}^{(j)\s[i]} \rmv\rmv\rmv  \sum_{q_{n}^{(j)}=1}^{N_q} \rmv\rmv\rmv\rmv \psi (q_n^{(j)}) \rmv\rmv\rmv  \nonumber\\ 
	& \hspace{5mm}\times \sum_{a_{n}^{(j)}=0}^{M_n^{(j)}} \rmv\rmv 
	\bar{g}(\bm{z}_{n}^{(j)}; \bm{p}_{n}^{\prime\s[i]},\rmv\rmv \bm{y}^{\prime\s (j)  \s [i]}_{n} \rmv\rmv, \rmv a^{(j)}_{n}\rmv\rmv\rmv\rmv, \rmv q_n^{(j)})\\[-2pt]
	\wyprimeprime{}_{n}^{(j)\s [i]}  &=  \wxprime{}_{n}^{(j)\s [i]}  \rmv\rmv\rmv  \sum_{q_{n}^{(j)}=1}^{N_q} \rmv\rmv\rmv\rmv \psi(q_n^{(j)}) \rmv \rmv\rmv \nonumber\\ &\hspace{5mm}\times \sum_{a_{n}^{(j)}=0}^{M_n^{(j)}} \rmv\rmv 
	\bar{g}(\bm{z}_{n}^{(j)}; \bm{p}_{n}^{\prime\s[i]},\rmv\rmv \bm{y}^{\prime\s (j)  \s [i]}_{n} \rmv\rmv, \rmv a^{(j)}_{n}\rmv\rmv\rmv\rmv, \rmv q_n^{(j)}) \, .
	\end{align}
	An approximation of the message $\beta(q_n^{(j)}) $ in \eqref{eq:mesageend} is given as
	\begin{align}
		 \beta(q_n^{(j)}) &\approx \sum_{i=1}^{I}  \wxprime{}_{n}^{[i]}  \wyprime{}_{n}^{(j)\s[i]} \rmv\rmv\rmv  \nonumber\\
		  &\hspace{5mm}\times \sum_{a_{n}^{(j)}=0}^{M_n^{(j)}}\rmv\rmv\rmv 
		\bar{g}(\bm{z}_{n}^{(j)}; \bm{p}_{n}^{\prime\s[i]},\rmv\rmv \bm{y}^{\prime\s (j)  \s [i]}_{n} \rmv\rmv, \rmv a^{(j)}_{n}\rmv\rmv\rmv\rmv, \rmv q_n^{(j)}) \, .
	\end{align}
	\item[iii)] \emph{Belief Calculation and State Estimation}: The above approximate messages are further used for calculating the non-normalized weights corresponding to the beliefs $  \breve{f}_{\bar{\text{\textbf{x}}}}(\bar{\bm{x}}_{n}) $ and $\breve{f}_{\text{\textbf{y}}}(\bm{y}_{n}^{(j)})$  for all $ j \rmv\in\rmv \mathcal{J}$  as 
	$%
		\wxhat{}_{n}^{[i]} = \wxprime{}_{n}^{[i]} \prod_{j = 1}^{J} \wxprimeprime{}_{n}^{(j)\s [i]}
	$%
	and
	$ \wyhat{}_{n}^{(j)\s [i]} = \wyprime{}_{n}^{(j)\s [i]} \wyprimeprime{}_{n}^{(j)\s [i]} $
	respectively.
	
	After normalization, i.e., $ \wxobar{}_{n}^{[i]} = \wxhat{}_{n}^{[i]}/\sum_{i=1}^{I}\wxhat{}_{n}^{[i]} $ and $ \wyobar{}_{n}^{(j)\s [i]} = \wyhat{}_{n}^{(j)\s[i]}/\sum_{i=1}^{I}\wyhat{}_{n}^{(j)\s[i]} $, an approximation of the \ac{mmse} state estimates $ \hat{\bar{\bm{x}}}^\text{MMSE}_{n} $ and $ \hat{\bm{y}}^{(j)\s\text{MMSE}}_{n} $ in \eqref{eq:mmse}, \eqref{eq:mmseAmpl} and \eqref{eq:mmseplos} is given as 
	$
		\hat{\bar{\bm{x}}}^\text{MMSE}_{n} \approx \sum_{i=1}^{I} \bar{\bm{x}}_n^{\prime\s [i]}  \, \wxobar^{[i]}%
	$ and 
	$
	\hat{\bm{y}}^{(j)\s\text{MMSE}}_{n} \approx \sum_{i=1}^{I} {\bm{y}}_n^{\prime\s (j)\s [i]}  \, \wyobar^{(j)\s [i]}
	$.	

To avoid particle degeneracy \cite{ArulampalamTSP2002}, a resampling step\footnote{We suggest to use ``systematic" resampling for efficiency \cite{ArulampalamTSP2002}.} is performed as a preparation for the next time step $ n+1 $ leading to equally weighted particles $\{ \bar{\RV{x}}_{n-1}^{[i]}, \wx{}_{n-1}^{[i]} = 1/I \}_{i=1}^{I} $ and  $\{ {\RV{y}}_{n-1}^{(j)\s [i]}, \wy{}_{n-1}^{(j)\s [i]}  = 1/I \}_{i=1}^{I} $ for all $ j \rmv\in\rmv \mathcal{J}$ representing the beliefs $\breve{f}_{\bar{\text{\textbf{x}}}}(\bar{\bm{x}}_{n})$ and $\breve{f}_{\text{\textbf{y}}}(\bm{y}_{n}^{(j)})$.
\end{enumerate}

The resulting problem complexity scales only linearly in the number of particles $I$ and in the number of measurements $M_n^{(j)}$. %
For computational efficiency of the particle-based implementation the \ac{los} \ac{lhf} of the normalized amplitude measurement \eqref{eq:los_amplitude} is approximated by a truncated Gaussian \ac{pdf}, i.e., \[	 f_\text{L}(\zu| u^{(j)}_{n}) = f_\text{TN}(\zu; \sigma_\mathrm{u} (u^{(j)}_{n}) ,u^{(j)}_{n}, \gamma). \]

\subsubsection{Initial State Distributions}\label{sec:init}
We assume the initial state distributions to factorize as
$\breve{f}_{\bar{\text{\textbf{x}}}}(\bar{\bm{x}}_{0}) =  %
\breve{f}_{{\text{\textbf{p}}}}({\bm{p}}_{0})
\breve{f}_{{\text{\textbf{v}}}}({\bm{v}}_{0}) \breve{f}_{\mathrm{\gamma}_\text{r}}(\gamma_{\text{r}\s 0})$ and 
$\breve{f}_{\text{\textbf{y}}}(\bm{y}_{0}^{(j)}) = \breve{f}_{\text{u}}(u_{0}^{(j)}) \breve{f}_{\mathrm{\sdnrp}}(\sdnrp_{0}^{(j)}) \breve{f}_{\text{b}}(b_{0}^{(j)})  \breve{f}_{\mathrm{\gamma}_\text{f}}(\gamma_{\text{f}\s 0}^{(j)})$.
We propose to initialize the \ac{nlos} shape parameters as $\breve{f}_{\mathrm{\gamma}_\text{r}}(\gamma_{\text{r}\s 0}) =  f_\text{U}(\gamma_{\text{r}\s 0},0,d_\text{max})$,  
$\breve{f}_{\text{b}}(b_{0}^{(j)}) =  f_\text{U}(b_{0}^{(j)}\rmv\rmv\rmv\rmv,0,d_\text{max})$ and  $\breve{f}_{\mathrm{\gamma}_\text{f}}(\gamma_{\text{f}\s 0}^{(j)}) =  f_\text{U}(\gamma_{\text{f}\s 0}^{(j)}\rmv\rmv\rmv\rmv,0,d_\text{max})$.
The \ac{los} \acp{pmf} are initialized as a discrete uniform \ac{pmf} $\tilde{f}_{\text{q}\s 0}^{(j)}(q_0^{(j)}) =  f_\text{UD}(q_0^{(j)}\rmv\rmv\rmv\rmv,\mathcal{Q})$ taking all values of $\mathcal{Q}$ with equal probability.
We assume the velocity vector ${\bm{v}}_{0}$ to be zero mean, Gaussian, with covariance matrix $\sigma_\text{v}^2\,  \bm{I}_2 $ and $\sigma_\text{v} = 6\, \text{m/s}$, as we do not know in which direction we are moving.

The remainder of the states are initialized heuristically, by assuming an initial measurement vector $\V{z}_0$ containing $M^{(j)}_0$ measurements to be available. The normalized amplitude \acp{pdf} are initialized as %
$\tilde{f}_{\mathrm{u}\s 0}^{(j)}(u_0^{(j)}) = f_\text{TRayl}(u_0^{(j)};\, \zuZeroMax,$ $\, 0.05 \, \zuZeroMax, \gamma)$ where $\zuZeroMax$ is the maximum normalized amplitude measurement in $\V{z}_0^{(j)}$. %
The position state is initialized as $f(\bm{p}_0) \sim \prod_{j=1}^{J}  \prod_{m=1}^{M^{(j)}_0}  f_\text{L}(\zdZero | \bm{p}_{\s \text{init}}, \zuZeroMax)\, f(\bm{p}_{\s \text{init}}) $, where the proposal distribution $f(\bm{p}_{\s \text{init}})$ is drawn uniformly on two-dimensional discs around each anchor $j$, which are bounded by the maximum possible distance $d_\text{max}$ and a sample is drawn from each of the $J$ discs with equal probability. %
The \ac{dnr} \acp{pdf} are initialized as $\breve{f}_{\mathrm{\sdnrp}}(\sdnrp_{0}^{(j)}) = f_\text{TRayl}(\sdnrp_{0}^{(j)};\, \sdnrp_\mathrm{init}^{(j)},\, 0.05\, \sdnrp_\mathrm{init}^{(j)},\gamma)$, where %
$\sdnrp_\mathrm{init}^{(j)}$ is determined as described in the supplementary material \mref{Sec.}{sec:app_inr_init}.

\subsubsection{Normalization of the NLOS Distance Likelihood}\label{sec:numerical1}

As discussed in Sec.~\ref{sec:nlos_model}, the \ac{nlos} \ac{lhf} in \eqref{eq:nlos_delay_lhf} must be normalized by $Q_0(\bm{p}_n,\bm{\zeta}_n^{(j)})$. However, $Q_0(\bm{p}_n,\bm{\zeta}_n^{(j)})$ cannot be determined analytically and, being a function of $\bm{p}_n$ and $\bm{\zeta}_n^{(j)}$, it needs to be calculated for each individual particle (see Sec.~\ref{sec:particle_filter}). Thus, we need an efficient numerical approximation. For details see the supplementary material \mref{Sec.}{sec:app_numerical}.

\section{Results}\label{sec:results}

\newcommand{\myopa}{1}
\newcommand{\myopb}{1}
\newcommand{\myopc}{1}
\newcommand{\myopd}{1}
\newcommand{\myope}{1}
\newcommand{\myopf}{1}
\newcommand{\myopg}{1}
\newcommand{\myoph}{1}
\newcommand{\myopi}{0}
\newcommand{\myopj}{1}
\newcommand{\myopk}{1}
\newcommand{\myopl}{1}
\newcommand{\myopm}{1}
\newcommand{\myopn}{1}

\newcommand{\shadesofgraysynthetic}{Different shades of gray represent different numbers of anchors in OLOS according to Fig.~\ref{fig:track_geometric}. }
\newcommand{\shadesofgrayreal}{Different shades of gray represent different numbers of anchors in OLOS according to Fig.~\ref{fig:track_nxp}. }

\newcommand{\resultcaption}[2]{Performance of #1 in terms of the \ac{rmse} of the estimated agent position (a), (c) as a function of the discrete observation time $n$, and (b), (d) as the cumulative frequency in inverse logarithmic scale, determined #2. }

\newcommand{\mseploti}[3]{
\centering
\setlength{\abovecaptionskip}{0pt}
\setlength{\belowcaptionskip}{0pt}
\setlength{\figurewidth}{0.42\textwidthav}
\setlength{\figureheight}{0.15\textwidthav}
\captionsetup[subfloat]{captionskip=-2mm} 
\def\datapath{./figures/mse_along_path_#1}
\subfloat[\label{fig:mse_along_path_#1}]{\scalebox{1}{\hspace{-2mm}\includegraphics{./pdf_figures/RobustPositioningJournal_V9-figure#2.pdf}}}
\vspace*{-3mm}

\def\datapath{./figures/mse_cdf_#1}
\subfloat[\label{fig:mse_cdf_#1}]{\scalebox{1}{\includegraphics{./pdf_figures/RobustPositioningJournal_V9-figure#3.pdf}}}
}

We validate the proposed model and analyze the performance gain caused by the {features} of the proposed algorithm using both synthetic data obtained using numerical simulation and real radio measurements. The performance is compared with the \ac{pcrlb} and that of the \ac{pdaai}. For synthetic measurements with geometry related multipath\footnote{Note that for measurements involving stochastic multipath as in Sec.~\ref{sec:performance_synthetic_dense}, the system model of the \ac{mpslam} algorithm is not suited, leading to divergence of the track.}%
, we also compare to the \ac{mpslam} algorithm presented in \cite{LeitingerTWC2019,LeitingerICC2019}. 

\subsection{Common Analysis Setup}\label{sec:simulation_model}
The following setup and parameters are commonly used for all analyses presented.

The \ac{pdf} of the joint agent state $\bar{\RV{x}}_n$ is factorized as $f(\bar{\bm{x}}_n |\bar{\bm{x}}_{n\minus 1}) = f({\bm{x}}_n |{\bm{x}}_{n\minus 1})\, f(\gamma_{\text{r}\s n} |\gamma_{\text{r}\s n\minus 1})$, where the agent motion, i.e. the state transition \ac{pdf} $f(\bm{x}_n|\bm{x}_{n-1})$ of the agent state $\RV{x}_n$, is described by a linear, constant velocity and stochastic acceleration model\cite[p.~273]{BarShalom2002EstimationTracking}, given as $\RV{x}_n = \bm{A}\, \RV{x}_{n\minus 1} + \bm{B}\, \RV{w}_{n}$,
with
the acceleration process $\rv{w}_n$ being i.i.d. across $n$, zero mean, and Gaussian with covariance matrix ${\sigma_{\text{a}}^2}\, \bm{I}_2$, ${\sigma_{\text{a}}}$ %
is the acceleration standard deviation, and $\bm{A} \in \mathbb{R}^{\text{4x4}}$ and $\bm{B} \in \mathbb{R}^{\text{4x2}}$ are defined according to \cite[p.~273]{BarShalom2002EstimationTracking}, with sampling period $\Delta T$.
The state transition of the rise distance $\rv{\gamma}_{\text{r}\s n}$, i.e., the state transition \ac{pdf} $ f(\gamma_{\text{r}\s n} |\gamma_{\text{r}\s n\minus 1})$, is $\rv{\gamma}_{\text{r}\s n} = \rv{\gamma}_{\text{r}\s n\minus 1} + \rv{\epsilon}_{\gamma_{\text{r}\s n}}$, where the noise $\rv{\epsilon}_{\gamma_{\text{r}\s n}}$ is i.i.d. across $n$, zero mean, Gaussian, with variance $\sigma^2_{\mathrm{\gamma}_\text{r}}$.
Similarly, the state transition model of the joint anchor state $\RV{y}_n^{(j)}$, i.e. the state transition \ac{pdf} $f(\bm{y}_n|\bm{y}_{n-1})$, is chosen as $\RV{y}_n^{(j)} = \RV{y}_{n\minus 1}^{(j)} + \RV{\epsilon}_{\text{\textbf{y}}\s n}^{(j)}$
, where the noise vector $\RV{\epsilon}_{\text{\textbf{y}}\s n}^{(j)}$ is i.i.d. across $n$ and $j$, zero mean, jointly Gaussian, with covariance matrix $\mathrm{diag}\{[\sigma^2_\text{u}\, \sigma^2_\mathrm{\sdnrp}\, \sigma^2_\text{b}\, \sigma^2_{\mathrm{\gamma}_\text{f}}]\}$ and the individual \ac{stv} $\sigma^2_\text{u}$, $\sigma^2_\mathrm{\sdnrp}$, $\sigma^2_\text{b}$ and $\sigma^2_{\mathrm{\gamma}_\text{f}}$.
Unless noted differently the \ac{stv} are set as 
$\sigma_a=2~\mathrm{m/s^2}$, 
$\sigma_\text{u} = 0.05\,\hat{u}_{n\minus 1}^{(j)\s \text{MMSE}}$, 
$\sigma_\mathrm{\sdnrp} = 0.05\, \hat{\sdnrp}_{n\minus 1}^{(j)\s \text{MMSE}}  $, 
$\sigma_\text{b} = 0.05\, \hat{b}_{n\minus 1}^{(j)\s \text{MMSE}} $, 
$\sigma_{\mathrm{\gamma}_\text{f}} =0.05\, \hat{\gamma}_{\text{f}\s {n\minus 1}}^{(j)\s \text{MMSE}}$, 
$\sigma_{\mathrm{\gamma}_\text{r}} = 0.5\, \hat{\gamma}_{\text{r}\s {n\minus 1}}^{\text{MMSE}} $. While $\sigma_a$ is set according to the maximum agent acceleration \cite{BarShalom2002EstimationTracking}, for the \ac{stv} of all other parameters we use values relative to the \ac{rmse} estimate of the previous time step $n \minus 1$ as a heuristic. Note that this choice allows \textit{no tuning of the \ac{stv}} to be required for all experiments presented, even though the propagation environments are considerably different.
 We used $5\cdot 10^4$ particles before the first resampling operation and $5000$ particles for inference during the track. We set the detection threshold as low as $\gamma = 1.77$ ($5\,\mathrm{dB}$) for all simulations, which allows the algorithm to facilitate low-energy \acp{mpc} (this choice is further discussed in Sec.~\ref{sec:performance_synthetic}). %
The set of possible \ac{los} probabilities is chosen as $\mathcal{Q} = \{0.01, 0.33, 0.66, 1\}$. The state transition matrix $\bm{Q}^{(j)} \triangleq \bm{Q}$ is set as follows: $[\bm{Q}]_{1,1}=0.9$, $[\bm{Q}]_{4,4}=0.95$, $[\bm{Q}]_{2,1}=0.1$ and $[\bm{Q}]_{3,4}=0.05$. For $2\leq k \leq 3$, $[\bm{Q}]_{k,k} = 0.85$, $[\bm{Q}]_{k-1,k} = 0.05$ and $[\bm{Q}]_{k+1,k} = 0.1$. For all other tuples $\{i,k\}$, $[\bm{Q}]_{i,k} = 0$ in order to encourage high LOS probabilities \cite{SoldiTSP2019}. 
For the numerical approximation of $Q_0(\bm{p}_n,\bm{\zeta}_n^{(j)})$ as discussed in Sec.~\ref{sec:numerical1}, we used $K_\text{T} = 30$. %
The results are shown in terms of the \ac{rmse} of the estimated agent position $e_{n}^{\text{RMSE}}~=~\sqrt{\E{\norm{\hat{\bm{p}}^{\text{MMSE}}_n -\bm{p}_n}{2}}}$, evaluated using a numerical simulation with 500 realizations.
For each of the scenarios investigated, we consistently analyze the influence of the individual features of our algorithm according to Table.~\ref{tbl:algorithms}. It shows the algorithm variants implemented and the corresponding features that are enabled for an algorithm (x) or not ( ). 
\begin{table}[b] \vspace*{-3mm}
	\renewcommand{\baselinestretch}{1}\small\normalsize
	\setlength{\tabcolsep}{2pt} %
	\renewcommand{\arraystretch}{1} %
	\scriptsize
	\centering
	\renewcommand{\myopi}{1}
	\caption{Algorithm variants investigated for different scenarios}\label{tbl:algorithms}
	\begin{tabular}{r | c c c c c | c | c c }
		&{AL1} &AL2 &AL3 &AL4 &AL5 &{AL6} &AL4$^\prime$ & AL5$^\prime$\\
		&\includegraphics{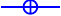} &\includegraphics{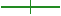} &\includegraphics{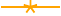} &\includegraphics{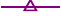} &\includegraphics{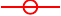} &\includegraphics{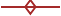}
		&\includegraphics{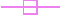}  &\includegraphics{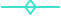} \\
		\midrule
		{$q_n^{(j)}$ tracking}       &   &    & x  & x  & x &  & x & x\\
		{Non-uniform $f_\text{NL}$}  &   & x  &    & x  & x & & x & x\\
		{Decoupled SPA}               &   & x  & x  &    & x & &   & x\\
		{$5 \cdot 10^4$ particles}   &   &    &    &    &   & {x} & x & x\\
		{\ac{mpslam}}   &   &    &    &    &   & {x} &  & \\
		\midrule
		Shown in Figs. & \multicolumn{5}{c|}{{\ref{fig:mse_along_path_cluster2}, \ref{fig:mse_cdf_cluster2}, \ref{fig:results_geometric}}} &{%
			 \ref{fig:mse_along_path_geometric}, \ref{fig:mse_cdf_geometric}} & \multicolumn{2}{c}{{\ref{fig:mse_along_path_cluster2}, \ref{fig:mse_cdf_cluster2}}}\\
	\end{tabular}
	\renewcommand{\myopi}{0}
	\vspace{-2.5mm}
\end{table}
When ``$q_n^{(j)}$ tracking'' is deactivated, we set $q_n^{(j)} = 0.999$ for all $n$, $j$. When we use ``decoupled SPA", the suboptimal message passing scheme presented in Sec.~\ref{sec:spa} is used. %
Not applying the ``non-uniform $f_\text{NL}$'' means \eqref{eq:nlos_scale_function} is replaced by $s^2_\text{u} \triangleq  1/2 $, and for AL4$^\prime$ and AL5$^\prime$ we use $5\cdot 10^4$ particles instead of $5000$. %
Note that AL1 represents a multi-sensor variant of the conventional \ac{pdaai}.
The \ac{mpslam} algorithm is implemented according to \cite{LeitingerICC2019, LeitingerTWC2019} using the measurements $\bm{z}_{m,n}^{(j)}$, i.e., distance and amplitude measurements, as an input%
. For consistency, the state transition \acp{pdf} and initial state distributions of the agent state and normalized amplitude state are set as described in Sec.~\ref{sec:simulation_model} and \ref{sec:init}. For convergence, we had to use $5\cdot 10^4$ particles and an anchor driving noise of $\sigma_\text{An} = 0.02~\mathrm{m}$, other parameters are $P_\text{s} = 0.999$, $\mu_\text{n,1} = 0.05$. The mean number of false alarms was approximated as $\mu_\text{FA} = N_\text{s}\, e^{-\gamma^2}$ (see \cite{LeitingerICC2019} for definitions).
For stability we increased the delay measurement variances of all virtual anchors (not the physical anchors) by a factor of $2$ with respect to the Fisher information-based value.
\begin{figure}[t]
	
	\centering
	\setlength{\abovecaptionskip}{0pt}
	\setlength{\belowcaptionskip}{0pt}
	
	\setlength{\figurewidth}{0.28\textwidthav}
	\setlength{\figureheight}{0.28\textwidthav}

	\scalebox{1}{\includegraphics{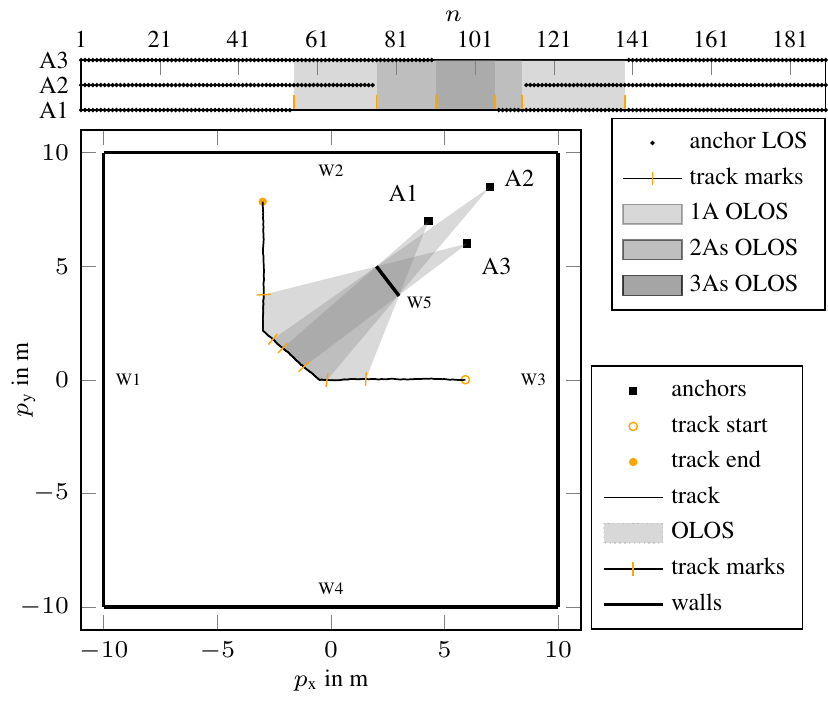}}
	\vspace{-1mm}
	\caption{Simulated trajectory and anchor positions for all synthetic experiments (Sec.~\ref{sec:performance_synthetic}) and environment setup (walls {and resulting obstructions}) for Geometry-related synthetic experiments (Sec.~\ref{sec:performance_synthetic_geometry}). %
	}\label{fig:track_geometric}
\end{figure}
\begin{figure*}[t]
	
	\centering
	\setlength{\abovecaptionskip}{0pt}
	\setlength{\belowcaptionskip}{0pt}
	
	\setlength{\figurewidth}{0.75\textwidth}
	\setlength{\figureheight}{0.23\textwidth}

	\vspace{-1mm}
	\def\datapath{./figures/state_space_cluster}
	\hspace{-3mm}\scalebox{1.0}{\includegraphics{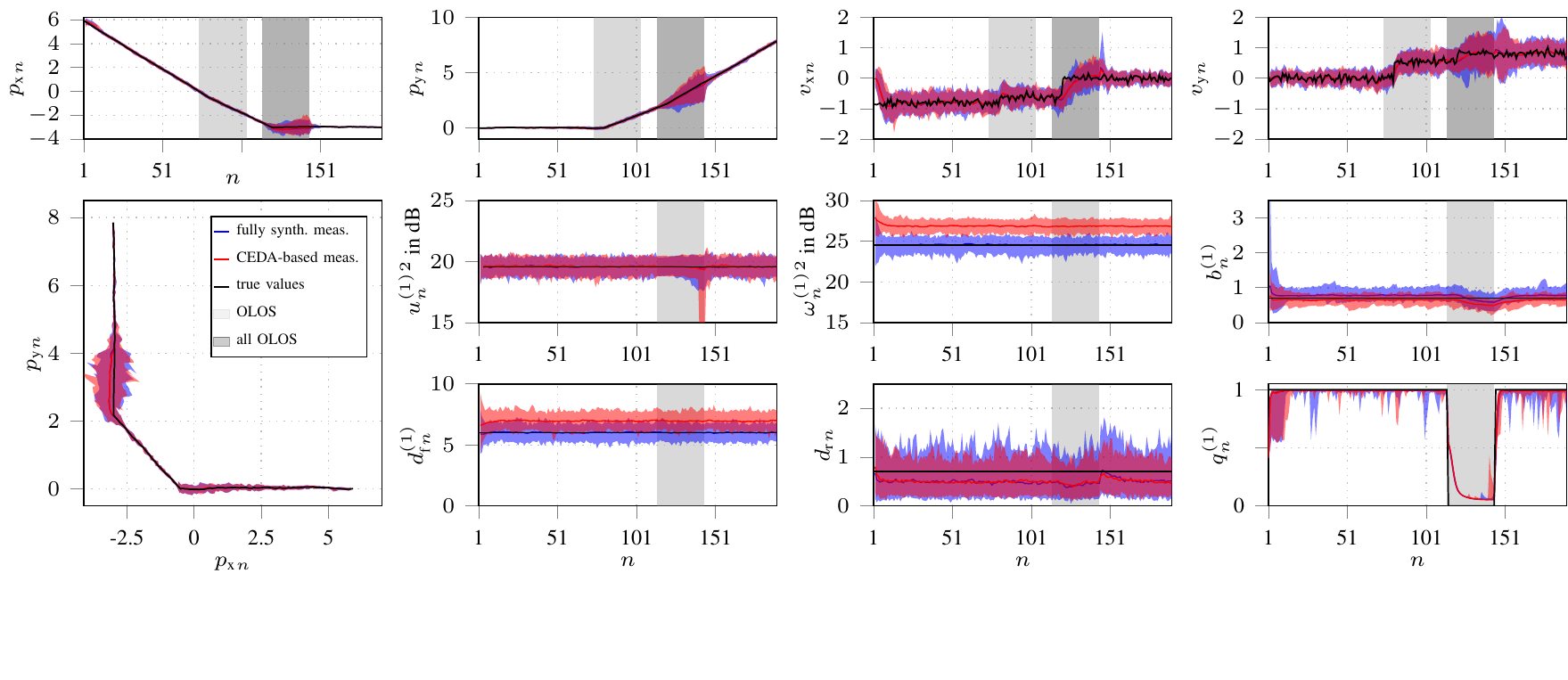}}
	\vspace{-1cm}
	\caption{\ac{mmse} estimates (determined using fully synthetic measurements and CEDA-based measurements) and true values of all state variables v.s. discrete time $n$ for the experiment described in Sec.~\ref{sec:performance_synthetic_dense}. We show the mean \ac{mmse} estimate and the corresponding range from minimum to maximum value. Anchor state variables are only shown for anchor A1. \shadesofgraysynthetic}\label{fig:state_space_cluster}
	\vspace*{-2mm}
\end{figure*}

As a performance benchmark, we provide the \ac{crlb} on the position error variance considering all visible \ac{los} measurements of a single time step $n$, which we refer to as the snapshot-based positioning CRLB (SP-CRLB) \cite{Jourdan2008, ShenTIT2010, WitrisalJWCOML2016, WinShenProc2018Bounds}. Furthermore, we provide the corresponding \acf{pcrlb} \cite{Tichavsky1998} that additionally considers the dynamic model of the agent state and the ``P-CRLB-LOS", which is the P-CRLB assuming the LOS component to all anchors is always available and, thus, provides a lower bound for the proposed estimator. See the supplementary material \mref{Sec.}{sec:app_crlb} for further details. 

\subsection{Analysis on Synthetic Measurements} \label{sec:performance_synthetic}
For the synthetic setup, we investigate the scenario shown in Fig.~\ref{fig:track_geometric}. The agent moves along a trajectory, with two distinct direction changes, where the agent velocity is set to vary around a magnitude of $0.8 \,\mathrm{m/s}$. %
It is observed at $N=190$ discrete time steps $n \in \{1, \,...\, , N\}$ at a constant observation rate of $\Delta T = 100\,\mathrm{ms}$, resulting in a continuous observation time of $19\,\mathrm{s}$.
We simulate three anchors, A1-A3, which are placed in close vicinity to each other. {The limited directional diversity of the anchors (corresponding to a poor geometric dilution of precision (GDOP) \cite{GodHaiBlu:TIT2010}), poses a challenging setup for delay measurement-based position estimation.} %
{Note that the environment setup shown in Fig.~\ref{fig:track_geometric}, i.e., walls \textit{and resulting obstructions}, are only used in Sec.~\ref{sec:performance_synthetic_geometry}.}
For all synthetic radio measurements involving the proposed \ac{ceda} (see \mref{Sec.}{sec:app_ceda}), we choose the transmitted complex baseband signal $s(t)$ to be of  root-raised-cosine shape with a roll-off factor of $0.6$ and a duration of $2\,\mathrm{ns}$ (bandwidth of $500\,\mathrm{MHz}$). The signal is critically sampled, i.e., $T_\text{s} = 1.25\,\mathrm{ns}$, with a total number of $N_\text{s} = 161$ samples, amounting to a maximum distance of $d_\text{max} = 60\,\mathrm{m}$.

\subsubsection{Synthetic Measurements with Stochastic Multipath} \label{sec:performance_synthetic_dense}
In this section we present results using synthetic measurements generated by simulating the \acp{mpc} as zero mean stochastic process. More specifically, we compare results obtained by simulating the radio signal according to \eqref{eq:signal_model_sampled_stochastic} and applying the \ac{ceda} to results obtained using fully synthetic measurements, which are generated according to Sec.~\ref{sec:system_model} without involving the \ac{ceda}.
For fully synthetic measurements the average number of \ac{nlos} measurements per time $n$ and anchor $j$ prior to the simulated detection process was approximated as $N_\text{s}$. Detection further reduces the prior number of \ac{nlos} events by the mean \ac{nlos} detection probability.
We simulate two \ac{olos} situations clearly separated in time, a partial one at $n \in [75 , 104 ]$, where only the \ac{los} to anchor $A2$ is blocked, and a full one at $n \in [115 ,144 ]$, where the \ac{los} to all anchors is blocked.
The following true system parameters are used, which are set constant for all time steps $n$ and anchors $j$: The normalized amplitude is set to $\tilde{\V{u}}_n =  [\sqrt{19.5}\,\mathrm{dB}~\sqrt{20.0}\,\mathrm{dB}~\sqrt{20.5}\,\mathrm{dB}]^\text{T}$ and the parameters of the \ac{dps} are set to $\sdnrt{2} = \sqrt{25} \,\mathrm{dB}$, $ \tilde{\gamma}_{\text{r}\s n} = 0.7\,\mathrm{m}$, $\tilde{\gamma}_{\mathrm{f}\s n}^{(j)} =6\,\mathrm{m}$, $\tilde{b}_n^{(j)} = 0.7 \,\mathrm{m}$.

\begin{figure}[t]
	\centering
	\setlength{\abovecaptionskip}{0pt}
	\setlength{\belowcaptionskip}{0pt}
	
	\setlength{\figurewidth}{0.39\textwidth}
	\setlength{\figureheight}{0.26\textwidthav}
	
	\def\datapath{./figures/nlos_statistic_cluster}
	\hspace{-1mm}\scalebox{1.1}{\includegraphics{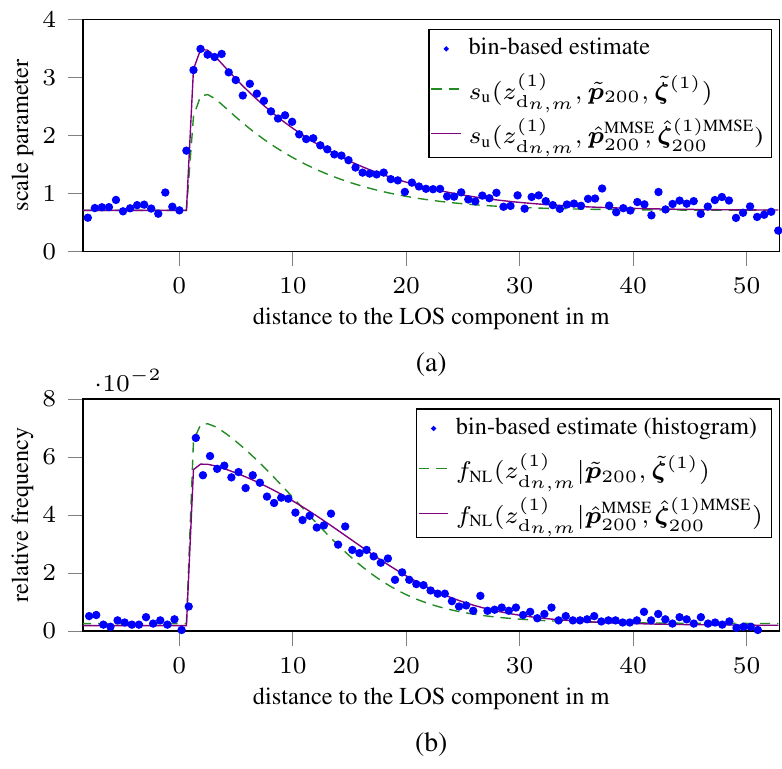}}
	\vspace{-1mm}
	\caption{\Acfp{npe} of (a) the Rayleigh scale parameter of the amplitude measurements and (b) the relative frequency of the distance measurements compared to (a) the \ac{nlos} scale function and (b) the \ac{nlos} distance \ac{lhf}. All values are shown as a function of the difference of the distance measurement and the corresponding LOS component distance, given as $\zdOne - d_{\text{LOS}\s}^{(1)}(\tilde{\bm{p}}_{200})$ for anchor A1. 
	}\label{fig:nlos_statistic_cluster}
	\vspace{-4mm}
\end{figure}
\begin{figure*}[t]
	\hspace{-2mm}
	\parbox[c]{0.5\textwidth}{\mseploti{cluster}{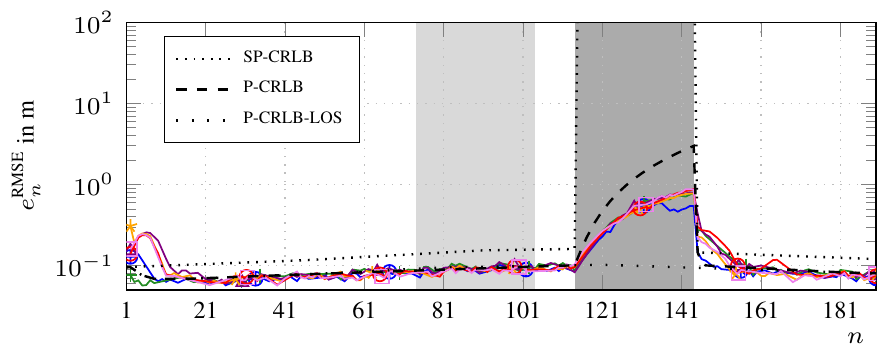}{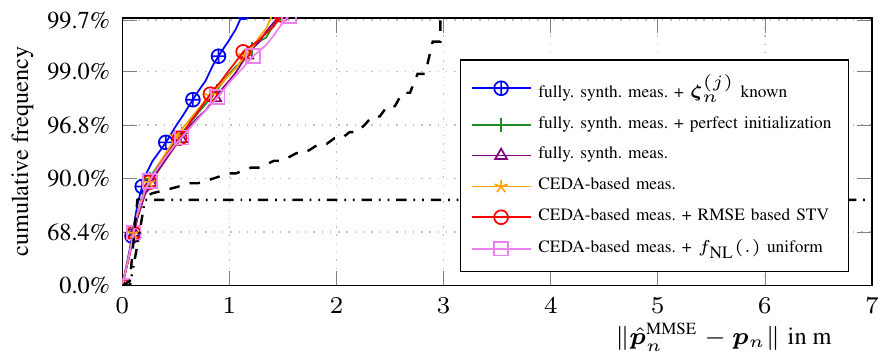}}
	\parbox[c]{0.5\textwidth}{\mseploti{cluster2}{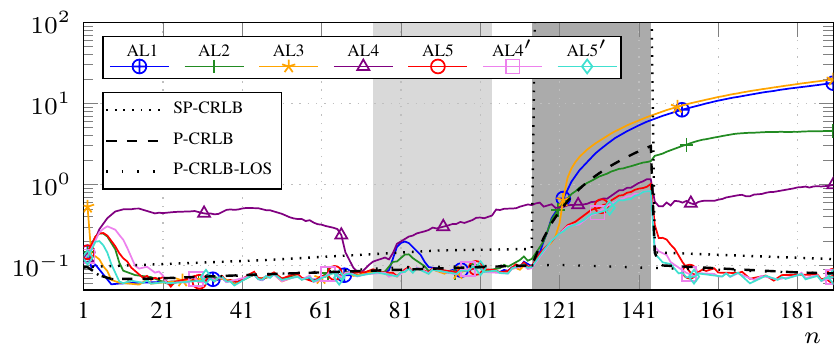}{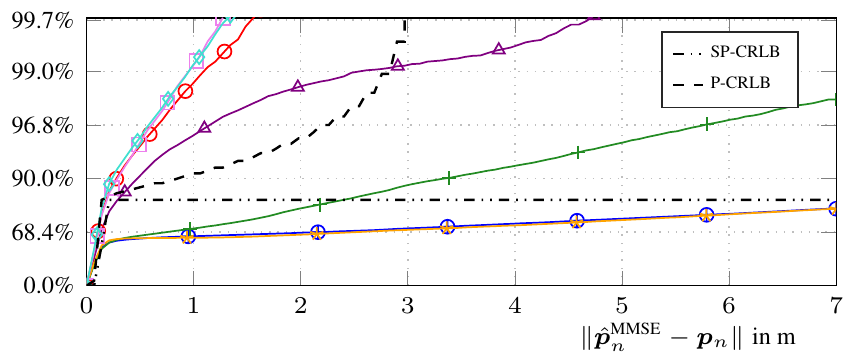}}
	
	\caption{\resultcaption{different algorithm variants}{from numerical simulation of stochastic multipath according to Sec.~\protect\ref{sec:performance_synthetic_dense}}\shadesofgraysynthetic
	}\label{fig:results_cluster}
\end{figure*}
We start by validating the system model presented in Sec.~\ref{sec:system_model}.
For this experiment the relatively defined \ac{stv} are set with respect to the true values instead of the RMSE values, given as
$\sigma_\text{u}= 0.05\,\tilde{u}_n^{(j)}$, 
$\sigma_\mathrm{\sdnrp} = 0.05\, \tilde{\sdnrp}_n^{(j)}  $, 
$\sigma_\text{b} = 0.05\, \tilde{b}_n^{(j)} $, 
$\sigma_{\mathrm{\gamma}_\text{f}} =0.05\, \tilde{\gamma}_{\text{f}\s n}^{(j)}$, 
$\sigma_{\mathrm{\gamma}_\text{r}} = 0.5\, \tilde{\gamma}_{\text{r}\s n} $.
Fig.~\ref{fig:state_space_cluster},~\ref{fig:nlos_statistic_cluster}, \ref{fig:mse_along_path_cluster} and \ref{fig:mse_cdf_cluster} show the results of the performed numerical simulations. 
Fig.~\ref{fig:state_space_cluster} shows \ac{mmse} estimates of all state variables as a function of time $t^\prime$ and compares to the respective true values. The \ac{mmse} estimates are determined according to \eqref{eq:mmse}-\eqref{eq:mmseplos} using both fully synthetic measurements and \ac{ceda}-based measurements.
Fig.~\ref{fig:nlos_statistic_cluster} compares distance-model-agnostic, \acfp{npe} of scale parameter and relative measurement frequency with the presented model functions, i.e., with the NLOS scale function \eqref{eq:nlos_scale_function} and the NLOS distance \ac{lhf} \eqref{eq:nlos_delay_lhf}. Each of the functions is determined both ways, using the \ac{mmse} estimates of $\RV{\zeta}_{200}^{(1)}$ of the last time step, given as $\hat{\bm{\zeta}}_{200}^{(1) \text{MMSE}}$ and using the respective true values used for simulation $\tilde{\bm{\zeta}}^{(1)}$. %
The \acp{npe} are determined using all \ac{nlos} measurements (the \ac{los} measurements are removed) of the last 20 time steps, given as $\rmv\rmv \{ \bm{z}_{n,m}^{(1)} \,|\,{m \rmv\rmv \in \rmv\rmv \mathcal{M}_n^{(1)} \rmv\rmv\rmv \setminus   \rmv\rmv \tilde{a}_n^{1}  ,  \rmv n \rmv\rmv \in  \rmv\rmv \{180,\, ...\, ,200\} } \} $. For details about the \acp{npe} see the supplementary material \mref{Sec.}{sec:app_nonparametric}.
This analysis is complemented by Figs.~\ref{fig:mse_along_path_cluster} and \ref{fig:mse_cdf_cluster} which show the position RMSE $e_n^{\mathrm{RMSE}}$ in two ways. First, as a function of the discrete observation time $n$ and, second, as the cumulative frequency of the \ac{rmse} evaluated over the whole time span. %
Fig.~\ref{fig:state_space_cluster} demonstrates that using CEDA-based measurements the MMSE estimates of the parameters of the \ac{nlos} \ac{lhf} %
 (i.e., the MMSE estimates corresponding to $\RV{\zeta}_n^{(1)}$ \vspace{0.3mm}) 
 are slightly biased, in particular the \ac{dnr} estimate $\hat{\sdnrp}_n^{(1)\s \text{MMSE}}$. This effect is a consequence of the asymptotic bandwidth assumption used in the derivation of the NLOS likelihood model (see \mref{Sec.}{sec:app_nlos_model}).  However, %
 as in Fig.~\ref{fig:state_space_cluster} the model functions parameterized with the MMSE values accurately fit the \acp{npe}, the MMSE estimate of the agent position $\hat{{\bm{p}}}^{\text{MMSE}}_n$ in Fig.~\ref{fig:state_space_cluster} remains unbiased and, thus, the positioning performance in Figs.~\ref{fig:mse_along_path_cluster} and \ref{fig:mse_cdf_cluster} using ``CEDA-based measurements" is identical to the performance using ``fully synthetic measurements" up to random deviations.

In addition, in Figs.~\ref{fig:mse_along_path_cluster} and \ref{fig:mse_cdf_cluster} we compare to fully synthetic measurements with (i) known initial state distributions, slightly lowering the RMSE around $n = 0$, and (ii) assuming the parameters of $\RV{\zeta}_n^{(j)}$ to be known constants, leading to a significant increase of performance at the end of the full \ac{olos} situation as the bias information does not vanish over discrete time $n$. With \ac{ceda}-based measurements we also compare to results where (i) we calculate the relatively defined \ac{stv} using the RMSE values of the respective last time step $n\minus 1$ according to Sec.~\ref{sec:simulation_model} and where (ii) we use a uniform delay intensity function $f_\text{NL}(\zd) = 1/d_\text{max}$ showing no significant degradation of performance. The latter result suggests that for low values of $\gamma$, the information provided in \eqref{eq:nlos_delay_lhf} is insignificant (c.f. Fig.~\ref{fig:nlos_like_gamma}). Therefore, in what follows, we \textit{keep the uniform delay intensity function} leading to a considerable reduction of runtime since $Q_0(\bm{p}_n,\bm{\zeta}_n^{(j)})$ does not need to be calculated (see also Sec.~\ref{sec:numerical1} and Sec.~\ref{sec:execution_time}).
Next, we investigate the influence of the individual features of our algorithm as described in Sec.~\ref{sec:simulation_model} and Table~\ref{tbl:algorithms}. 
Figs.~\ref{fig:mse_along_path_cluster2} and \ref{fig:mse_cdf_cluster2} show the \ac{rmse} of this experiment as a function of $t^\prime$ as well as the cumulative frequency of the \ac{rmse}. 
\begin{figure*}[t!]
	
	\centering
	\setlength{\abovecaptionskip}{0pt}
	\setlength{\belowcaptionskip}{0pt}
	
	\setlength{\figurewidth}{0.425\textwidthav}
	\setlength{\figureheight}{0.26\textwidthav}

	\def\datapath{./figures/meas_space_geometric}
	\hspace{-2mm}\scalebox{1.1}{\includegraphics{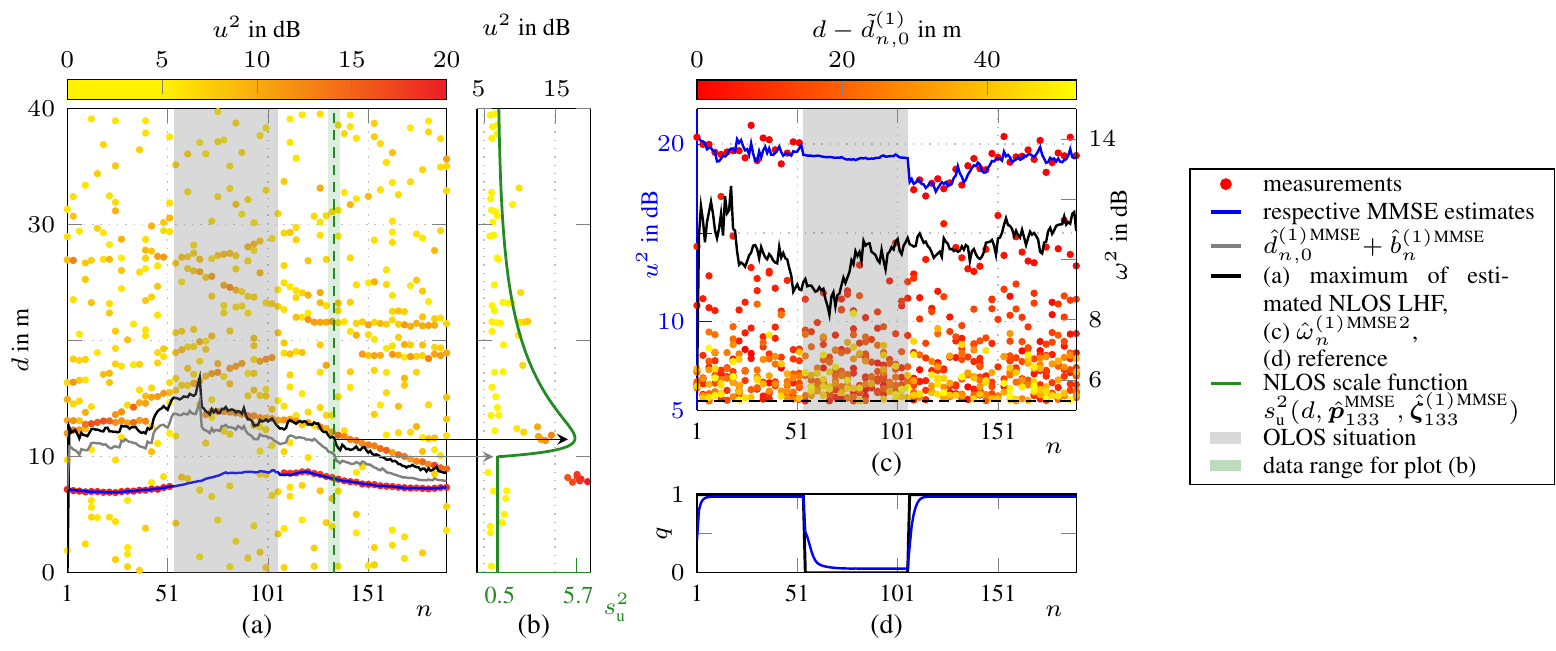}}
	\caption{A single measurement realization and the respective MMSE estimates using the proposed algorithm (AL5) for geometry-related synthetic measurements (see Sec.~\ref{sec:performance_synthetic_geometry}). We show \ac{mmse} estimates in (a) distance domain $\hat{d}_{n,0}^{(1)\s \text{MMSE}}$, (c) amplitude domain ($\hat{u}_{n}^{(1)\s \text{MMSE}\s 2}$, $\hat{\sdnrp{}}_{n}^{(1)\s \text{MMSE}\s 2}$), (d) LOS probability domain $\hat{q}_{n}^{(1)\s \text{MMSE}}$, and (b) amplitude as a function of distance domain for $n\in \{130,\, ... \, ,136\}$. The measurements in (b) correspond to the upper and in (c) to the left axis. Note that in (a) and (c) only measurements for every third time $n$ are shown (i,e., $n\in \{1,4,7,...\}$).}\label{fig:meas_space_geometric}
	\vspace*{-3mm}
\end{figure*}
The RMSE of the multi-sensor \ac{pdaai} (AL1) mostly attains the \ac{pcrlb} during \ac{los} and partial \ac{olos} situations. A reason for that is that the angle, which the remaining anchors A1 and A3 span with respect to the agent is sufficiently large to provide a reasonable position estimate. However, AL1 shows a slightly increased RMSE around $n = 80$ due to the agent direction change and significantly deviates from the very beginning of the full OLOS situation, losing the track in every single realization. 
Comparing the curves of AL2-AL5, one can conclude that every single algorithm feature investigated lowers the \ac{rmse} significantly when activated. 
The \ac{rmse} of the proposed algorithm AL5 constantly attains the \ac{pcrlb}, which indicates no lost track, even falling below the \ac{pcrlb} in full OLOS situations. This is possible as it leverages the additional position information contained inside the \acp{mpc} via the non-uniform \ac{nlos} \ac{lhf}, which is not considered by the \ac{pcrlb} model.
In contrast, AL2 loses a large percentage of tracks after the full \ac{olos} situation, because NLOS measurements significantly contribute to the LOS based position hypotheses due to the insufficient representation of the existence probability by the amplitude state particles (see Sec.~\ref{sec:pnl}).
While AL3 constantly attains the \ac{pcrlb} during the \ac{los} situation as well the partial \ac{olos} situation, it loses the track for every realization in full OLOS. After a short amount of time in which AL3 can maintain the agent position through the agent state transition model and the decreasing LOS probability, it identifies MPCs as the LOS component due to their coherent appearance and large amplitude, which is not covered by the uniform NLOS model, and loses the track.
AL4 shows a seemingly random performance degradation, which is due to the insufficient representation of the high dimensional joint state by the particle filter and some resulting lost tracks, which AL5 overcomes by decoupling the anchor states (see Sec.~\ref{sec:algorithm}). However, the discrepancy between AL4 and AL5 can be dissolved by using a sufficiently high number of particles (see AL4$^\prime$ and AL5$^\prime$), at the cost of significantly increasing the runtime (see Sec.~\ref{sec:execution_time}).

\subsubsection{Synthetic Measurements with Geometry-related Multipath} \label{sec:performance_synthetic_geometry}
In this section, we discuss results using synthetic measurements based on the simple floorplan shown in Fig.~\ref{fig:track_geometric}. The measurements are obtained by simulating a radio signal according to \eqref{eq:signal_model_sampled}, consisting of the \ac{los} component and \textit{specular} \acp{mpc}, and using the proposed \ac{ceda}. The \ac{mpc} delays are calculated out of the floorplan (i.e. W1-W5) using the mirror images (virtual anchors) up to the third order \cite{PedersenJTAP2018}. The \ac{snr} of the \ac{los} component as well as the \acp{mpc} \cite{LiTWC2022} are set to ${20}\,\mathrm{dB}$ at a distance of $1\,\mathrm{m}$ and are assumed to follow free-space path loss. The \ac{snr} of the individual \acp{mpc} are additionally attenuated by $3$~dB after each reflection (e.g., $6$ dB for a second-order reflection). As depicted in Fig.~\ref{fig:track_geometric}, for this experiment the anchors are obstructed by an obstacle (W5), which leads to partial and full \ac{olos} situations in the center of the investigated trajectory.
\begin{figure*}[t]
	
	\hspace{-2mm}
	\renewcommand{\myopi}{1}
	\parbox[c]{0.5\textwidth}{\mseploti{geometric}{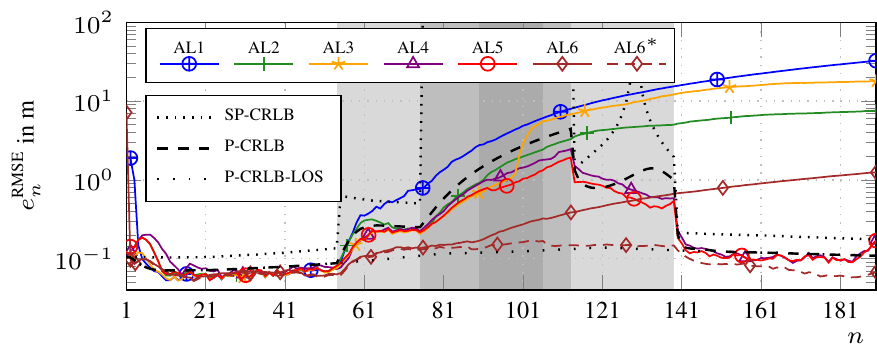}{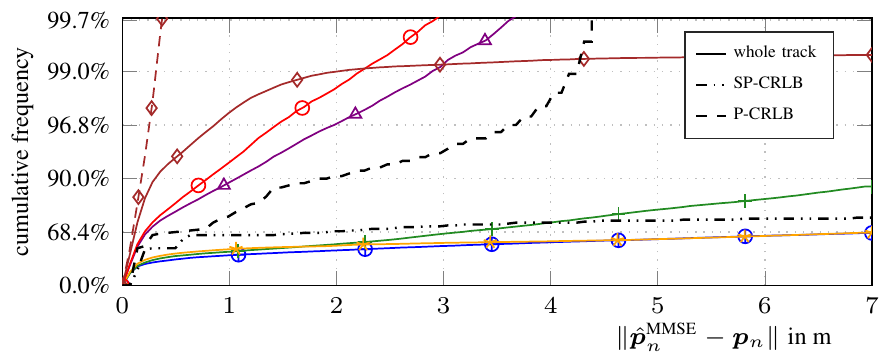}}
	\renewcommand{\myopi}{0}
	\parbox[c]{0.5\textwidth}{\mseploti{nxp}{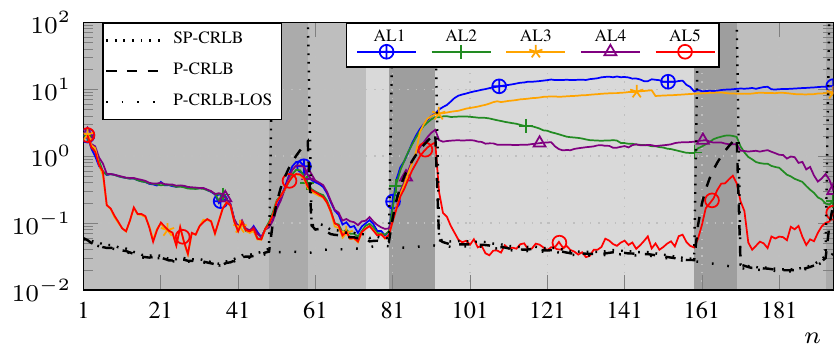}{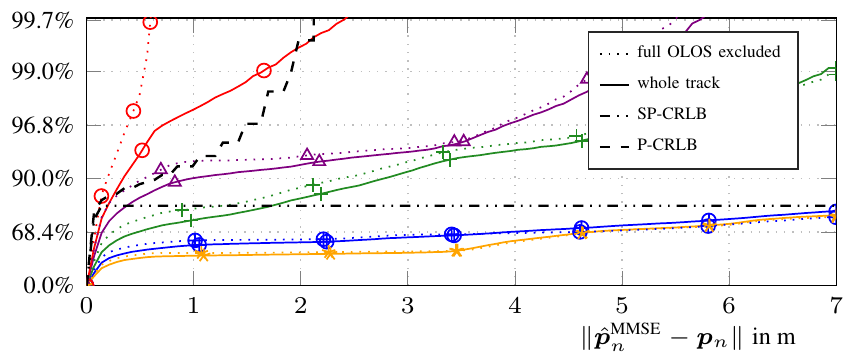}}
	
	\caption{\resultcaption{all algorithm variants of Table~\ref{tbl:algorithms}}{from numerical simulation of {specular} \acp{mpc} according to Sec.~\protect\ref{sec:performance_synthetic_geometry} in (a), (b) and using real radio measurements according to Sec.~\protect\ref{sec:performance_real} in (c),  (d)}\shadesofgraysynthetic
	}\label{fig:results_geometric}
	\vspace{-2mm}
\end{figure*}
Figs.~\ref{fig:meas_space_geometric}, \ref{fig:mse_along_path_geometric}, and  \ref{fig:mse_cdf_geometric} show results of the performed numerical simulation. Fig.~\ref{fig:meas_space_geometric} provides a graphical representation of the measurement space, showing a single measurement realization $\bm{z}$ together with the corresponding \ac{mmse} estimates of the proposed algorithm (AL5). The \ac{mmse} estimates are determined according to \eqref{eq:mmse}-\eqref{eq:mmseplos}. In particular, Fig.~\ref{fig:meas_space_geometric}a shows that (i) the MMSE estimate of the LOS delay $\hat{d}^{(j)\s \text{MMSE}}_{n,0} = d^{(j)}_{\text{LOS}} (\hat{\bm{p}}^{\text{MMSE}}_n)$ remains stable over the whole OLOS situation and that (ii) the maximum of the NLOS LHF follows the first \ac{mpc} available. We determine the shape of the NLOS LHF using the respective \ac{mmse} estimates of all \acp{rv} on which \eqref{eq:nlos_amplitude_lhf} depends. Fig.~\ref{fig:meas_space_geometric}c shows that the \ac{dnr} estimate $\hat{\sdnrp}_n^{(j)\s \text{MMSE}}$ accurately represents the dynamic behavior of the multipath energy, deceasing rapidly when the strongest, first \ac{mpc} is covered, while the \ac{snr} estimate $\hat{u}_n^{(j)\s \text{MMSE}}$ remains stable. For visualization, Fig.~\ref{fig:meas_space_geometric}b shows the NLOS scale function at time $n=133$ parametrized with the respective MMSE estimates of all NLOS function parameters. Fig.~\ref{fig:meas_space_geometric}d shows the LOS existence probability $q_n^{(1)}$ well representing the \ac{olos} situation.
Figs.~\ref{fig:mse_along_path_geometric} and  \ref{fig:mse_cdf_geometric} show the \ac{rmse} as a function of the discrete observation time $n$ as well as the cumulative frequency of the \ac{rmse}. 
Again, we investigate the influence of the individual features of our algorithm according to Sec.~\ref{sec:simulation_model} and Table~\ref{tbl:algorithms}.
Comparing the presented curves, we again observe AL5 to significantly outperform AL1-AL5, with the qualitative performance differences being almost identical to those of Sec.~\ref{sec:performance_synthetic_dense}. The only significant dissimilarity is the seemingly smaller deviation between AL4 and AL5. This is because AL4 does not lose any tracks during initialization, as the average energy and distance to the LOS component of the measurements of the first time step $n = 0$ are significantly lower in this scenario, leading to a better coverage of the state space by the particle filter. Thus, we only observe a slightly more unstable local behavior of AL4.
The \ac{mpslam} algorithm (AL6) achieves a significantly reduced RMSE during the first part of the OLOS situation, due to geometric information provided by the specular MPCs, outperforming the proposed method (AL5). However, the investigated scenario is geometrically ambiguous as there is little directional change in the agent movement \cite{KrekovicTSP2020}. Also there are many low-SNR components, which disappear and reappear, due to the obstacle (W5). This is why AL6 follows ambiguous paths for many realizations (i.e., it loses the track), leading to a significantly reduced performance after the full OLOS situation. We additionally added AL6$^{*}$, which represents the numerical results after removing $20.6\%$ (103 realizations) of diverged tracks. This result demonstrates the dramatically increased accuracy that can be obtained using \ac{mpslam}. 
\vspace{-1mm}
 \subsection{Performance for Real Radio Measurements} \label{sec:performance_real}

\begin{figure}[t]

	\centering
	
	\setlength{\abovecaptionskip}{0mm}
	\setlength{\belowcaptionskip}{0mm}

	\subfloat[\label{fig:car_ant}]{\includegraphics[height=0.20\textwidth]{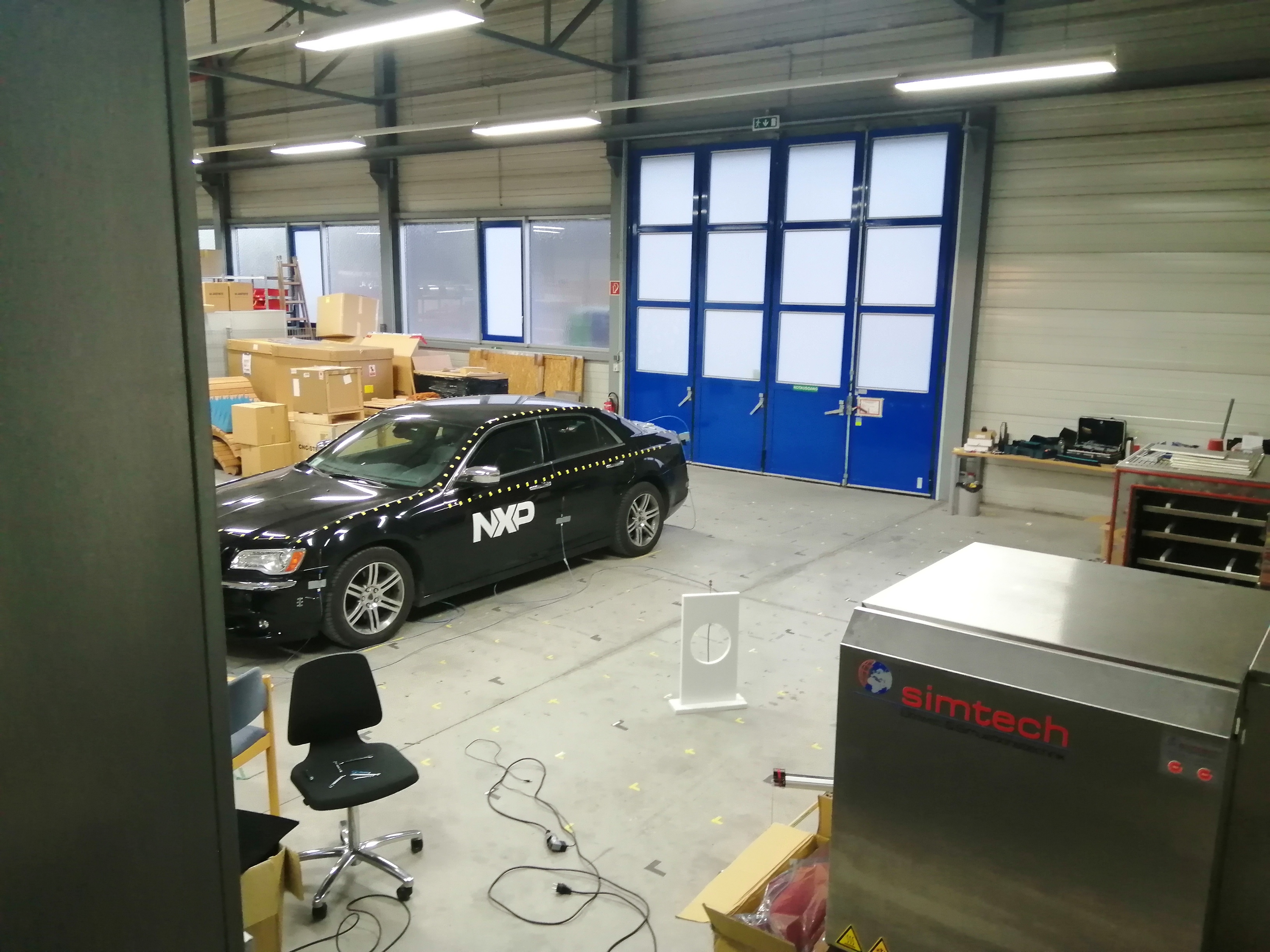}}\vspace{2mm}
	\subfloat[\label{fig:absorber}]{\includegraphics[height=0.20\textwidth]{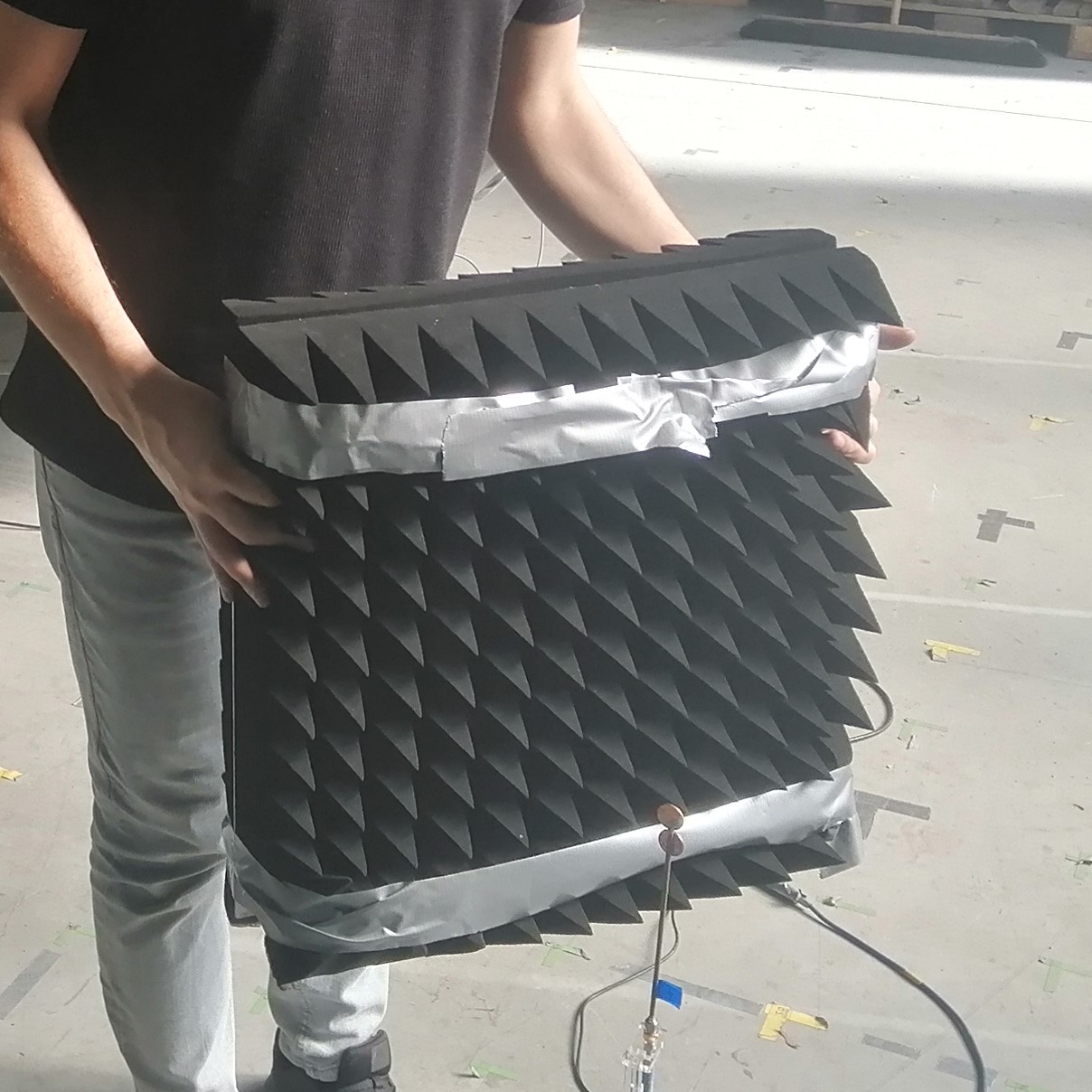}}
	\vspace{-3mm}
	\setlength{\figurewidth}{0.28\textwidth}
	\setlength{\figureheight}{0.28\textwidth}
	\captionsetup[subfloat]{captionskip=-3mm} 
	\subfloat[\label{fig:track_nxp}]{\hspace{-5mm}\includegraphics{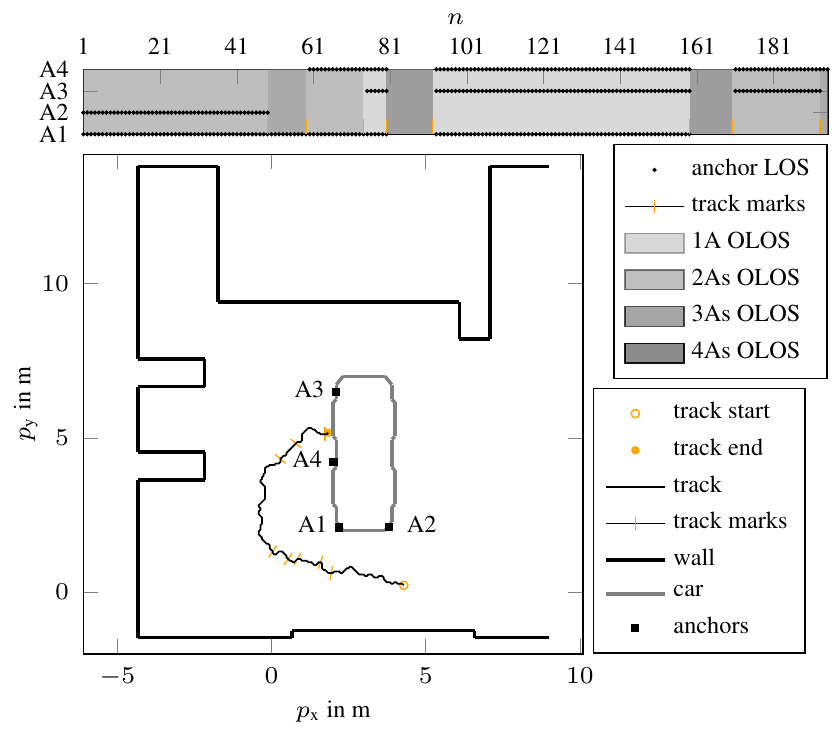}}

	\caption{Measurement setup for real radio-signal experiments described in Sec.~\ref{sec:performance_real}. We show pictures of (a) the overall scenario and (b) the OLOS setup used, as well as (c) the abstracted floorplan and trajectory.}\label{fig:measurement_setup_nxp}
	\vspace{-3mm}
\end{figure}

 For further validation of the proposed algorithm, we use real radio measurements collected in a laboratory hall of {NXP Semiconductors, Gratkorn, Austria}. The hall, shown in Fig.~\ref{fig:car_ant},  features a wide, open space and includes a demonstration car (Lancia Thema 2011), furniture, and metallic surfaces, thereby representing a typical multipath-prone industrial environment. An agent is assumed to move along a pseudo-random trajectory (selected out of a grid of agent positions), obtained in a static measurement setup. We selected $N=195$ measurements, assuming an observation rate of $\Delta T = 170\,\mathrm{ms}$. The agent velocity is set to vary around a magnitude of $0.35\,\mathrm{m/s}$. This leads to a corresponding continuous observation time $33.15\,\mathrm{s}$.
 At each selected position, a radio signal was transmitted from the assumed agent position, which was received by 4 anchors. 
 The agent was represented by a polystyrene build, while the anchor antennas were mounted on the demonstration car. The agent as well as the anchors were equipped with a dipole antenna with an approximately uniform radiation pattern in the azimuth plane and zeros in the floor and ceiling directions. The radio signal was recorded by an M-sequence correlative channel sounder with frequency range $3-10\,\mathrm{GHz}$. Within the measured band, the actual signal band was selected by a filter with root-raised-cosine impulse response $s(t)$, with a roll-off factor of $0.6$, a two-sided 3-dB bandwidth of $B=499.2\,\mathrm{MHz}$ and a center frequency of $7.9872\,\mathrm{GHz}$ (corresponding to channel 9 of IEEE 802.15.4a),
 and critically sampled with $T_\text{s} = 1/(1.6 \, B)$.
 We used $N_\text{s} = 161$ samples, amounting to a maximum distance of $d_\text{max} = 60\,\mathrm{m}$ for the \ac{ceda}.
 We created two full \ac{olos} situations at $n \in [80 , 92]$ and $n \in [159 , 170]$ using an obstacle consisting of a metal plate covered with attenuators as shown in Fig.~\ref{fig:absorber}. A floor plan showing the track, the environment (i.e, the car, other reflecting objects and walls), the antenna positions, and \ac{olos} conditions with respect to all antennas is shown in Fig.~\ref{fig:track_nxp}.
 The metal surface of the car strongly reflected the radio signal, leading to a radiation pattern of $270^\circ$ for A1 and A2 and $180^\circ$ for A3 and A4. Thus, during large parts of the trajectory the \ac{los} of 2 or 3 out of 4 anchors is not available. Moreover, the pulse reflected by the car surface strongly interferes with the \ac{los} pulse, leading to significant fluctuations of the amplitudes. {In addition, this leads to the channel estimator being prone to produce a high \ac{snr} component just after the \ac{los} component. As this violates our signal model, we processed the \ac{ceda} measurements attenuating all components, where $\zd \in \tilde{d}_{n,0}^{(j)} +\text[0,\; 2\, c\, T_\text{p}]$, except for the highest component.}
As only two antennas (A1 and A2) are visible at the track starting point, the position estimate obtained by trilateration is ambiguous. In the scenario presented, the relative antenna position with respect to the car can be assumed to be known. Thus, for this experiment, we used the antenna pattern as prior information for initialization of the position state. 
For the numerical evaluation presented, we added \ac{awgn} to the real radio signal obtained. 
We set $\norm{\bar{\V{r}}_{\mathrm{raw}}^{(j)}}{2}/\sigma^{(j)\s 2} = 20\,\mathrm{dB}$, where $\norm{\bar{\V{r}}_{\mathrm{raw}}^{(j)}}{2}$ is the average energy of the real measured signal per anchor $j$. 
Figs.~\ref{fig:mse_along_path_nxp} and \ref{fig:mse_cdf_nxp} show the \ac{rmse} as a function of the discrete observation time $n$ as well as the cumulative frequency of the \ac{rmse}. Again, we analyze the influence of the individual features of our algorithm according to Sec.~\ref{sec:simulation_model} and Table~\ref{tbl:algorithms} and observe AL5 to significantly outperform the other algorithm variants. Different to Sec.~\ref{sec:performance_synthetic_geometry} all presented algorithms fail to reach the \ac{pcrlb} over parts of the track. The exact consistency in progression of the \ac{rmse} curves suggests unmodeled effects (e.g. diffraction at the vehicle body) as well as inaccuracies in the reference as a probable reason.

\vspace{-1mm}
\subsection{Runtime} \label{sec:execution_time}
Table~\ref{tbl:execution_times} shows the average runtime of the proposed algorithm (A5) and compares it to the runtime of the multi-sensor \ac{pdaai} (AL1) and that of the \ac{mpslam} algorithm (AL6). All runtimes are estimated using Matlab implementations executed on an AMD Ryzen Threadripper 1900X 8-Core Processor with up to $4\,\text{GHz}$ for all scenarios investigated. We also show the average number of measurements (over all anchors and time steps) ${M}_\text{mean}$, the number of anchors $J$ and the number of particles, which determine the algorithm complexity per time step. The runtime of our algorithm (AL5) is of the same order of magnitude than that of the multi-sensor \ac{pdaai} (AL1), which is in the range of tens of milliseconds for all scenarios investigated. 
In contrast, the runtime of the \ac{mpslam} algorithm (AL6) is significantly higher, since it requires joint data association between all map features \cite{LeitingerTWC2019} and a higher number of particles for numerical stability. \vspace{-2.5mm}
\begin{table}[h]
	\renewcommand{\baselinestretch}{1}\small\normalsize
	\setlength{\tabcolsep}{3pt} 
	\renewcommand{\arraystretch}{1} 
	\centering
	\footnotesize
	
	\caption{Algorithm runtimes and characteristic values of all investigated scenarios.}\label{tbl:execution_times}
	\begin{tabular}{ r c c c c} 
		\toprule
		 & \textbf{particles} $I$ &  \textbf{Sec.~\ref{sec:performance_synthetic_dense}} &  \textbf{Sec.~\ref{sec:performance_synthetic_geometry}} &  \textbf{Sec.~\ref{sec:performance_real}} \\
		\midrule

		\multicolumn{1}{r}{\textbf{proposed (AL5)}} &5000& $ 53\,\mathrm{ms}$  & $ 34\,\mathrm{ms}$ & $30\,\mathrm{ms}$ \\
		\multicolumn{1}{r}{\textbf{\ac{pdaai} (AL1)}} &5000& $ 40\,\mathrm{ms}$  & $ 27\,\mathrm{ms}$ & $ 23\,\mathrm{ms}$ \\
		\multicolumn{1}{r}{\textbf{\ac{mpslam} (AL6)}} &30000& n.a. 
		  & $ 1.6\,\mathrm{s}$ & n.a. \\
		\midrule 
		 ${M}_\text{mean}\times J$ & & $28 \times 3 $ & $12 \times 3$ & $7 \times 4$ \\
		\bottomrule
	\end{tabular}
	\vspace{-2.5mm}
\end{table}

\section{Conclusion}\label{sec:conclusion}

 We have presented a particle-based \acf{spa} that sequentially estimates the position of a mobile agent %
using range and amplitude measurements provided by a snapshot-based \acf{ceda}. 
We introduced a novel \acf{nlos} model that is adapted to the \acf{dps} of the multipath radio channel. %
We analyzed the performance of the proposed algorithm using both numerically simulated and real measurements in different channel conditions and showed that the additional information provided by the \ac{nlos} model can support the estimation of the agent position. Our algorithm significantly outperformed the conventional \ac{pdaai} filter and consistently attained the \ac{pcrlb} in partial \ac{olos} situations (i.e., no lost tracks). %
While \acf{mpslam} can naturally outperform our algorithm in channels showing resolved, specular \acp{mpc}, we demonstrate the proposed algorithm to offer a significantly smaller number of lost tracks at reduced execution time {in a geometrically ambiguous scenario}. 
A possible direction for future research includes extending the model to multiple biases with respect to several \acp{mpc} using joint probabilistic data association and dynamic \ac{mpc} initialization \cite{MeyerProc2018, LiTWC2022} or to several \ac{mpc} clusters by using data association with extended objects \cite{MeyWilJ21}.
\renewcommand{\baselinestretch}{0.95}\small\normalsize 
\bibliographystyle{IEEEtran}
\bibliography{IEEEabrv,References,TempRefs}

\acrodef{mimo}[MIMO]{multiple input multiple output}
\acrodef{awgn}[AWGN]{additive white Gaussian noise}
\acrodef{crlb}[CRLB]{Cram\'er-Rao lower bound}
\acrodef{dmc}[DMC]{dense multipath component}
\acrodef{los}[LOS]{line-of-sight}
\acrodef{ml}[ML]{maximum likelihood}
\acrodef{mpc}[MPC]{multipath component}
\acrodef{nlos}[NLOS]{non-\acl{los}}
\acrodef{pdf}[PDF]{probability density function}
\acrodef{smc}[SMC]{specular multipath component}
\acrodef{snr}[SNR]{signal-to-noise-ratio}
\acrodef{sinr}[SINR]{signal-to-interference-plus-noise-ratio}
\acrodef{tdoa}[TDOA]{time difference of arrival}
\acrodef{toa}[TOA]{time-of-arrival}
\acrodef{aoa}[AOA]{angle-of-arrival}
\acrodef{uwb}[UWB]{ultra-wideband}
\acrodef{mse}[MSE]{mean squared error}
\acrodef{dps}[DPS]{delay power spectrum}
\acrodef{glrt}[GLRT]{generalized likelihood ratio test}
\acrodef{mse}[MSE]{mean squared error}
\acrodef{rmse}[RMSE]{root mean squared error}
\acrodef{nnlike}[NNLIKE]{normalized noise-free likelihood}
\acrodef{stdv}[STDV]{standard deviation}
\acrodef{rv}[RV]{random variable}
\acrodef{pda}[PDA]{probabilistic data association}
\acrodef{pmf}[PMF]{probability mass function}
\acrodef{pdaf}[PDAF]{probabilistic data association filter}
\acrodef{pdaai}[AIPDA]{amplitude-information probabilistic data association}
\acrodef{olos}[OLOS]{obstructed \ac{los}}
\acrodef{spa}[SPA]{sum-product algorithm}
\acrodef{mmse}[MMSE]{minimum mean-squared error}
\acrodef{lhf}[LHF]{likelihood function}
\acrodef{fa}[FA]{false alarm}
\acrodef{ceda}[CEDA]{channel estimation and detection algorithm} 
\acrodef{pcrlb}[P-CRLB]{posterior Cram\'er-Rao lower bound}
\acrodef{slam}[SLAM]{simultaneous localization and mapping}
\acrodef{mpslam}[MP-SLAM]{multipath-based SLAM}
\acrodef{dnr}[DNR]{dense-multipath-to-noise ratio}
\acrodef{stv}[STV]{state-transition variances}
\acrodef{npe}[BBE]{bin-based estimate}

 \begin{IEEEbiography}[{\includegraphics[width=25mm,height=32.15mm,clip,keepaspectratio]{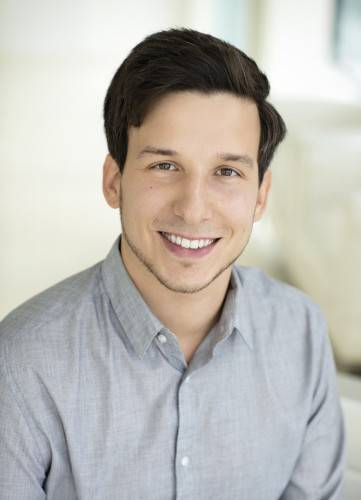}}]{Alexander~Venus} (S'20) received his B.Sc. and  Dipl.-Ing.\ (M.Sc.\ ) degrees (with highest honors) in biomedical engineering and information and communication engineering from Graz University of Technology, Austria in 2012 and 2015, respectively. He was a research and development engineer at Anton Paar GmbH, Graz from 2014 to 2019. He is currently a project assistant at Graz University of Technology, where he is pursuing his Ph.D. degree.
	
His research interests include radio-based localization and navigation, statistical signal processing, estimation/detection theory, machine learning and error bounds. 
\end{IEEEbiography}

\begin{IEEEbiography}[{\includegraphics[width=25mm,height=32.15mm,clip,keepaspectratio]{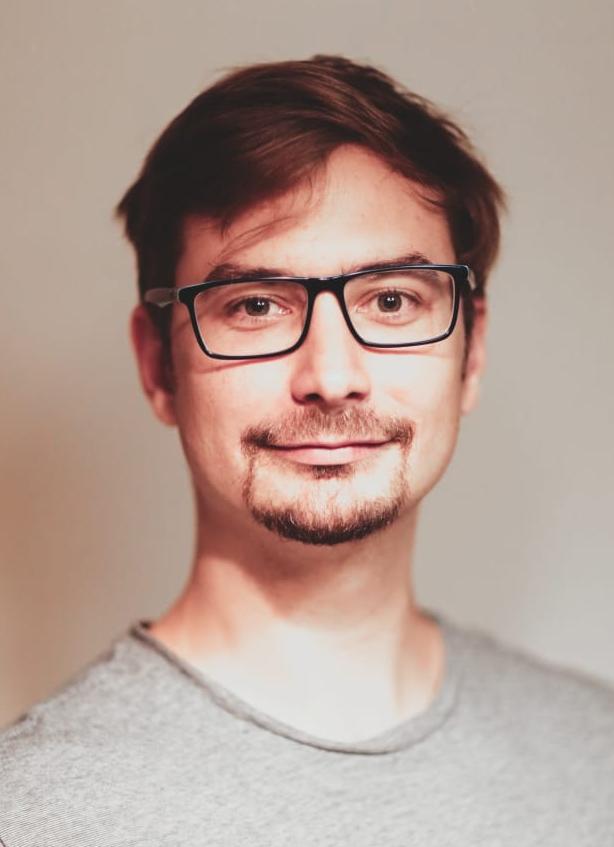}}]{Erik~Leitinger} (S'12--M'16) received his Dipl.-Ing.\ (M.Sc.\ ) and Ph.D.\ degrees (with highest honors) in electrical engineering from Graz University of Technology, Austria in 2012 and 2016, respectively. He was postdoctoral researcher at the department of Electrical and Information Technology at Lund University from 2016 to 2018. He is currently a University Assistant at Graz University of Technology.
	
	Dr. Leitinger served as co-chair of the special session "Synergistic Radar Signal Processing and Tracking" at the IEEE Radar Conference in 2021. He is co-organizer of the special issue "Graph-Based Localization and Tracking" in the Journal of Advances in Information Fusion (JAIF). Dr.\ Leitinger received an Award of Excellence from the Federal Ministry of Science, Research and Economy (BMWFW) for his Ph.D.\ Thesis. He is an Erwin Schr\"odinger Fellow. 
	
	His research interests include inference on graphs, localization and navigation, multiagent systems, stochastic modeling and estimation of radio channels, and estimation/detection theory. 
\end{IEEEbiography}

\begin{IEEEbiography}[{\includegraphics[width=25mm,height=32.15mm,clip,keepaspectratio]{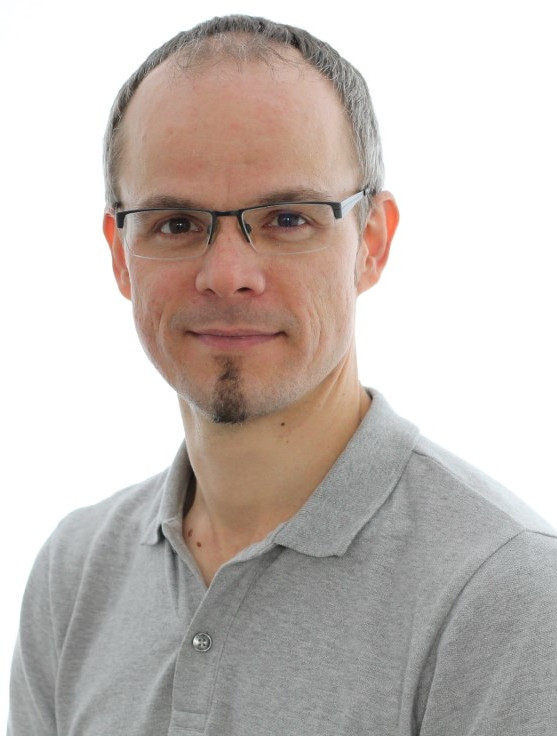}}]{Stefan~Tertinek} received the Dipl.-Ing. degree in electrical engineering from Graz University  of Technology, Graz, Austria, in 2007, and the Ph.D. degree in electrical engineering from University College Dublin, Dublin, Ireland, in 2011. From 2011 to 2018 he was with Danube Mobile Communications Engineering GmbH \& Co KG (majority owned by Intel Austria GmbH), Linz, Austria, as a RF System Engineer involved in research and product development of multiple generations of cellular RF transceiver and modem platforms. In 2018 he joined NXP Semiconductors Austria GmbH \& Co KG as a RF System Architect in the Product Line Secure Car Access, where he works on ultra-wideband (UWB) and Bluetooth radio technologies with a focus on localization, radar and machine learning.
\end{IEEEbiography}

\begin{IEEEbiography}[{\includegraphics[width=1in,height=1.25in,clip,keepaspectratio]{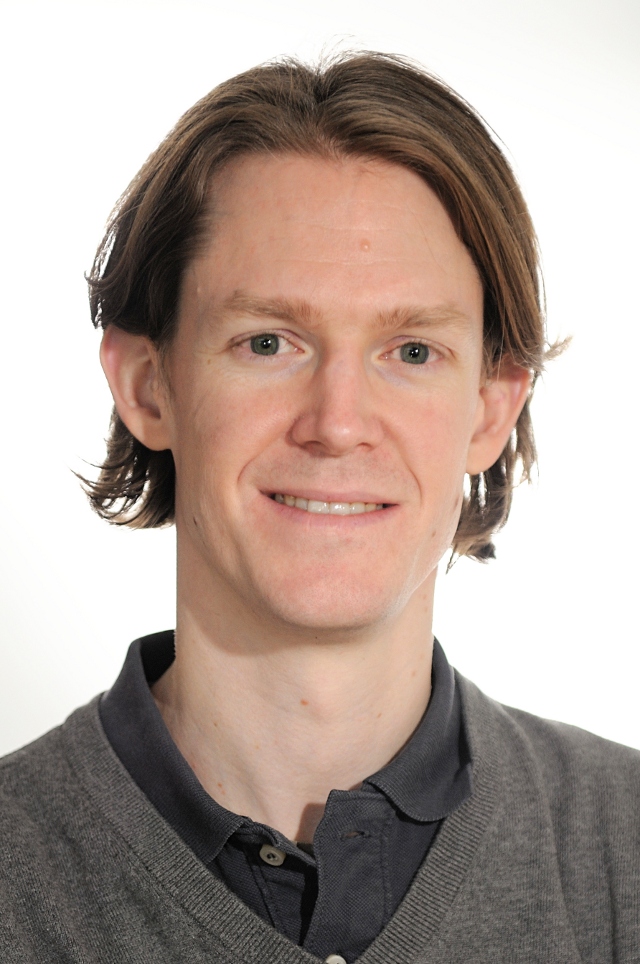}}]{Klaus~Witrisal} (S'98--M'03) received the Ph.D. degree (cum laude) from Delft University of Technology, Delft, The Netherlands, in 2002, and the Habilitation from Graz University of Technology in 2009. He is currently an Associate Professor at the Signal Processing and Speech Communication Laboratory (SPSC) of Graz University of Technology and head of the Christian Doppler Laboratory for Location-aware Electronic Systems. His research interests are in signal processing for wireless communications, propagation channel modeling, and positioning. Klaus Witrisal served as an associate editor of IEEE Communications Letters, co-chair of the TWG ``Indoor" of the COST Action IC1004, cochair of the EWG ``Localisation and Tracking" of the COST Action CA15104, leading chair of the IEEE Workshop on Advances in Network Localization and Navigation (ANLN), and TPC (co)-chair of the Workshop on Positioning, Navigation and Communication (WPNC). 
\end{IEEEbiography}

\end{document}


\title{\huge{A Graph-based Algorithm for Robust Sequential Localization Exploiting Multipath for Obstructed-LOS-Bias Mitigation:\\ Supplementary Material}}
\author{\large{Alexander Venus, Erik Leitinger, Stefan Tertinek, and  Klaus Witrisal}\\[2mm]\small{June 2022}}
\maketitle
\frenchspacing

\noindent This manuscript provides additional analysis for the publication ``A Graph-based Algorithm for Robust Sequential Localization Exploiting Multipath for Obstructed-LOS-Bias Mitigation'' by the same authors \cite{Main}.
\vspace{-2mm}

\newcommand{\sn}{\ensuremath{s_{\mathrm{N}}({\tau,\sigma)}}}
\newcommand{\snq}{\ensuremath{s^2_{\mathrm{N}}({\tau  ,\sigma)}}}
\newcommand{\fun}{\ensuremath{ f({u}_\text{N};{ \mathcal{H}_{1}, \tau, \sigma)}}}
\section{Derivation of the NLOS likelihood function} \label{sec:app_nlos_model}
Previous work \cite{Yu2020, VenusRadar2021} %
presents non-uniform \acf{nlos} models for delay/distance measurements consisting of a weighted mixture of two distribution functions: First, a uniform distribution modeling false-alarm measurements, and, second, different types of exponentially decaying functions modeling \acp{mpc}. Thereby, the \acp{mpc} are represented by the typical exponential path loss of expected received power as observable in radio propagation channels \cite{MolischBook2012}. %
More specifically, \cite{Yu2020} uses an approximate convolutive model for delay measurements only, where a single exponential kernel is convolved with a Gaussian function, while \cite{VenusRadar2021} approximates the distribution of delay measurements using a double exponential function and a conventional \ac{awgn} Rayleigh model for the amplitude measurements. Both of these models are heuristically motivated and lack of an accurate description of the joint \ac{nlos} \ac{lhf} for delay and amplitude measurements of a \ac{ceda}.
Hence, we seek to find an accurate model for the \ac{nlos} \ac{lhf}, which captures the \ac{mpc}-related statistic of the multipath radio channel.

We want to determine the statistic of the stochastic, \ac{nlos} fraction of our signal model in \meqref{eq:signal_model_sampled_stochastic}. To this end, we define the \ac{nlos}-only model
\vspace{-1mm}
\begin{equation}\vspace{-1mm} \label{eq:noise_sampled_dmc}
	\RV{r}_{\text{N}\s n}^{(j)} =  \int \rmv \rmv  \V{s}(\tau) \rv{\nu}_{\text{D}\s n}^{(j)}(\tau) \, \mathrm{d} \tau + \RV{\noise{}}_{n}^{(j)}
\end{equation}
with $\V{s}(\tau)$, $\rv{\nu}_{\text{D}\s n}^{(j)}(\tau)$ and $\RV{\noise{}}_{n}^{(j)}$ defined in accordance to the main text (see \mref{Sec.}{sec:signal_model}). Thus, $\RV{r}_{\text{N}\s n}^{(j)}$ is a zero-mean circularly-symmetric complex Gaussian random vector, with covariance matrix corresponding to \meqref{eq:cov_dmc}. %

Since we only consider a single radio signal snapshot in the current analysis, in the following we omit the indices for time $n$ and anchor $j$ for brevity of notation.

We base our analysis on the optimum estimation and detection method of a {single signal component} in \ac{awgn}\footnote{Note that the amplitude model of the commonly used \ac{pdaai}, which is the basis for the \ac{los} model presented in \mref{Sec.}{sec:los_model}, is derived using the same approach\cite{LerroACC1990}}: The \ac{ml} estimator for the normalized amplitude reads \cite{Kay1993}
\begin{align} \label{eq:ml_normalized_amplitude}
u_{\text{ML}}(\V{r}) &=  \argmax_{u, \tau, \sigma} f_\text{CN} ({u}; \V{r}^{\text{H}} \bm{s}({\tau}) ,  \sigma ) \nonumber \\ 
&= \max_{\tau, \sigma} \frac{ |\V{r}^{\text{H}} \bm{s}({\tau})|}{\sigma \; \norm{\bm{s}({\tau})}{}  } %
\end{align}
with $f_\text{CN}(\rv{x}; \mu, \sigma)$ being a circular-symmetric complex Gaussian distribution with mean $\mu$ and standard deviation $\sigma$.
Accordingly, the \ac{glrt}\cite{Kay1998} can be defined using \eqref{eq:ml_normalized_amplitude} as 
\vspace*{-1mm}
\begin{equation} \label{eq:glrt} 
u_{\text{ML}}(\V{r})  \overset{\mathcal{H}_{1}}{\underset{\mathcal{H}_{0}}{\gtrless}} \gamma 
  \vspace{-1mm}
\end{equation}
with the test statistic being equivalent to $u_{\text{ML}}(\V{r})$.

We are interested in the statistic of the normalized amplitude estimate when the \ac{glrt} decides $\mathcal{H}_{1}$%
, i.e., we consider %
\begin{equation} \label{eq:rv_trans}
f( \rv{u}_\text{N} ; \mathcal{H}_{1}) ~~\text{with}~~	\rv{u}_\text{N} = {u}_{\text{ML}}(\RV{r}_{\text{N}})\,.
\end{equation}
A joint \ac{ceda} in the sense of \mref{Sec.}{sec:channel_estimation} attempts to decompose the received signal into a finite number of individual, decorrelated components. 
Consider that \eqref{eq:rv_trans} implies the assumption that the \ac{ceda} is not able to decompose elements of the convolution in \eqref{eq:noise_sampled_dmc}, i.e., it is not able to decorrelate the elements of $\RV{r}_{\text{N}}$ and to reduce the cross-terms of the noise covariance matrix \meqref{eq:cov_dmc}. {This assumption holds in good approximation, as the integral in \eqref{eq:noise_sampled_dmc} equivalently models a sum of a non-countable infinity of signal components with infinitesimal spacing and, thus, the influence of the decomposition procedure onto the resulting statistic is negligible.

Now we assume that %
delay ${\tau}$ and noise standard deviation $\sigma$ are known, neglecting the influence of jointly estimating these parameters along with the normalized amplitude ${u}$ in \eqref{eq:ml_normalized_amplitude}. Then $\rv{u}_\text{N} ; \mathcal{H}_{1}$ follows a truncated Rayleigh distribution \meqref{eq:truncated_rayleigh_pdf} cut off at the detection threshold $\gamma$, given as
\begin{equation} \label{eq:full_ampl_model}
\fun = f_\text{TRayl} ({u}_\text{N};	\sn , \, \gamma )
\end{equation}
with the squared Rayleigh scale parameter being
\begin{equation} \label{eq:nlos_scale_function_full}
\snq =  \frac{1}{2} \Big( \frac{C_\text{S}({\tau})}{{\sigma}^2 \norm{\bm{s}({\tau})}{2}} + 1 \Big) \,. 
\end{equation}
$C_\text{S}({\tau})$ denotes the covariance function of the zero-mean, Gaussian inner product $\RV{r}_{\text{N}}^{\text{H}} \; \bm{s}({\tau})$, given as%
\begin{align} \label{eq:cov_projection}
	&C_\text{S}({\tau}) \triangleq \E{|\RV{r}_{\text{N}}^{\text{H}}\; \bm{s}({\tau})|^2} \nn\\ &= \int  \bm{s}({\tau})^{\text{H}}\bm{s}(\tau^\prime)\, \bm{s}(\tau^\prime)^\text{H} \bm{s}({\tau})\, S_{\mathrm{D}}(\tau^\prime%
	)\, \mathrm{d} \tau^\prime  + \norm{\bm{s}(\tau)}{2} {\sigma}^2
\end{align}
with the \ac{dps} $S_{\mathrm{D}}(\tau)$, defined according to \meqref{eq:dps_fun} (see the main text, \mref{Sec.}{sec:dps_model} for further discussion).
Assuming bandwidth and respective sampling time to approach infinity, the inner product $\bm{s}({\tau})^{\text{H}}\bm{s}(\tau^\prime)$ in \eqref{eq:cov_projection} approaches $\norm{\bm{s}({\tau})}{2}\, \delta(\tau^\prime - {\tau})$, where $\delta(\cdot)$, denotes the Dirac delta distribution. Inserting into   \eqref{eq:nlos_scale_function_full} yields
\begin{equation} \label{eq:nlos_scale_function_approx}
\snq = {\frac{1}{2}} \Big( \frac{\norm{\bm{s}({\tau})}{2} S_{\mathrm{D}}  ({\tau})}{{\sigma}^2 } + 1 \Big) %
\end{equation}%
which is equivalent to \meqref{eq:nlos_scale_function} and, thus, \eqref{eq:full_ampl_model} becomes equivalent to \meqref{eq:nlos_amplitude_lhf}.

While the effect of estimating ${\sigma}$ becomes negligible\footnote{For unknown $\sigma$, the statistic of two times the squared normalized amplitude $2\, \rv{u}_\text{N}^{2}$ is described by a Fisher distribution \cite[Ch. 15.10.3]{AdlerTaylor2007RandomFieldsGeometry} with numerator degrees of freedom equal to $2$ and denominator degrees of freedom equal to $2\, {N_{\text{s}}}$. For large ${N_{\text{s}}}$ the statistic of  $2\, \rv{u}_\text{N}^{2}$ can be well approximated by a $\chi^2$ distribution \cite[Ch. 2.2]{Kay1998} and, therefore, the statistic of $\rv{u}_\text{N}$ by the Rayleigh distribution described in the main text.} for a large number of samples ${N_{\text{s}}}$, 
which is true for wideband ranging applications in general\cite{MolischTAP2006}, estimating ${\tau}$ leads to a small but visible bias in the scale parameter\cite{LeitingerAsilomar2020}. %
Following the steps of \cite{LeitingerAsilomar2020}, this bias, which is usually ignored in amplitude models \cite{LerroACC1990,LeitingerICC2019,LiAsilomar2020,KirubarajanTAES1996}, can be represented by replacing the truncated Rayleigh distribution of \eqref{eq:full_ampl_model} with a truncated Rician distribution \meqref{eq:truncated_rice_pdf} with non-centrality parameter of $\sqrt{0.5}$.
However, different from \cite{LeitingerAsilomar2020} the expected power of the stochastic process in \eqref{eq:noise_sampled_dmc} is not a constant with respect to the elements of $\RV{r}_{\text{N}}$ and, thus, neither is the corresponding scale parameter \eqref{eq:nlos_scale_function_full} with respect to ${\tau}$. To take this behavior into account, we model the expected amplitude of the stochastic process to be \textit{constant in the local environment}, i.e., we modify the non-centrality parameter of $\sqrt{0.5}$ by the ratio of the current scale parameter to the scale parameter for \ac{awgn}, given as $\sn/\sqrt{0.5}$. We get %
\begin{equation} \label{eq:full_ampl_model_bias}
	\fun \rmv  = \rmv  f_\text{TRice} ({u}_\text{N}; \sn ,	\sn  ,  \gamma   )\, .
\end{equation}

We validate the above model by a numerical simulation study. The results are provided in Fig.~\ref{fig:spread_approximation}.
We show \eqref{eq:nlos_scale_function_full} and  \eqref{eq:nlos_scale_function_approx} together with estimates obtained by applying the proposed \ac{ceda} (see Sec.~\ref{sec:app_ceda}) to simulated radio signals. The simulated radio signals are obtained using numerical simulation according to \eqref{eq:noise_sampled_dmc}, i.e., they consist of a stochastic process only. The \ac{dps} parameters are set to constant values, chosen in line with \mref{Sec.}{sec:performance_synthetic_dense} for simulation. We used $\gamma = 0$ dB in order accept all \ac{ceda} estimates. For simplicity, the distance to the \ac{los} component was assumed to be equal to zero. For this experiment, we use two versions of the \ac{ceda} that solve the optimization problem in Sec.~\ref{sec:app_ceda}, \eqref{eq:ml_tau} in two ways: First, we use the normal variant as suggested in Sec.~\ref{sec:app_ceda} involving continuous unconstrained optimization \cite{Lagarias1998} (\acs{npe} Rayleigh opt, \acs{npe} Rice opt.) and, second, we use grid-based optimization only, where the grid values correspond to the sampling grid of the radio signal (\acs{npe}~Rayleigh~grid). Out of the estimates provided by the \ac{ceda} we then estimate the scale parameter by grouping the estimates into delay/distance bins and applying, respectively, the maximum likelihood estimator for a truncated Rayleigh distribution (\acs{npe} Rayleigh opt, \acs{npe} Rayleigh grid), or truncated Rice distribution (\acs{npe} Rice opt.); see Sec.~\ref{sec:app_nonparametric} for details abut the bin-based estimation process. We simulated $600$ signals amounting to approximately $500$ samples per estimation bin (empirically determined value) at the signal parameters and detection threshold configured. 

Analyzing Fig.~\ref{fig:spread_approximation} one can observe that the simplified, asymptotic model \eqref{eq:nlos_scale_function_approx} significantly underestimates the spread parameter as it neglects correlations in $C_\text{S}({\tau})$ occurring due to finite signal bandwidth. The correlation-aware model in \eqref{eq:full_ampl_model} accurately represents the spread parameter for grid-based estimates of ${\tau}$ (\acs{npe}~Rayleigh~grid). However, continuous optimization of ${\tau}$ (\acs{npe} Rayleigh opt) leads to an offset, which can be considered using the Rician model \eqref{eq:full_ampl_model_bias} instead (\acs{npe} Rice opt.). However, as we assumed the noise level to be constant in the local environment, \eqref{eq:full_ampl_model_bias} cannot represent the influence of finite signal bandwidth with respect to estimation of ${\tau}$. This effect is visible at lag $0$, due to the rapid change of variance in the adjacent region.

\begin{figure}[t]
	\centering
	\setlength{\figurewidth}{0.425\textwidthav}
	\setlength{\figureheight}{0.2\textwidthav}
	\setlength{\abovecaptionskip}{0pt}

	\def\datapath{./figures/spread_approximation}
	\includegraphics{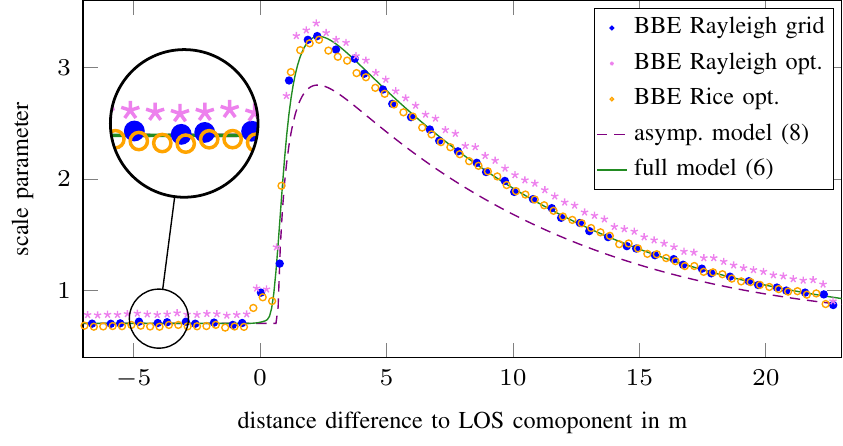}
	\caption{\Acfp{npe} of the scale parameter of the respective truncated Rayleigh or Rician distribution v.s. the \ac{nlos} scale function \meqref{eq:nlos_scale_function} as a function of the difference of a distance measurement to the LOS component distance. The non-parametric estimates are determined as shown in Sec.~\ref{sec:app_nonparametric}.}
	\vspace*{-4mm}
	\label{fig:spread_approximation}
\end{figure}

The above analysis showed that the approximate model consisting of \eqref{eq:full_ampl_model} and \eqref{eq:nlos_scale_function_approx} is insufficient as a model for \textit{generating measurements}. However, in \mref{Sec.}{sec:performance_synthetic_dense} we demonstrate it to suffice as a model for \textit{inference}: We obtain no loss in performance\footnote{in terms of the \ac{rmse} of the agent position estimate} of the proposed algorithm when comparing results with data generated according to \eqref{eq:nlos_scale_function_approx} to results using the stochastic radio signal model \meqref{eq:signal_model_sampled_stochastic} and the proposed \ac{ceda} (see Sec.~\ref{sec:app_ceda}) for generating measurements\footnote{Indeed, preliminary simulations using \eqref{eq:full_ampl_model} even showed significantly worse performance of the proposed algorithm, due to local minima introduced by the oscillating nature of $s(t)$ leading to unstable behavior of the particle-based implementation.}. In contrast, the runtime of the overall algorithm using the approximate model is orders of magnitude lower than using the full model consisting of \eqref{eq:full_ampl_model_bias} and \eqref{eq:nlos_scale_function_full}, especially since the latter requires numerical approximation of the convolutions in \eqref{eq:cov_projection}.
See \mref{Sec.}{sec:performance_synthetic_dense} for further details.

\section{\Acf{dnr} Initialization} \label{sec:app_inr_init}
First we determine the ML estimator for a set of i.i.d. samples $\mathcal{X} = \{x_1,...,x_{|\mathcal{X}|}\}$ following a truncated Rayleigh distribution, i.e. we solve $\argmax_{s} \prod_{x\in \mathcal{X}} f_\text{TRayl}(x; s , \lambda)$. This can be done in a straightforward manner by calculating the first derivative of \meqref{eq:truncated_rayleigh_pdf} and equating to zero. We find 
\begin{equation} \label{eq:ml_tunc_rayleigh}
	s^2_\text{ML}(\mathcal{X},\lambda) =\frac{1}{2 |\mathcal{X}_{>\lambda}|} \sum_{x \in \mathcal{X_{>}\lambda}} x^2 - \lambda^2 ,
\end{equation}
where $\mathcal{X}_{>\lambda} = \{x \in \mathcal{X} \, |\, x> \lambda \}$.

Next, we determine the integral of the scale parameter $s^2_\text{u} (\zd,\V{p}_n,\V{\zeta}_n^{(j)})$ from \meqref{eq:nlos_scale_function} over $\zd$ as
\begin{equation} 
\int_0^{d_\text{max}} \rmv\rmv\rmv\rmv s^2_\text{u} (d) \, \mathrm{d} d =  \frac{1}{2} \,\Big(\, {\sdnrp}_\mathrm{init}^{(j)\s 2} \rmv\rmv \int_0^{d_\text{max}} \rmv\rmv\rmv\rmv \bar{S}_{\text{D}}(d) \, \mathrm{d} d  \, + \rmv\rmv \int_0^{d_\text{max}} \rmv\rmv\rmv\rmv\rmv\rmv  \mathrm{d} d \, \Big)
\end{equation}
dropping the dependence on $\V{p}_n$, ${\V{\zeta}}_n^{(j)}$ and $\bar{\V{\zeta}}_n^{(j)}$.
Evaluating the integrals on the right-hand side and reordering yields
\begin{equation} \label{eq:omega_int}
	{\sdnrp}_\mathrm{init}^{(j)} = 2\, \int_{0}^{d_\text{max}} \rmv s^2_\text{u} (d) \,  \mathrm{d} d - d_\text{max} . %
\end{equation}	
Next, we assign all $M_0^{(j)}$ normalized amplitude measurements of the initial time step $\zuZero$, except the measurement with the largest normalized amplitude\footnote{Note that when the SNR is high the LOS measurement tends to show the largest normalized amplitude, which, if not excluded, biases the \ac{dnr} estimate significantly. When the SNR is low, its influence is negligible.} into $k \in \{ 1 ,\, ... \,, N_{\sdnrp}^{(j)}\}$ equally spaced bins, depending on the value of their corresponding distance measurements $\zdZero$ (see also \mref{Sec.}{sec:channel_estimation}). The discussed bins are given as the sets $\mathcal{U}^{(j)}_{\sdnrp\s k} = \{ \zuZero \;|\; m \in \mathcal{M}_0^{(j)} \rmv\rmv\setminus  m^{(j)}_\text{max} , d_{\mathrm{\omega}\s  k\minus 1} \rmv\rmv\leq\rmv\rmv \zdZero{} \rmv\rmv\leq\rmv\rmv  d_{\mathrm{\omega}\s  k} \}$ with $ m^{(j)}_\text{max} = \argmax_{m\in \mathcal{M}_0^{(j)}} \zuZero $, $d_{\mathrm{\omega}\s k}=k\, \frac{d_\text{max}}{N_\sdnrp^{(j)}}$ %
and $N_\sdnrp^{(j)} = 2 + \lfloor M_0^{(j)}/3 \rfloor$, where the divisor $3$ and the offset of $2$ were set empirically. %
We numerically approximate the integral in \eqref{eq:omega_int} by individually estimating the scale parameter for each bin using \eqref{eq:ml_tunc_rayleigh} and summing the rectangles formed by each bin, i.e.,
\vspace{-2mm}
\begin{equation}
	{\sdnrp}_\mathrm{init}^{(j)} \approx 2\, \frac{d_\text{max}}{N_\sdnrp^{(j)}} \sum_{k = 1}^{N_\sdnrp^{(j)}} \rmv s^2_\text{ML}(\mathcal{U}^{(j)}_{\sdnrp\s k},\lambda) - d_\text{max}.
\end{equation}	

\section{Normalization of the NLOS Distance Likelihood} \label{sec:app_numerical}

As discussed in \mref{Sec.}{sec:numerical1}, the normalization constant $Q_0(\bm{p}_n,\bm{\zeta}_n^{(j)})$ in \meqref{eq:nlos_delay_lhf} cannot be found analytically. For computational efficiency, we approximate the integral using the trapezoid rule\cite{Atkinson1989} as $Q_0(\bm{p}_n,\bm{\zeta}_n^{(j)}) \approx \sum_{k=1}^{K_\text{T}} \frac{f_\text{int}(d^{(j)}_{n,k-1}) + f_\text{int}(d^{(j)}_{n,k})}{2} \Delta d^{(j)}_{n,k}$ where $f_\text{int}(d) = \text{exp}({{-\gamma^2 / (2\, s_\text{u}^2 (d,\bm{p}_n,\bm{\zeta}_n^{(j)}) }} ) )$ with $\Delta d^{(j)}_{n,k} = d^{(j)}_{n,k} - d^{(j)}_{n,k-1}$ and supporting points chosen non-uniformly at $d^{(j)}_{n,0} = 0$, $d^{(j)}_{n,1} = d_{\text{LOS}\s}^{(j)}(\bm{p}_n)+b_n^{(j)}$,  $d^{(j)}_{n,K_\text{T}} = d_\text{max}$ and, for $ 2\leq k \leq K_\text{T}-1$, 
$d^{(j)}_{n,k} = \text{exp}(\frac{\text{ln}(d^{(j)}_{n,K_\text{T}})-\text{ln}(d^{(j)}_{n,1})}{K_\text{T}-1}\, (k-1)+\text{ln}(d^{(j)}_{n,1}))$.  Fig.~\ref{fig:cluster_normalization}~visualizes the discussed approximation scheme.

\section{Bin-based estimates (BBE)} \label{sec:app_nonparametric}
This section discusses the bin-based estimation of the statistic of the \ac{nlos} process, as needed for evaluation in \mref{Sec.}{sec:performance_synthetic_dense} (esp. \rmv \mref{Fig.}{fig:nlos_statistic_cluster} )
and Sec.~\ref{sec:app_nlos_model} (esp. Fig.~\ref{fig:spread_approximation}). %
We assume all selected measurements to be %
denoted by $\hat{u}_{\text{BB}\s m}$ and $\hat{d}_{\text{BB}\s m}$, respectively, with $m \in \mathcal{M}_\text{BB} = \{ 1 , \, ... \,  , M_\text{BB}\}$ and $M_\text{BB} = |\mathcal{M}_\text{BB} |$ being the number of selected measurements.
Similar to Sec.~\ref{sec:app_inr_init}, we assign all selected measurements %
to $k \in \mathcal{N}_\text{BB} =  \{ 1 ,\, ... \, , N_\text{BB}\}$ equally spaced bins. For the normalized amplitudes, we define $\mathcal{U}_{\text{BB}\s k} = \{\hat{u}_{\text{BB}\s m} \;|\; m \in \mathcal{M}_\text{BB} \,,  d_{\text{L}\s k\minus 1} \leq \hat{d}_{\text{BB}\s m} \leq d_{\text{L}\s k}  \}$ and for the distances $\mathcal{D}_{\text{BB}\s k} = \{\hat{d}_{\text{BB}\s m} \;|\; m \in \mathcal{M}_\text{BB}\,, L_{k-1} \leq \hat{d}_{\text{BB}\s m} \leq L_{k} \}$ with $ d_{\text{L}\s k} = k\, \frac{d_\text{max}}{N_\text{BB}}$ and $N_\text{BB} = 2\, N$. $d_\text{max}$ is the maximum observable distance from \mref{Sec.}{sec:signal_model} and the factor of $2$ is an empirically chosen constant for visualization. Additionally, we define the bin centers as $d_{\text{C}\s k}=(k+ \frac{1}{2})\, \frac{d_\text{max}}{N_\text{BB}}$. 
We visualize the bin-based statistics in terms of the squared \textit{Rayleigh scale parameter} as
\begin{equation}
\{\, (\, d_{\text{C}\s k},\, s^2_\text{ML}(\mathcal{U}_{\text{BB}\s k},\lambda)\,  ) ~~|~~ k \in \mathcal{N}_\text{BB} \, \}
\end{equation}
where $(\cdot,\,\cdot)$ denotes a couple and $s^2_\text{ML}$ is determined according to \eqref{eq:ml_tunc_rayleigh}. 
The squared \textit{Rician scale parameter} is visualized in the same manner as 
\vspace{-2mm}
\begin{equation}
\{\, (\, d_{\text{C}\s k},\, \bar{s}^2_\text{ML}(\mathcal{U}_{\text{BB}\s k},\lambda)\,  ) ~~|~~ k \in \mathcal{N}_\text{BB} \, \}
\end{equation}
where $\bar{s}^2_\text{ML}(\mathcal{X},\lambda) $ is the ML estimator for a set of i.i.d. samples $\mathcal{X} = \{x_1,...,x_{|\mathcal{X}|}\}$ following a truncated Rician distribution, given as $\bar{s}^2_\text{ML}(\mathcal{X},\lambda) = \argmax_{s} \prod_{x\in \mathcal{X}} f_\text{TRice}(x; u , s , \lambda)$, with $u \triangleq {1/2}\, s$ according to \eqref{eq:full_ampl_model_bias}. As there is no straightforward analytical solution to this optimization problem, we solve numerically using a grid search.
Finally, the \textit{relative frequency} is visualized as
\begin{equation}
 \Big\{\, \Big(\, d_{\text{C}\s k},\, \Big(\sum_{k^\prime \in \mathcal{N}_\text{BB}} \sum_{d \in \mathcal{D}_{\text{BB}\s k^\prime }} d \Big)^{-1} \sum_{d \in \mathcal{D}_{\text{BB}\s k}} d ~~\Big|~~ k \in \mathcal{N}_\text{BB} \, \Big\}\, .
\end{equation}

\begin{figure}[t]
	\centering
	\setlength{\figurewidth}{0.425\textwidthav}
	\setlength{\figureheight}{0.1\textwidthav}
	
	\includegraphics{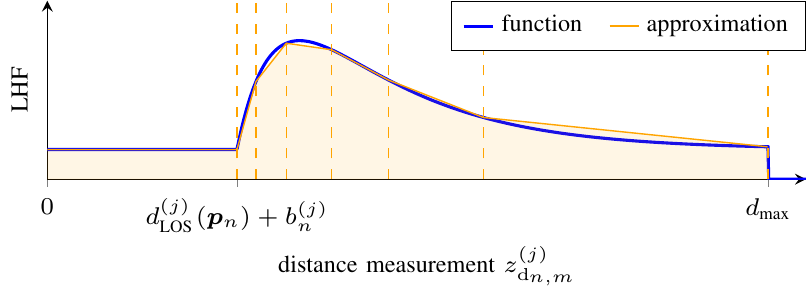}%
\caption{Trapezoid integral with non-uniformly spaced support points. We set $K_\text{T} = 7$ for demonstration purposes.}\label{fig:cluster_normalization}
\vspace{-3mm}
\end{figure}

\section{Channel Estimation and Detection Algorithm (CEDA)}  \label{sec:app_ceda}

We start by redefining the discrete-time specular signal vector \meqref{eq:signal_model_sampled} for notational convenience as
\begin{align}\label{eq:stack_recvgen}
 \RV{r}_n^{(j)} = \vm{S}(\tilde{\vm{\tau}}_n^{(j)})\tilde{\vm{\alpha}}_n ^{(j)}+ \RV{w}_n^{(j)},%
\end{align}
where $\tilde{\vm{\alpha}}_n^{(j)} = [\tilde{\alpha}_{n,0}^{(j)} \; ... \; \tilde{\alpha}_{n, \tilde{K}_n^{(j)}}^{(j)} ]^{\text{T}}$ are the complex amplitudes and $\tilde{\vm{{\tau}}}_n^{(j)} = [\tilde{\tau}_{n,0}^{(j)} \; ...  \; \tilde{\tau}_{n,\tilde{K}_n^{(j)}}^{(j)}   ]^{\text{T}}$ are the delays of all $\tilde{K}_n^{(j)} + 1$ signal components, including the \ac{los} component and $\tilde{K}_n^{(j)}$ \acp{mpc} and $\vm{S}(\tilde{\vm{\tau}}_{n}^{(j)}) = [\vm{s}({\tilde{\tau}}_{n,0}^{(j)}) \, ...  \, \vm{s}({\tilde{\tau}}_{\tilde{K}_n^{(j)} }^{(j)} )]$ is the signal matrix. Since the proposed \ac{ceda} operates independently on each radio signal snapshot, we omit the indices for time $n$ and anchor $j$ in the following for brevity of notation.

Using \eqref{eq:stack_recvgen} the model \ac{lhf} of a single signal snapshot can be written as  
\begin{align}
f(\vm{r};\vm{\tau},\vm{\alpha},\sigma^2) &= \frac{e^{-(\vm{r} - \vm{S}(\vm{\tau})\vm{\alpha})^\text{H} (\vm{r} - \vm{S}(\vm{\tau})\vm{\alpha})\sigma^{-2}}}{(\pi\sigma^2)^{N_{\text{s}}}}. \label{eq:likelihood}
\end{align}
Based on \eqref{eq:likelihood} we formulate a deterministic maximum likelihood (ML) estimator for delays of multiple components, with the complex amplitudes and the noise variance as nuisance parameters.
\begin{algorithm}[t!]\label{alg:a1}\footnotesize%
	\textbf{Initialization:}
	\begin{itemize}\setlength\itemsep{1mm}
		\item $m = 0$ and $\hat{\vm{\tau}}_0 = [\,]$\;
	\end{itemize}
	\textbf{Iterations:} \\
	\Do{$ u_{\text{ML}}(\vm{r}_{\mathrm{res}}) < \gamma$}
	{
		$m~\leftarrow~m + 1$\;
		\lIf{m=1}{set $\vm{r}_{\mathrm{res}} \leftarrow \vm{r}$}
	    \lElse{
	    	compute $\vm{r}_{\mathrm{res}} = \vm{r} -\vm{S}(\hat{\vm{\tau}}_{m-1})\s\hat{\vm{\alpha}}_{m-1}$}%
		add component using $\hat{\tau}_{m} = \argmax\limits_{\tau_m} \frac{|\vm{r}_{\mathrm{res}}^{\text{H}}\vm{s}(\tau_m)|^2}{\vm{s}(\tau_m)^{\text{H}}\vm{s}(\tau_m)}$\;
		$\hat{\vm{\tau}}_{m}~\leftarrow$ prepend $\hat{\tau}_{m}$ to $\hat{\vm{\tau}}_{m-1}$\;\vspace{0.7mm}
		compute $\hat{\sigma}^2 = \frac{1}{{N_{\text{s}}}-1} \norm{\vm{r}_{\mathrm{res}}}{2}$\;\vspace{0.5mm}
		compute $\hat{\V{\alpha}}_m$ using $\hat{\bm{\tau}}_{m}$ in \eqref{eq:alpha_hat}\;\vspace{0.5mm}
	}%
	\caption{Snapshot-based \ac{ceda}}
\end{algorithm}

Taking the natural logarithm of \eqref{eq:likelihood} enables formulating the maximization problem as \cite{Kay1993}
\begin{equation}
\{\hat{\vm{\tau}},\hat{\vm{\alpha}},\hat{\sigma}^2\} = \argmax_{\vm{\tau},\vm{\alpha},\sigma^2}\Big( \rmv \minus {N_{\text{s}}} \ln{\sigma^2} - \frac{\|\vm{r} - \vm{S}(\vm{\tau})\vm{\alpha}\|^2}{\sigma^2}\Big)\rmv\rmv\label{eq:log_likelihood}
\end{equation}
where a ``hat" denotes ML parameter estimates.
Taking the gradient with respect to $\vm{\alpha}$ we obtain a closed form solution %
given as \cite{ZiskindTASSP1988} \vspace{-0.5mm}
\begin{align}
\hat{\vm{\alpha}}%
= (\vm{S}(\vm{\tau})^{\text{H}}\vm{S}(\vm{\tau}))^{-1} \vm{S}(\vm{\tau})^{\text{H}}\vm{r}\label{eq:alpha_hat}
\end{align}
only depending on the delays.
Inserting \eqref{eq:alpha_hat} into \eqref{eq:log_likelihood} removes the amplitude dependency from the {(log-)likelihood}.
The maximization problem becomes
\begin{align}
\{\hat{\vm{\tau}},\hat{\sigma}^2\} &= \argmax_{\vm{\tau},\sigma^2}\big( -{N_{\text{s}}} \ln{\sigma^2} - \|\vm{r}\|^2 {\sigma}^{-2} \nonumber \\
&\quad+ \vm{r}^{\text{H}}\vm{S}(\vm{\tau})(\vm{S}(\vm{\tau})^{\text{H}}\vm{S}(\vm{\tau}))^{-1}\vm{S}(\vm{\tau}) ^{\text{H}}\vm{r} {\sigma}^{-2} \big). \label{eq:concentrated_log_likelihood}
\end{align}
To solve for  $\V{\tau}$ we simplify \eqref{eq:concentrated_log_likelihood}, by assuming the individual signal components to be uncorrelated \cite{Fleury1999}, i.e., $\bm{S}(\vm{\tau})^\text{H}\, \bm{S}(\vm{\tau}) = \mathrm{diag}\{[\norm{\V{s}(\tau_0)}{2}\, ... \, \norm{\V{s}(\tau_{K})}{2}]\}$. 
Thus, we can decompose the optimization problem with respect to $\V{\tau}$ into individual terms.
Following an expectation maximization scheme similar to \cite{Fleury1999}, we can solve the equation iteratively, in a bottom-up manner. The expectation term for iteration $m$ reads
\begin{equation}
\vm{r}_{\mathrm{res}} = \vm{r} -\vm{S}(\hat{\vm{\tau}}_{m-1})\,\hat{\vm{\alpha}}_{m-1}%
\end{equation}
and the maximization terms are
\begin{equation} \label{eq:ml_tau}
\hat{\tau}_m = \argmax_{\tau_m}  \frac{ |\vm{r}_{\mathrm{res}}^{\text{H}} \bm{s}(\tau_m)|^2}{\norm{\bm{s}(\tau_m)}{2} }
\end{equation}
and
\begin{equation}
 \hat{\sigma}^{2} = \frac{1}{{N_{\text{s}}}-1} \norm{\vm{r}_{\mathrm{res}}}{2} .
\end{equation}

We solve \eqref{eq:ml_tau} by successively performing grid-based optimization with the grid set to $T_\text{s} / 3$ and applying a continuous unconstrained optimizer \cite{Lagarias1998}.
Following \cite{NadlerTSP2011}, we search for components until the \ac{glrt} for a single signal component in noise, as given in \eqref{eq:ml_normalized_amplitude}, falls below the detection threshold $\gamma$, which is a constant to be chosen. Note that the maximum in \eqref{eq:ml_normalized_amplitude} is approximated using the current estimates $\hat{\tau}_m$ and $\hat{\sigma}$. See \cite{LeitingerAsilomar2020} on how to determine $\gamma$ out of a fixed value for the false alarm probability per signal snapshot.

An overview of the resulting algorithm is shown in Algorithm~\ref{alg:a1}, which represents a search-and-substract approach in the sense of \cite{RichterPhD2005}. Note that the presented scheduling is suboptimal with respect to the joint update of $\vm{\alpha}$ in \eqref{eq:alpha_hat} but offers the advantage of the  execution time being in the range of tens of milliseconds even with a large number of detected signal components.%

\section{Snapshot-Based \acl{crlb} (SP-CRLB) and Posterior \acs{crlb} (P-CRLB)}\label{sec:app_crlb}
\acused{pcrlb}
Here, we provide the expressions for what we refer to as the ``snapshot-based positioning CRLB (SP-CRLB)" and the corresponding posterior CRLB (P-CRLB) \cite{Tichavsky1998}, which are used as a performance benchmark in the main text \mref{Sec.}{sec:results}). They provide lower bounds on the \ac{rmse} of the position estimate, given as \cite{ShenTIT2010}
\begin{equation}
	e_n^{\text{RMSE}} \geq \sqrt{\text{tr}\{  \bm{J}_{\bm{p}\s \text{S} \s  n}^{-1}  \}} \geq \sqrt{\text{tr}\{  \bm{J}_{\bm{p}\s \text{P} \s n}^{-1}  \}}  %
\end{equation}
with $\bm{J}_{\bm{p}\s \text{S} \s  n}$ and  $\bm{J}_{\bm{p}\s \text{P} \s  n}$ being the respective Fisher information matrices.
In particular, the SP-CRLB considers the information contained in the signal waveforms recorded by all $J$ anchors at a \textit{single} time step $n$. %
We use the results from \cite[Eq. 14]{WitrisalJWCOML2016} as the model used fits our signal model in \meqref{eq:signal_model_sampled_stochastic}. We get
\vspace{-2mm}
\begin{equation} \vspace{-1mm} \label{eq:spcrlb}
  \bm{J}_{\bm{p} \s \text{S}\s n} = \frac{8 \pi^2 \beta_\text{bw}^2}{c^2} \sum_{j=1}^{J}  \tilde{u}_{n}^{(j)\s 2} \V{D}_{\text{r} \s n}^{(j)} 1_{\mathbb{V}_n^{(j)}}
\end{equation}
where $\V{D}_{\text{r} \s n}^{(j)} = [\cos(\tilde{\phi}_n^{(j)})\, \sin(\tilde{\phi}_n^{(j)})]\, [\cos(\tilde{\phi}_n^{(j)})\,  \sin(\tilde{\phi}_n^{(j)})]^\text{T} $ is the ranging direction matrix \cite{ShenTIT2010}, with the (true) angle of arrival $\tilde{\phi}_n^{(j)} = \mathrm{atan2}(p_{\text{Ax}}^{(j)}-\tilde{p}_{\text{x}\s n}, p_{\text{Ay}}^{(j)}-\tilde{p}_{\text{y}\s n} )$, and $\mathbb{V}_n^{(j)}$ is the set containing all time step indices $n$ %
where the \ac{los} component is visible. $c$, $\beta_\text{bw}$, $\tilde{u}_{n}^{(j)}$, $p_{\text{Ax}}^{(j)}$, $\tilde{p}_{\text{x}\s n}$, $p_{\text{Ay}}^{(j)}$, $\tilde{p}_{\text{y}\s n}$ and $1_{\mathbb{V}_n^{(j)}}$ are defined in accordance to the main text (see \mref{Sec.}{sec:signal_model} and \mref{Sec.}{sec:system_model}). 
The P-CRLB additionally considers the information provided by the state transition model of the {agent state} $\RV{x}_n$. Following \cite[Sec. III]{Tichavsky1998}, we get
\begin{equation} \label{eq:pcrlb}
	\bm{J}_{\bm{p}\s \text{P} \s n} = ( \bm{A} \, \bm{J}_{\bm{p}\s \text{P} \s n\minus 1}^{-1} \, \bm{A}^\text{T} \, + \, {\sigma_{\text{a}}^2} \, \bm{B} \, \bm{B}^\text{T} )^{-1} + \bm{J}_{\bm{p}\s \text{S} \s n}
\end{equation}
which is a recursive equation corresponding to the covariance update equations of the Kalman filter \cite{ArulampalamTSP2002}. $\bm{A}$, $\bm{B}$ and ${\sigma_{\text{a}}}$ are defined in accordance to the main text (see \mref{Sec.}{sec:simulation_model}). Since we initialize the {agent state} $\RV{x}_n$ using an initial measurement $\bm{z}_0$ (see \mref{Sec.}{sec:init}), we accordingly calculate $\bm{J}_{\bm{p}\s \text{P} \s 0}$ using \eqref{eq:spcrlb} with the corresponding true values $\tilde{u}_{0}^{(j)\s 2}$ and $\tilde{\bm{p}}_0$.

While \eqref{eq:spcrlb} considers the (inevitable asymptotic) loss of SNR related to the stochastic process  $\rv{\nu}_{\text{D}\s n}^{(j)}(\tau)$ in \meqref{eq:signal_model_sampled_stochastic}, it does {not model} the additional information provided by coupling the \acp{mpc} with the \ac{los} object via the NLOS bias $\tilde{b}_n^{(j)}$ in \meqref{eq:delta_fun}. This allows the RMSE of the proposed algorithm to \textit{fall below} the provided CRLB, demonstrating the additional information leveraged using the proposed NLOS model. However, in contrast to mapping approaches \cite{LeitingerTWC2019,GentnerTWC2016,KimTWC2020}, which can facilitate multipath information via estimated map features (virtual anchors \cite{PedersenJTAP2018,Meissner2015Diss}), our model just allows to mitigate the NLOS bias between MPC-related distance measurements and the LOS component distance. Thus, a strict lower bound can be obtained by assuming the LOS component to be available at all times $n$, i.e., setting $1_{\mathbb{V}_n^{(j)}}\vspace{0.23mm} \triangleq 1$ in \eqref{eq:spcrlb}. We refer to the corresponding P-CRLB as P-CRLB-LOS in the main text.

\acrodef{mimo}[MIMO]{multiple input multiple output}
\acrodef{awgn}[AWGN]{additive white Gaussian noise}
\acrodef{crlb}[CRLB]{Cram\'er-Rao lower bound}
\acrodef{dmc}[DMC]{dense multipath component}
\acrodef{los}[LOS]{line-of-sight}
\acrodef{ml}[ML]{maximum likelihood}
\acrodef{mpc}[MPC]{multipath component}
\acrodef{nlos}[NLOS]{non-\acl{los}}
\acrodef{pdf}[PDF]{probability density function}
\acrodef{smc}[SMC]{specular multipath component}
\acrodef{snr}[SNR]{signal-to-noise-ratio}
\acrodef{sinr}[SINR]{signal-to-interference-plus-noise-ratio}
\acrodef{tdoa}[TDOA]{time difference of arrival}
\acrodef{toa}[TOA]{time-of-arrival}
\acrodef{aoa}[AOA]{angle-of-arrival}
\acrodef{uwb}[UWB]{ultra-wideband}
\acrodef{mse}[MSE]{mean squared error}
\acrodef{dps}[DPS]{delay power spectrum}
\acrodef{glrt}[GLRT]{generalized likelihood ratio test}
\acrodef{mse}[MSE]{mean squared error}
\acrodef{rmse}[RMSE]{root mean squared error}
\acrodef{nnlike}[NNLIKE]{normalized noise-free likelihood}
\acrodef{stdv}[STDV]{standard deviation}
\acrodef{rv}[RV]{random variable}
\acrodef{pda}[PDA]{probabilistic data association}
\acrodef{pmf}[PMF]{probability mass function}
\acrodef{pdaf}[PDAF]{probabilistic data association filter}
\acrodef{pdaai}[AIPDA]{amplitude-information probabilistic data association}
\acrodef{olos}[OLOS]{obstructed \ac{los}}
\acrodef{spa}[SPA]{sum-product algorithm}
\acrodef{mmse}[MMSE]{minimum mean-squared error}
\acrodef{lhf}[LHF]{likelihood function}
\acrodef{fa}[FA]{false alarm}
\acrodef{ceda}[CEDA]{channel estimation and detection algorithm} 
\acrodef{pcrlb}[P-CRLB]{posterior Cram\'er-Rao lower bound}
\acrodef{slam}[SLAM]{simultaneous localization and mapping}
\acrodef{mpslam}[MP-SLAM]{multipath-based SLAM}
\acrodef{dnr}[DNR]{dense-multipath-to-noise ratio}
\acrodef{stv}[STV]{state-transition variances}
\acrodef{npe}[BBE]{bin-based estimate}

\bibliographystyle{IEEEtran}
\bibliography{IEEEabrv,References,TempRefs}